\documentclass[aps,prb,twocolumn,showpacs,10pt,floatfix,superscriptaddress,floatfix]{revtex4-1}
\usepackage{epsf}
\usepackage{bm}
\usepackage{hyperref}
\usepackage{amsfonts}
\usepackage{amssymb}
\usepackage{amsmath,mathtools}
\usepackage{array}
\usepackage{enumerate,dsfont}
\usepackage{dcolumn,multirow}
\usepackage{bbm}
\usepackage[utf8]{inputenc}
\usepackage{latexsym}

\newcommand{\vk}{\vec k}

\newcommand{\vb}{\vec b}

\newcommand{\vc}{\vec c}

\newcommand{\vn}{\vec n}

\renewcommand{\vr}{\vec r}

\newcommand{\vB}{\vec B}
\newcommand{\vC}{\vec C}

\newcommand{\ZZ}{\mathbb{Z}}

\newcommand{\vGamma}{\mbox{\boldmath $\Gamma$}}

\newcommand{\ket}[1]{{| #1 \rangle}}
\newcommand{\bra}[1]{{\langle #1 |}}

\newcommand{\RR}{\mathbbm{R}}
\newcommand{\CC}{\mathbbm{C}}

\newcommand{\be}{\begin{equation}}
\newcommand{\ee}{\end{equation}}

\renewcommand{\vec}[1]{\mathbf{#1}}

\begin{document}

\title{Second-order topological insulators and superconductors with an order-two crystalline symmetry}

\author{Max Geier}
\affiliation{Dahlem Center for Complex Quantum Systems and Physics Department, Freie Universit\"at Berlin, Arnimallee 14, 14195 Berlin, Germany}
\author{Luka Trifunovic}
\affiliation{Dahlem Center for Complex Quantum Systems and Physics Department, Freie Universit\"at Berlin, Arnimallee 14, 14195 Berlin, Germany}
\author{Max Hoskam}
\affiliation{Dahlem Center for Complex Quantum Systems and Physics Department, Freie Universit\"at Berlin, Arnimallee 14, 14195 Berlin, Germany}
\affiliation{Department of Applied Physics, Eindhoven University of Technology, 5600 MB Eindhoven, The Netherlands}
\author{Piet W. Brouwer}
\affiliation{Dahlem Center for Complex Quantum Systems and Physics Department, Freie Universit\"at Berlin, Arnimallee 14, 14195 Berlin, Germany}

\date{\today}

\begin{abstract}
Second-order topological insulators and superconductors have a gapped
excitation spectrum in bulk and along boundaries, but protected zero modes at
corners of a two-dimensional crystal or protected gapless modes at hinges of a
three-dimensional crystal. A second-order topological phase can be induced by
the presence of a bulk crystalline symmetry. Building on Shiozaki and Sato's
complete classification of bulk crystalline phases with an order-two
crystalline symmetry [Phys.\ Rev.\ B {\bf 90}, 165114 (2014)], 
such as mirror reflection, twofold rotation, or inversion
symmetry, we classify all corresponding second-order topological insulators and
superconductors. The classification also includes antiunitary symmetries and
antisymmetries.
\end{abstract}
\maketitle

\section{Introduction}

In comparison to conventional ``first-order'' topological insulators and superconductors, which combine a gapped bulk with topologically protected gapless boundary states,\cite{hasan2010,bernevig2013,qi2011} the protected gapless states in a second-order topological insulator or superconductor exist in one dimension lower:\cite{schindler2018} A two-dimensional second-order topological insulator or superconductor has zero-energy states at corners of the crystal\cite{benalcazar2014,benalcazar2017,benalcazar2017b,peng2017,langbehn2017} and a three-dimensional topological insulator or superconductor has gapless modes along crystal edges or ``hinges''.\cite{sitte2012,langbehn2017,schindler2018,zhang2013,volovik2010,song2017,fang2018} Second-order topological insulator and superconductor phases have been proposed to exist in a (first-order) topological insulator in three dimensions to which a suitable time-reversal-breaking perturbation is applied,\cite{sitte2012,zhang2013} in the superfluid $^3$He-B phase,\cite{volovik2010} or in crystals with rotation or mirror symmetries.\cite{benalcazar2014,benalcazar2017,benalcazar2017b,peng2017,langbehn2017,schindler2018,song2017,fang2018} 

A complete classification of first-order topological insulators and superconductors has been developed, accounting for the presence or absence of non-spatial symmetries.\cite{schnyder2008,schnyder2009,kitaev2009} The three fundamental non-spatial symmetry operations time-reversal ${\cal T}$, particle-hole ${\cal P}$, and ${\cal C} = {\cal PT}$, known as ``chiral symmetry'', define the ten Altland-Zirnbauer symmetry classes,\cite{altland1997} see Table \ref{tab:AZ}. For each Altland-Zirnbauer class, the number and type of protected boundary states is uniquely rooted in the topology of the bulk band structure, so that topological classifications of gapped bulk band structure and gapless boundary states are essentially identical, a feature known as ``bulk-boundary correspondence''. Complete classifications for all Altland-Zirnbauer classes with additional spatial symmetries exist only for the order-two crystalline symmetries,\cite{shiozaki2014} such as mirror symmetry,\cite{chiu2013,morimoto2013,trifunovic2017} order-two rotation symmetry, inversion symmetry,\cite{lu2014} and non-symmorphic order-two crystalline symmetries.\cite{shiozaki2016} In parallel, a wealth of symmetry-based indicators has been identified for topological phases with other crystalline symmetries.\cite{fang2012,fang2013,slager2013,jadaun2013,liu2014b,alexandradinata2014,dong2016,kruthoff2017,po2017,bradlyn2017,shiozaki2017,khalaf2018} With crystalline symmetries, the bulk-boundary correspondence --- {\em i.e.}, the one-to-one correspondence between bulk topology and the number and type of gapless boundary states --- only applies to boundaries which are invariant with respect to the crystalline symmetry operation; non-symmetric boundaries are generically gapped.

In this article we consider the classification problem for second-order topological insulators. We identify the type and number of zero-energy states at corners or gapless modes at hinges and relate this classification of corner states and hinge modes to the topology of the bulk band structure. This program is carried out for all ten Altland-Zirnbauer classes with one additional order-two spatial symmetry, for which the classification of the bulk band structure is known.\cite{shiozaki2014} 

In contrast to first-order topological insulators, for which the number and
type of protected boundary states depends on the topology of the bulk band
structure only, the occurrence of zero-energy corner states or gapless hinge
modes may also depend on properties of the boundary, {\em i.e.}, on the lattice
termination. Correspondingly, the classification of corner states and hinge
modes of second-order topological insulators and superconductors has to
distinguish between termination-dependent and termination-independent
properties of corner states and hinge modes. This naturally leads to an
``intrinsic'' topological classification, in which crystals that differ by a
lattice termination only are considered topologically equivalent, and an
``extrinsic'' classification, which accounts for termination effects and
defines topological equivalence with respect to continuous transformations that
preserve both bulk and boundary gaps. 

\begin{figure}
\includegraphics[width=0.8\columnwidth]{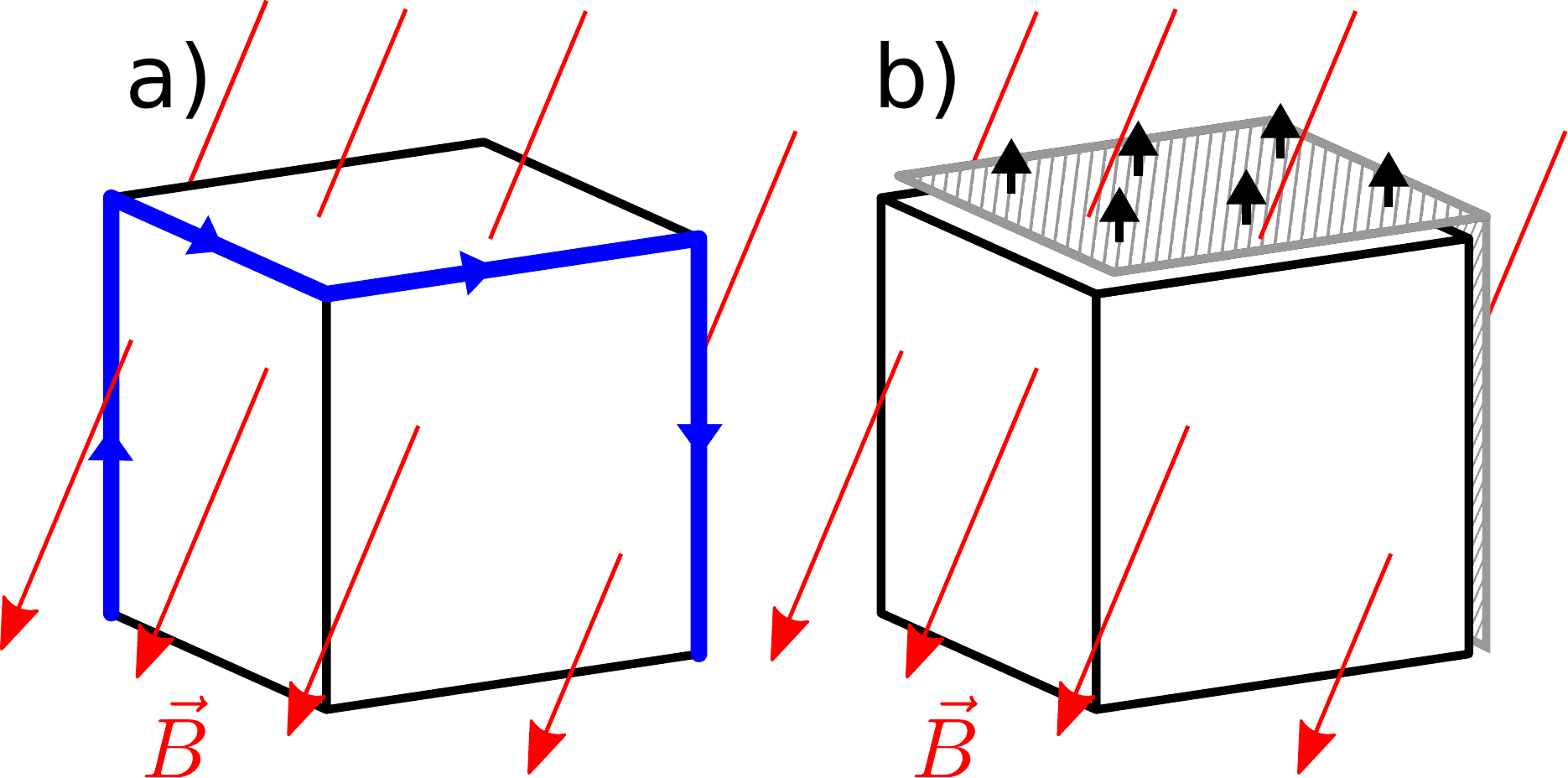}
\caption{\label{fig:qhe} Schematic picture of an ``extrinsic'' second-order
topological insulator consisting of a three-dimensional topological insulator placed in a
magnetic field in a generic direction, as proposed by Sitte {\em et al.}\cite{sitte2012} (a). Each surface has a finite flux and there
are chiral modes along hinges that touch two faces with opposite sign of the
magnetic flux. The gapless hinge modes may be removed by exchange-coupling some of the crystal faces to a two-dimensional ferromagnetic insulator (b).}
\end{figure}

An example of an ``extrinsic'' second-order topological insulator is a
three-dimensional topological insulator (without further crystalline
symmetries) placed in a magnetic field in a generic direction, such that there
is a finite magnetic flux through all surfaces,\cite{sitte2012,zhang2013} see
Fig.~\ref{fig:qhe}. Such a crystal has chiral modes along hinges that connect
faces with an inward and outward-pointing magnetic fluxes. The chiral modes are
stable with respect to continuous transformation of the Hamiltonian that
preserve bulk and surface gaps. They may be removed, however, upon exchange
coupling the crystal faces with an inward magnetic flux to ferromagnetic
insulating films, with a magnetization direction chosen such that the exchange
field reverses the effect of the applied magnetic field. 

An ``intrinsic'' second-order topological insulator or superconductor, for
which the presence of corner or hinge states does not depend on the lattice
termination, requires the presence of additional crystalline symmetries.
Examples that have been identified in the literature include mirror-reflection
symmetry,\cite{schindler2018,langbehn2017} rotation
symmetries,\cite{schindler2018,song2017,fang2018} or more general point group
symmetries.\cite{benalcazar2017,benalcazar2017b,khalaf2018} In these cases corner states
continue to exist under continuous transformations of the Hamiltonian that
close the boundary gap, provided the bulk gap is not closed and the lattice
termination remains compatible with the crystalline symmetry.

\begin{figure}
\includegraphics[width=0.95\columnwidth]{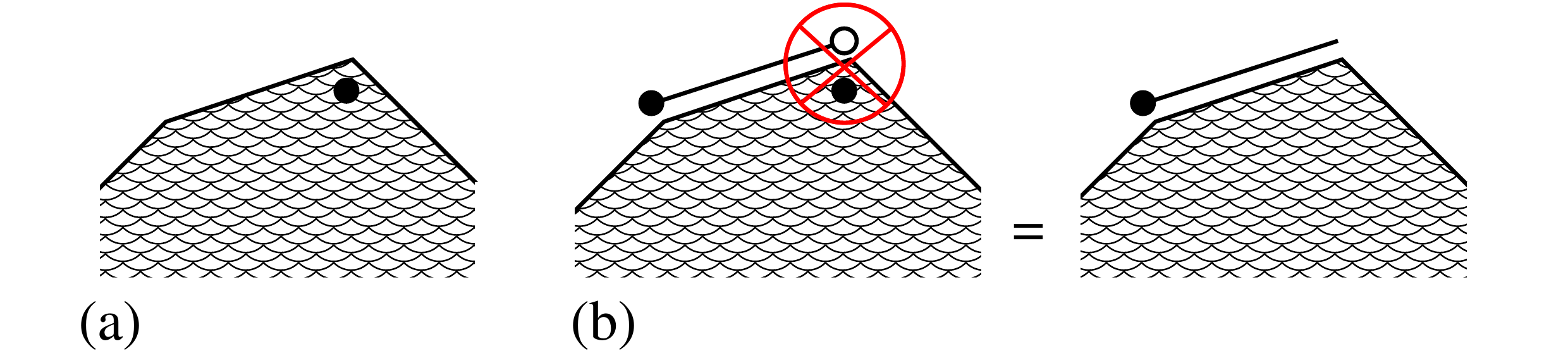}
\caption{\label{fig:generic} (a) Schematic picture of a generic corner of a two-dimensional crystal. A generic corner may host a protected zero-energy state if and only if the corresponding Altland-Zirnbauer class in $d-1$ dimensions is nontrivial. (b) Zero-energy corner states in a generic corner can always be moved to a different corner by a suitable change of the lattice termination. For the example shown here, a one-dimensional topological insulator or superconductor with two end states is ``glued'' to one of the crystal faces adjacent to the top corner, such that its end state and the original zero-energy corner state mutually gap out. As a result, the corner state has moved to the corner on the left.}
\end{figure}

In the presence of a crystalline symmetry, a classification of corner states
and hinge modes must also distinguish between corners and hinges that are
themselves invariant with respect to the crystalline symmetry, and generic
non-symmetric corners or hinges. The classification of zero-energy states and
gapless modes at a generic, non-symmetric corner or hinge (schematically shown
in Fig.\ \ref{fig:generic}a) equals that of a generic codimension-one defect,
which is the same as the regular classification of topological phases but with
the dimension shifted by one,\cite{teo2010} see Table \ref{tab:AZ}. This simple
result also follows from the observation that the absence of gapless boundary
states implies that the bulk is essentially topologically trivial, so that a
corner or hinge may be seen as a junction between two ``stand-alone''
topological edges or surfaces.\cite{langbehn2017} Note that this classification
of corner states or hinge modes at a generic corner or hinge is an {\em
extrinsic} classification: Any corner state or hinge mode at a generic corner
or hinge can be moved away from that corner or hinge by a suitable change of
the crystal boundary, without affecting the bulk, see Fig.\
\ref{fig:generic}b.\cite{schindler2018,khalaf2018} 
Hence, the {\em intrinsic} classification of corner states or hinge modes at a
generic corner or hinge is always trivial. 

A classification of zero-energy states and gapless modes at mirror-symmetric corners and hinges is given in Sec.\ \ref{sec:4}. 
In addition to providing the intrinsic (termination-independent) and extrinsic
(termination-dependent) classifications, we consider the effect of
perturbations that locally break mirror symmetry at corners and hinges, to
account for the experimental reality that corners and
hinges are more prone to defects and disorder than crystal faces.  The
intrinsic classification of zero-energy states and gapless modes at
mirror-symmetric corners and hinges coincides with the classification of bulk
topological crystalline phases in two and three
dimensions,\cite{chiu2013,morimoto2013,shiozaki2014,trifunovic2017}
respectively, after removal of the first-order topological phases.  This
``corner-to-bulk correspondence'' (or ``hinge-to-bulk correspondence, for
three-dimensional topological crystalline insulators and superconductors) not
only confirms that every topological class of the bulk band structure is
associated with a unique configuration of zero-energy corner states or gapless
hinge modes, but also that for every possible configuration of mirror-symmetric
zero-energy corner states or hinge modes, there is a topological crystalline
phase that produces it. 

\begin{table}
  \begin{tabular*}{\columnwidth}{c @{\extracolsep{\fill}} ccc}
    \hline\hline
    Cartan & (anti)symmetries & $d=2$ & $d=3$ \\ \hline
    A & - & 0 & $\ZZ$ \\
    AIII & ${\cal C}$ & $\ZZ$ & 0 \\ \hline
    AI & ${\cal T}^+$ & 0 & 0 \\
    BDI & ${\cal T}^+$, ${\cal P}^+$ & $\ZZ$ & 0 \\
    D & ${\cal P}^+$ & $\ZZ_2$ & $\ZZ$ \\
    DIII & ${\cal T}^-$, ${\cal P}^+$ & $\ZZ_2$ & $\ZZ_2$ \\
    AII & ${\cal T}^-$ & 0 & $\ZZ_2$ \\
    CII & ${\cal T}^-$, ${\cal P}^-$ & $2\ZZ$ & 0 \\
    C & ${\cal P}^-$ & 0 & $2\ZZ$ \\
    CI & ${\cal T}^+$, ${\cal P}^-$ & 0 & 0 \\ \hline \hline
  \end{tabular*}
\caption{\label{tab:AZ} The ten Altland-Zirnbauer classes are defined according to the presence or absence of time-reversal symmetry (${\cal T}$), particle-hole antisymmetry (${\cal P}$), and chiral antisymmetry (${\cal C}$). The superscript $\pm$ indicates the square of the time-reversal or particle-hole conjugation operation. The presence of chiral antisymmetry ${\cal C} = {\cal P T}$ is automatic for Altland-Zirnbauer classes with both time-reversal symmetry and particle-hole antisymmetry. The third and fourth column give the classification of stable zero-energy states at generic corners of two-dimensional crystals $(d=2)$ or hinges of three-dimensional crystals ($d=3$). This is an ``extrinsic'' classification, in the sense that the number of corner states or hinge modes at a generic corner or hinge is not a bulk property and can be changed by a change of the lattice termination. Its topological protection is with respect to all continuous transformations that preserve both bulk and boundary gaps.}
\end{table}

With rotation or inversion symmetry there are no symmetry-invariant corners or
hinges for two- and three-dimensional crystals, respectively. Hence, each corner or hinge in a crystal with rotation symmetry or
inversion symmetry is a ``generic'' corner or hinge, described by the extrinsic
classification of Table~\ref{tab:AZ}. Zero-energy corner states or gapless hinge
modes at a given corner or hinge can always be removed by changing the lattice
termination. Nevertheless, as we show in Sec.\ \ref{sec:5}, the role of the bulk crystalline symmetry, combined with the
requirement that lattice termination is symmetry-compatible, is to impose a $\ZZ_2$ {\em
sum rule} to the total number of corner or hinge states, which is an odd
multiple of two for the nontrivial phases and an even multiple of two otherwise.
(For Altland-Zirnbauer classes with a time-reversal or particle-hole symmetry
squaring to $-1$ one should count pairs of corner states/hinge modes.)

In Refs.\ \onlinecite{schindler2018} and \onlinecite{langbehn2017} is the 
construction of a nontrivial intrinsic second-order phase out of a nontrivial bulk
mirror-reflection-symmetric phase made use of the bulk-boundary correspondence,
according to which a nontrivial topological crystalline bulk phase implies a
gapless boundary mode at a boundary that is
left invariant under mirror reflection. The existence of protected corner states
or hinge modes was then concluded upon noting that mass terms that are generated
upon tilting the boundary away from the mirror-invariant direction
have a different sign at mirror-related boundaries, such that a corner
separating mirror-related boundaries represents a domain wall and, hence, hosts
a zero-energy state or a gapless hinge mode. The same procedure can be applied
to a three-dimensional crystal with a twofold rotation symmetry, because these,
too, allow for symmetry-invariant faces. It fails, however, for a
two-dimensional crystal with twofold rotation symmetry or a three-dimensional
crystal with inversion symmetry, because these have no symmetry-invariant
surface. To derive the existence of a second-order topological phase
with zero-energy corner states in a two-dimensional crystal with twofold rotation symmetry or of gapless hinge modes in a three-dimensional crystal with inversion symmetry, we employ a dimensional reduction scheme, making use of the existence of symmetry-invariant faces for
crystals with the same order-two crystalline symmetry in one dimension higher.
Our results are consistent with nontrivial second-order topological phases
predicted recently by Fang and Fu\cite{fang2018} and by Khalaf {\em et
al.}\cite{khalaf2018} for three-dimensional inversion-symmetric crystals.

This article is organized as follows: In Sec.~\ref{sec:2} we introduce the
relevant symmetry classes for an order-two crystalline symmetry 
coexisting with time-reversal symmetry, particle-hole symmetry, or chiral
symmetry and we review Shiozaki and Sato's classification of the
crystalline bulk phases. The dimensional reduction map
is outlined in Sec.~\ref{sec:3}. A classification of mirror-symmetric corners
and hinges follows in Sec.~\ref{sec:4}; Section~\ref{sec:5} discusses twofold
rotation and inversion symmetry. A few representative examples of tight-binding models realizing
second-order topological phases are discussed in
Sec.~\ref{sec:6}. We conclude in Sec.~\ref{sec:7}. The appendices contain a
detailed discussion of the dimensional reduction scheme as well as a brief
discussion of all relevant crystalline symmetry classes that are not considered in the main text.

\section{Shiozaki-Sato symmetry classes}
\label{sec:2}

We consider a Hamiltonian $H_d(\vk)$ in $d$ dimensions, with $\vk =
(k_1,k_2,\ldots,k_d)$. In addition to the crystalline order-two symmetry, to be
discussed in detail below, the Hamiltonian $H_d$ possibly satisfies a
combination of time-reversal (${\cal T}$) symmetry, particle-hole (${\cal P}$)
antisymmetry, and/or chiral (${\cal C}$) antisymmetries.\footnote{Although
${\cal P}$ and and ${\cal C}$ are commonly referred to as ``particle-hole
symmetry'' and ``chiral symmetry'', we will refer to these as antisymmetries,
because they connect $H$ to $-H$, see Eq.~(\ref{eq:chiralsymmetry}).} These take
the form\cite{chiu2016}
\begin{align}
  H_d(\vk) &= U_{\cal T}^{\dagger} H_d(-\vk)^* U_{\cal T}, \nonumber \\
  \label{eq:chiralsymmetry}
  H_d(\vk) &= - U_{\cal P}^{\dagger} H_d(-\vk)^* U_{\cal P}, \\
  H_d(\vk) &= - U_{\cal C}^{\dagger} H_d(\vk) U_{\cal C}, \nonumber
\end{align}
where $U_{\cal T}$, $U_{\cal P}$, and $U_{\cal C}$ are $\vk$-independent unitary
matrices. If time-reversal symmetry and particle-hole symmetry are both
present, $U_{\cal C} = U_{\cal P} U_{\cal T}^*$. Further, the unitary matrices
$U_{\cal T}$, $U_{\cal P}$, and $U_{\cal C}$ satisfy $U_{\cal T} U_{\cal T}^* =
{\cal T}^2$ and $U_{\cal P} U_{\cal P}^* = {\cal P}^2$ and we require that
$U_{\cal C}^2= {\cal C}^2 = 1$ and $U_{\cal P} U_{\cal T}^* = {\cal T}^2 {\cal
P}^2 U_{\cal T} U_{\cal P}^*$. Throughout we use the symbols ${\cal T}^{\pm}$
and ${\cal P}^{\pm}$ to refer to a time-reversal symmetry or particle-hole
antisymmetry squaring to one ($+$) or minus one ($-$).  The ten
Altland-Zirnbauer classes defined by the presence or absence of three
non-spatial symmetry operations ${\cal T}$, ${\cal P}$, and ${\cal C}$ are
separated in two ``complex'' classes, which do not have antiunitary symmetries
or antisymmetries, and eight ``real'' classes, which have at least one
antiunitary symmetry or antisymmetry. Following common practice in the field, we
use Cartan labels to refer to the ten Altland-Zirnbauer symmetry classes, see
Table~\ref{tab:AZ}.

In addition to the non-spatial (anti)symmetries ${\cal T}$, ${\cal P}$, and
${\cal C}$, the Hamiltonian $H_d(\vk)$ satisfies an order-two crystalline
symmetry or antisymmetry. The ``spatial type'' of the symmetry operation is
determined by number $d_{\parallel}$ of spatial degrees of freedom that are
inverted: Mirror reflections have $d_{\parallel} = 1$, twofold rotations have
$d_{\parallel} = 2$, and inversion has $d_{\parallel} = 3$. (In two dimensions,
the spatial operations of inversion and twofold rotation are formally
identical. We will refer to this operation as a twofold rotation.) We will use
the symbol ${\cal S}$ to denote a general unitary order-two crystalline
symmetry, replacing ${\cal S}$ by ${\cal M}$, ${\cal R}$, or ${\cal I}$ for
considerations that apply specifically to mirror reflection, twofold rotation,
or inversion symmetry, respectively. For a general antiunitary symmetry,
antiunitary antisymmetry, and unitary antisymmetry we use the composite symbols
${\cal T}^{\pm} {\cal S}$, ${\cal P}^{\pm} {\cal S}$ and ${\cal CS}$,
respectively, again replacing ${\cal S}$ by ${\cal M}$, ${\cal R}$, ${\cal I}$
when appropriate. Without loss of generality we may require that the symmetry
operation ${\cal S}$ squares to one.\footnote{For spin $1/2$ electrons often
spatial symmetries squaring to $-1$ are used. Multiplication by $i$ then gives
a symmetry operation squaring to $1$. Note, however, that multiplication with
$i$ turns a symmetry that commutes with ${\cal T}$ or ${\cal P}$ into a
symmetry that anticommutes with ${\cal T}$ or ${\cal P}$ and vice versa.}
Following Refs.\ \onlinecite{chiu2013,morimoto2013,shiozaki2014}, to further
characterize the (anti)symmetry operation, we specify the signs $\eta_{{\cal
T},{\cal P},{\cal C}}$ indicating whether it commutes ($\eta = +$) or
anticommutes ($\eta=-$) with time-reversal ${\cal T}$, particle-hole
conjugation ${\cal P}$, or the chiral operation ${\cal C}$. 

Unitary symmetry and antisymmetry operations ${\cal S}$ and ${\cal CS}$ are represented by unitary matrices $U_{\cal S}$ and $U_{\cal CS}$ (with ${\cal S}$ being replaced by ${\cal M}$, ${\cal R}$, or ${\cal I}$ as needed), respectively, such that
\begin{align}
  \lefteqn{H_d(k_1,\ldots,k_{d_{\parallel}},k_{d_{\parallel}+1},\ldots,k_d)} &
  \nonumber \\ & =
    U_{\cal S} H_d(-k_1,\ldots,-k_{d_{\parallel}},k_{d_{\parallel}+1},\ldots,k_d) U_{\cal S}^{-1},
  \label{eq:Usym}
\end{align}
if $H_d$ satisfies a unitary symmetry, and
\begin{align}
  \lefteqn{H_d(k_1,\ldots,k_{d_{\parallel}},k_{d_{\parallel}+1},\ldots,k_d)} &
  \nonumber \\ & =
    - U_{\cal CS} H_d(-k_1,\ldots,-k_{d_{\parallel}},k_{d_{\parallel}+1},\ldots,k_d) U_{\cal CS}^{-1},
\end{align}
if $H_d$ satisfies a unitary antisymmetry. The matrices $U_{\cal S}$ and $U_{\cal CS}$ satisfy $U_{\mathcal{S,CS}}^2 = 1$, $U_{\mathcal{S,CS}} U_{\cal T} = \eta_{\cal T} U_{\cal T} U_{\mathcal{S,CS}}^*$, $U_{\mathcal{S,CS}} U_{\cal P} = \eta_{\cal P} U_{\cal P} U_{\mathcal{S,CS}}^*$, and $U_{\mathcal{S,CS}} U_{\cal C} = \eta_{\cal C} U_{\cal C} U_{\mathcal{S,CS}}$. Similarly, antiunitary symmetry and antisymmetry operations ${\cal T}^{\pm} {\cal S}$ and ${\cal P}^{\pm} {\cal S}$ are represented by unitary matrices $U_{\cal TS}$ and $U_{\cal PS}$, with
\begin{align}
  \lefteqn{H_d(k_1,\ldots,k_{d_{\parallel}},k_{d_{\parallel}+1},\ldots,k_d)} &
  \nonumber \\ &=
  U_{\cal TS} H_d(k_1,\ldots,k_{d_{\parallel}},-k_{d_{\parallel}+1},\ldots,-k_d)^* U_{\cal TS}^{-1},
\end{align}
if $H_d$ satisfies an antiunitary symmetry, and
\begin{align}
  \lefteqn{H_d(k_1,\ldots,k_{d_{\parallel}},k_{d_{\parallel}+1},\ldots,k_d)} &
  \nonumber \\ &=
  - U_{\cal PS} H_d(k_1,\ldots,k_{d_{\parallel}},-k_{d_{\parallel}+1},\ldots,-k_d)^* U_{\cal PS}^{-1},
  \label{eq:Aantisym}
\end{align}
if $H_d$ satisfies an antiunitary antisymmetry. The matrices $U_{\cal TS}$ and $U_{\cal PS}$ satisfy the conditions $U_{\mathcal{TS,PS}}^2 = \pm 1$, $U_{\mathcal{TS,PS}} U_{\cal T}^* = \eta_{\cal T} U_{\cal T} U_{\mathcal{TS,PS}}^*$, $U_{\mathcal{TS,PS}} U_{\cal P}^* = \eta_{\cal P} U_{\cal P} U_{\mathcal{TS,PS}}^*$, and $U_{\mathcal{TS,PS}} U_{\cal C}^* = \eta_{\cal C} U_{\cal C} U_{\mathcal{TS,PS}}$.

\begin{table}[t]
\begin{tabular*}{\columnwidth}{l @{\extracolsep{\fill}} ccccc}
\hline\hline 
AZ class & $s$ & $t$ &  symmetry operations  & representative \\ \hline
A & $0$ & $0$ & $\,_+U$ & ${\cal S}$ \\
AIII & $1$ & $0$ & $\,_\alpha U_+$ & ${\cal S}_+$ \\ \hline
A & $0$ & $1$ & $\,_-U$ & ${\cal CS}$ \\
AIII & $1$ & $1$ & $\,_\alpha U_-$ & ${\cal S}_-$ \\ \hline\hline
\end{tabular*}
\caption{Shiozaki-Sato equivalence classes of unitary symmetry and antisymmetry operations for the Altland-Zirnbauer classes A and AIII. The symbol $\,_{\sigma}U_{\eta_{\cal C}}$ is used to denote unitary symmetry ($\sigma=+$) and antisymmetry ($\sigma=-$) operations that commute ($\eta_{{\cal C}} = +$) or anticommute ($\eta_{{\cal C}}=-$) with the chiral symmetry, if applicable. The last column lists a unitary crystalline symmetry ${\cal S}_{\eta_{\cal C}}$ or the product of a unitary symmetry operation ${\cal S}$ and the chiral operation ${\cal C}$ as a crystalline symmetry operation representative of the Shiozaki-Sato class $(s,t)$.\label{tab:1}}
\end{table}

As pointed out in Ref.~\onlinecite{shiozaki2014}, the characterization of
unitary and antiunitary symmetry and antisymmetry operations by means of the
signs $\eta_{{\cal T},{\cal P},{\cal C}}$ and the square (in case of
antiunitary symmetries) is partially redundant, because symmetry operations
that are characterized differently may be mapped onto each other using
non-spatial symmetries of the Hamiltonian $H_d$. For example, if a
time-reversal symmetric Hamiltonian $H_d$ satisfies a crystalline unitary
symmetry ${\cal S}$, then it also satisfies the antiunitary symmetry ${\cal T
S}$. Using such equivalences, Shiozaki and Sato group the (anti)symmetries into
$44$ ``equivalence classes'', which, together with the Altland-Zirnbauer class
of Table~\ref{tab:AZ}, are labeled by one integer $s$ or by two integers $s$
and $t$. These equivalence classes are defined in
Tables~\ref{tab:1}--\ref{tab:3} for the complex Altland-Zirnbauer classes with
unitary symmetries and antisymmetries, the complex Altland-Zirnbauer classes
with antiunitary symmetries and antisymmetries, and the real Altland-Zirnbauer
classes. For each of these Shiozaki-Sato classes, the tables also list a
representative (anti)symmetry operation, consisting of a unitary symmetry
${\cal S}$ squaring to one or a product of a unitary symmetry and one of the
fundamental non-spatial symmetry operations ${\cal T}$, ${\cal P}$, or ${\cal
C}$, with indices $\eta_{{\cal T},{\cal P},{\cal
C}}$ specifying the fundamental commutation or anticommutation relations with
the non-spatial symmetries ${\cal T}$, ${\cal P}$, and ${\cal C}$, if present.
We implicitly assume that (anti)symmetry operations ${\cal T}$, ${\cal P}$, and ${\cal C}$ used for the construction of the representative (anti)symmetry operation commute with the crystalline symmetry operation
${\cal S}$. With these assumptions, the indicated square
of ${\cal T}$ and ${\cal P}$ (in Table \ref{tab:2}) 
and the commutation relations of ${\cal S}$ with
${\cal C}$ (in Tables \ref{tab:1} and \ref{tab:2}) or with ${\cal T}$ or ${\cal P}$ (in Table \ref{tab:3}) fix the algebraic properties of the representative (anti)symmetry operations ${\cal TS}$, ${\cal PS}$, and ${\cal CS}$.

\begin{table}[t]
\begin{tabular*}{\columnwidth}{l @{\extracolsep{\fill}} cccc}
\hline\hline 
AZ class & $s$ &  symmetry operations  & representative \\ \hline
A & $0$ & $\,_+A^+$ & ${\cal T}^+ {\cal S}$ \\
AIII & $1$ & $\,_\alpha A^+_+$ & ${\cal T}^+ {\cal S}_+$ \\ 
A & $2$ & $\,_-A^+$ & ${\cal P}^+ {\cal S}$ \\
AIII & $3$ & $\,_\alpha A^{-\alpha}_{-}$ & ${\cal T}^-{\cal S}_-$ \\
A & $4$ & $\,_+A^-$ & ${\cal T}^- {\cal S}$ \\
AIII & $5$ & $\,_\alpha A^-_+$ & ${\cal T}^- {\cal S}_+$ \\ 
A & $6$ & $\,_-A^-$ & ${\cal P}^- {\cal S}$ \\
AIII & $7$ & $\,_\alpha A^{\alpha}_{-}$ & ${\cal T}^+{\cal S}_-$ 
\\ \hline\hline
\end{tabular*}
\caption{Shiozaki-Sato equivalence classes of antiunitary symmetry and
	antisymmetry operations for the Altland-Zirnbauer classes A and AIII.
	The symbol $\,_{\sigma}A_{\eta_{\cal C}}^{\pm}$ is used to denote
	antiunitary symmetry ($\sigma=+$) and antisymmetry ($\sigma=-$)
	operations that commute ($\eta_{{\cal C}} = +$) or anticommute
	($\eta_{{\cal C}}=-$) with the chiral symmetry, if applicable, and
	square to $\pm 1$. The last column lists the product of a unitary
	crystalline symmetry ${\cal S}$ (${\cal S}_{\eta_{\cal C}}$ for class
	AIII) and time-reversal ${\cal T}^{\pm}$ or particle-hole conjugation
	${\cal P}^{\pm}$ as a crystalline symmetry operation representative of
	the Shiozaki-Sato class $(s,t)$.
	\label{tab:2}}
\end{table}
\begin{table}[t]
\begin{tabular*}{\columnwidth}{l @{\extracolsep{\fill}} ccccc}
\hline\hline 
AZ class & $s$ & $t$ & symmetry operation & representative \\ \hline
AI & $0$ & $0$ &  $\,_+U^{\alpha}_{\alpha}$, $\,_+A^{+}_{\alpha}$ & ${\cal S}_+$ \\
BDI & $1$ & $0$ & $\,_\alpha U^{\beta}_{\beta,\beta}$, $\,_{\alpha}A^{+}_{\beta,\beta}$ & ${\cal S}_{++}$ \\
D & $2$ & $0$ & $\,_{+}U^{\alpha}_{\alpha}$, $\,_{-}A^{+}_{\alpha}$ &  ${\cal S}_+$ \\
DIII & $3$ & $0$ & $\,_\alpha U^{\alpha\beta}_{\beta,\beta}$, $\,_{\alpha}A^{-\alpha}_{\beta,\beta}$ & ${\cal S}_{++}$ \\
AII & $4$ & $0$ & $\,_+U^{\alpha}_{\alpha}$, $\,_+A^{-}_{\alpha}$ &  ${\cal S}_+$ \\
CII & $5$ & $0$ & $\,_\alpha U^{\beta}_{\beta,\beta}$, $\,_{\alpha}A^{-}_{\beta,\beta}$ & ${\cal S}_{++}$ \\
C & $6$ & $0$ & $\,_{+}U^{\alpha}_{\alpha}$, $\,_{-}A^{-}_{\alpha}$ &  ${\cal S}_+$ \\
CI & $7$ & $0$ & $\,_\alpha U^{\alpha\beta}_{\beta,\beta}$, $\,_{\alpha}A^{\alpha}_{\beta,\beta}$ & ${\cal S}_{++}$ \\ \hline
AI & $0$ & $1$ &   $\,_-U^{\alpha}_{-\alpha}$,  $\,_-A^{-}_{\alpha}$  & ${\cal CS}_-$ \\
BDI & $1$ & $1$ &  $\,_\alpha U^{\alpha \beta}_{\beta,-\beta}$,  $\,_\alpha A^{\alpha}_{\beta,-\beta}$  & ${\cal S}_{+-}$  \\
D & $2$ & $1$ &  $\,_{-}U^{\alpha}_{\alpha}$,  $\,_{+}A^{+}_{\alpha}$  &  ${\cal CS}_+$ \\
DIII & $3$ & $1$ &  $\,_\alpha U^{\beta}_{-\beta,\beta}$,  $\,_\alpha A^{+}_{\beta,-\beta}$  & ${\cal S}_{-+}$ \\
AII & $4$ & $1$ &  $\,_-U^{\alpha}_{-\alpha}$,  $\,_-A^{+}_{\alpha}$  & ${\cal CS}_-$ \\
CII & $5$ & $1$ &  $\,_\alpha U^{\alpha \beta}_{\beta,-\beta}$,  $\,_\alpha A^{-\alpha}_{\beta,-\beta}$  &  ${\cal S}_{+-}$ \\
C & $6$ & $1$ &  $\,_{-}U^{\alpha}_{\alpha}$,  $\,_{+}A^{-}_{\alpha}$  & ${\cal CS}_+$ \\
CI & $7$ & $1$ &  $\,_\alpha U^{\beta}_{-\beta,\beta}$,  $\,_\alpha A^{-}_{\beta,-\beta}$  &  ${\cal S}_{-+}$ \\ \hline
AI & $0$ & $2$ &  $\,_+U^{\alpha}_{-\alpha}$,  $\,_+A^{-}_{\alpha}$  & ${\cal S}_-$  \\
BDI & $1$ & $2$ &  $\,_{\alpha}U^{-\beta}_{\beta,\beta}$,  $\,_{\alpha}A^{-}_{\beta,\beta}$  &  ${\cal S}_{--}$ \\
D & $2$ & $2$ &  $\,_{+}U^{\alpha}_{-\alpha}$,  $\,_{-}A^{-}_{\alpha}$  &  ${\cal S}_-$ \\
DIII & $3$ & $2$ &  $\,_{\alpha}U^{-\alpha\beta}_{\beta,\beta}$,  $\,_{\alpha}A^{\alpha}_{\beta,\beta}$  &  ${\cal S}_{--}$ \\
AII & $4$ & $2$ &  $\,_+U^{\alpha}_{-\alpha}$,  $\,_+A^{+}_{\alpha}$  &  ${\cal S}_-$  \\
CII & $5$ & $2$ &  $\,_{\alpha}U^{-\beta}_{\beta,\beta}$,  $\,_{\alpha}A^{+}_{\beta,\beta}$  &  ${\cal S}_{--}$ \\
C & $6$ & $2$ &  $\,_{+}U^{\alpha}_{-\alpha}$,  $\,_{-}A^{+}_{\alpha}$  &  ${\cal S}_-$ \\
CI & $7$ & $2$ &   $\,_{\alpha}U^{-\alpha\beta}_{\beta,\beta}$,  $\,_{\alpha}A^{-\alpha}_{\beta,\beta}$  &  ${\cal S}_{--}$ \\ \hline
AI & $0$ & $3$ &  $\,_-U^{\alpha}_{\alpha}$,  $\,_-A^{+}_{\alpha}$ & ${\cal CS}_+$ \\
BDI & $1$ & $3$ &  $\,_{\alpha}U^{\alpha \beta}_{-\beta,\beta}$,  $\,_{\alpha}A^{-\alpha}_{-\beta,\beta}$  & ${\cal S}_{-+}$ \\
D & $2$ & $3$ &  $\,_{-}U^{\alpha}_{-\alpha}$,  $\,_{+}A^{-}_{\alpha}$  & ${\cal CS}_-$ \\
DIII & $3$ & $3$ &  $\,_{\alpha}U^{\beta}_{\beta,-\beta}$,  $\,_{\alpha}A^{-}_{\beta,-\beta}$   &  ${\cal S}_{+-}$\\
AII & $4$ & $3$ &   $\,_-U^{\alpha}_{\alpha}$,  $\,_-A^{-}_{\alpha}$  & ${\cal CS}_+$ \\
CII & $5$ & $3$ &  $\,_{\alpha}U^{\alpha \beta}_{-\beta,\beta}$,  $\,_{\alpha}A^{\alpha}_{-\beta,\beta}$  & ${\cal S}_{-+}$ \\
C & $6$ & $3$ &  $\,_{-}U^{\alpha}_{-\alpha}$,  $\,_{+}A^{+}_{\alpha}$  & ${\cal CS}_-$ \\
CI & $7$ & $3$ &  $\,_{\alpha}U^{\beta}_{\beta,-\beta}$,  $\,_{\alpha}A^{+}_{\beta,-\beta}$   & ${\cal S}_{+-}$ \\ \hline\hline
\end{tabular*}
\caption{Shiozaki-Sato equivalence classes of symmetry and antisymmetry
	operations for the eight real Altland-Zirnbauer classes. The symbols
	$\,_{\sigma}U^{\pm}_{\eta_{\cal T},\eta_{\cal P}}$ and
	$\,_{\sigma}A^{\pm}_{\eta_{\cal T},\eta_{\cal P}}$ are used to denote
	unitary symmetry (U, $\sigma=+$), unitary antisymmetry (U, $\sigma=-$),
	antiunitary symmetry (A, $\sigma=+$), and antiunitary antisymmetry (A,
	$\sigma=-$) operations that square to $\pm 1$ and commute ($\eta_{{\cal
	T},{\cal P}} = +$) or anticommute ($\eta_{{\cal T},{\cal P}}=-$) with
	time-reversal and particle-hole conjugation, if applicable. The last
	column lists a unitary crystalline symmetry ${\cal S}_{\eta_{\cal
	T},\eta_{\cal P}}$ or the product of a unitary crystalline symmetry and
	the chiral operation ${\cal C}$ as a representative of the equivalence
	class. \label{tab:3}}
\end{table}

Following this scheme, Shiozaki and Sato have classified all insulators and superconductors with a single crystalline order-two unitary or antiunitary symmetry or antisymmetry.\cite{shiozaki2014} 
Central to the classification of Ref.~\onlinecite{shiozaki2014} is a set of isomorphisms between the groups $K^{\CC}(s,t|d_{\parallel},d)$,
$K^{\CC}(s|d_{\parallel},d)$, and $K^{\RR}(s,t|d_{\parallel},d)$ classifying
$d$-dimensional Hamiltonians in the Shiozaki-Sato symmetry class $(s,t)$ or $s$
and with $d_{\parallel}$ inverted spatial dimensions. For the complex Altland-Zirnbauer classes
with unitary (anti)symmetry these isomorphisms are (with $d_{\parallel} < d$)
\begin{align}
  K^{\CC}(s,t|d_{\parallel},d) &= K^{\CC}(s,t+1|d_{\parallel}+1,d) \nonumber \\ &= K^{\CC}(s-1,t|d_{\parallel},d-1), \label{eq:K1}
\end{align}
with the integers $s$ and $t$ taken mod $2$. For the complex Altland-Zirnbauer classes with
antiunitary (anti)symmetry the isomorphisms read
\begin{align}
  K^{\CC}(s|d_{\parallel},d) &= K^{\CC}(s-2|d_{\parallel}+1,d) \nonumber \\ &= K^{\CC}(s-1|d_{\parallel},d-1), \label{eq:K2}
\end{align}
where the label $s$ is taken mod $8$. Finally, the isomorphisms for the real Altland-Zirnbauer classes are
\begin{align}
  K^{\RR}(s,t|d_{\parallel},d) &= K^{\RR}(s,t+1|d_{\parallel}+1,d) \nonumber \\ &= K^{\RR}(s-1,t|d_{\parallel},d-1), \label{eq:K3}
\end{align}
where the integers $s$ and $t$ are taken mod $8$ and mod $4$, respectively. When
applied repeatedly, these isomorphisms can be used to relate the classification
problem of $d$-dimensional Hamiltonians with an order-two crystalline symmetry
to a zero-dimensional classification problem, which can be solved with
elementary methods. The Shiozaki-Sato classification for two and three
dimensional crystals with a mirror reflection ${\cal M}$, twofold rotation
${\cal R}$, or inversion symmetry ${\cal I}$ is summarized in
tables~\ref{tab:SS1}-\ref{tab:SS3}. The corresponding classifying
groups  for complex and real
Altland-Zirnbauer classes without crystalline symmetries are denoted
$K^{\mathbb{C}}(s,d)$ and $K^\mathbb{R}(s,d)$, respectively. Since they are well known\cite{schnyder2008,schnyder2009,kitaev2009,stone2011,wen2012,abramovici2012,kennedy2016} we do not list them here explicitly; if needed, they can be inferred from Table \ref{tab:AZ}, which lists $K(s,d-1)$ for $d=2$ and $d=3$.

\begin{table}
\begin{tabular*}{\columnwidth}{l @{\extracolsep{\fill}} cccccccc}
\hline\hline 
class & $s$ & $t$ & \begin{tabular}{c} $d=2$\\ ${\cal M}$ \end{tabular} & \begin{tabular}{c} $d=2$\\ ${\cal R}$ \end{tabular} & \begin{tabular}{c} $d=3$\\ ${\cal M}$ \end{tabular} & \begin{tabular}{c} $d=3$\\ ${\cal R}$ \end{tabular} &  \begin{tabular}{c} $d=3$\\ ${\cal I}$ \end{tabular} \\ \hline
A$^{\cal S}$ & $0$ & $0$ & $0$ & $\ZZ^2\, (\ZZ)$ & $\ZZ$ & $0$ & $\ZZ$ \\
AIII$^{\cal S_+}$ & $1$ & $0$ & $\ZZ$ & $0$ & $0$ & $\ZZ^2\, (\ZZ)$ & $0$ \\ \hline
A$^{\cal CS}$ & $0$ & $1$ & $\ZZ^2\, (\ZZ)$ & $0$ & $0$ & $\ZZ$ & $0$ \\
AIII$^{\cal S_-}$ & $1$ & $1$ & $0$ & $\ZZ$ & $\ZZ^2\, (\ZZ)$ & $0$ & $\ZZ^2\, (\ZZ)$ \\ \hline\hline
\end{tabular*}
\caption{Classification of topological crystalline phases with an order-two crystalline symmetry or antisymmetry for the complex Altland-Zirnbauer classes, based on Ref. \onlinecite{shiozaki2014}. The symbols ${\cal M}$, ${\cal R}$, and ${\cal I}$ refer to mirror reflection ($d_{\parallel} = 1$), twofold rotation ($d_{\parallel} = 2$), and inversion ($d_{\parallel} = d = 3$), respectively. The entries in brackets give the purely crystalline component $K'^{\CC}(s,t|d_{\parallel},d)$ if different from the full group $K^{\CC}(s,t|d_{\parallel},d)$.\label{tab:SS1}}
\end{table}
\begin{table}
\begin{tabular*}{\columnwidth}{l @{\extracolsep{\fill}} ccccccc}
\hline\hline 
class & $s$ & \begin{tabular}{c} $d=2$\\ ${\cal M}$ \end{tabular} & \begin{tabular}{c} $d=2$\\ ${\cal R}$ \end{tabular} & \begin{tabular}{c} $d=3$\\ ${\cal M}$ \end{tabular} & \begin{tabular}{c} $d=3$\\ ${\cal R}$ \end{tabular} &  \begin{tabular}{c} $d=3$\\ ${\cal I}$ \end{tabular} \\ \hline
A$^{{\cal T}^+{\cal S}}$ & $0$ & $\ZZ\, (0)$ & $\ZZ_2$ & $0$ & $\ZZ_2$ & $0$ \\
AIII$^{{\cal T}^+{\cal S}_+}$ & $1$ & $\ZZ_2$ & $0$ & $\ZZ\, (0)$ & $\ZZ_2$ & $2\ZZ\, (0)$ \\
A$^{{\cal P}^+{\cal S}}$ & $2$ & $\ZZ_2$ & $2\ZZ\, (0)$ & $\ZZ_2$ & $0$ & $0$ \\
AIII$^{{\cal T}^-{\cal S}_-}$ & $3$ & $0$ & $0$ & $\ZZ_2$ & $2\ZZ\, (0)$ & $0$ \\
A$^{{\cal T}^-{\cal S}}$ & $4$ & $2\ZZ\, (0)$ & $0$ & $0$ & $0$ & $0$ \\
AIII$^{{\cal T}^-{\cal S}_+}$ & $5$ & $0$ & $0$ & $2\ZZ\, (0)$ & $0$ & $\ZZ\, (0)$ \\
A$^{{\cal P}^-{\cal S}}$ & $6$ & $0$ & $\ZZ\, (0)$ & $0$ & $0$ & $\ZZ_2$ \\
AIII$^{{\cal T}^+{\cal S}_-}$ & $7$ & $0$ & $\ZZ_2$ & $0$ & $\ZZ\, (0)$ & $\ZZ_2$ \\\hline\hline
\end{tabular*}
\caption{Classification of topological crystalline phases with an order-two antiunitary crystalline symmetry or antisymmetry for the complex Altland-Zirnbauer classes, based on Ref. \onlinecite{shiozaki2014}. The symbols ${\cal M}$, ${\cal R}$, and ${\cal I}$ refer to mirror reflection ($d_{\parallel} = 1$), twofold rotation ($d_{\parallel} = 2$), and inversion ($d_{\parallel} = d = 3$), respectively.  The entries in brackets give the purely crystalline component $K'^{\CC}(s|d_{\parallel},d)$ if different from the full group $K^{\CC}(s|d_{\parallel},d)$.\label{tab:SS2}}
\end{table}
\begin{table}
\begin{tabular*}{\columnwidth}{l @{\extracolsep{\fill}} cccccccc}
\hline\hline 
class & $s$ & $t$ & \begin{tabular}{c} $d=2$\\ ${\cal M}$ \end{tabular} & \begin{tabular}{c} $d=2$\\ ${\cal R}$ \end{tabular} & \begin{tabular}{c} $d=3$\\ ${\cal M}$ \end{tabular} & \begin{tabular}{c} $d=3$\\ ${\cal R}$ \end{tabular} &  \begin{tabular}{c} $d=3$\\ ${\cal I}$ \end{tabular} \\ \hline
    AI$^{\mathcal{S}_+}$ & $0$ & $0$ & $0$ & $2 \ZZ$ & $0$ & $0$ & $2 \ZZ$ \\
    BDI$^{\mathcal{S}_{++}}$ & $1$ & $0$ & $\ZZ$ & $0$ & $0$ & $2 \ZZ$ & $0$ \\
    D$^{\mathcal{S}_{+}}$ & $2$ & $0$ & $\ZZ_2$ & $\ZZ\, (0)$ & $\ZZ$ & $0$ & $0$  \\
    DIII$^{\mathcal{S}_{++}}$ & $3$ & $0$ & $\ZZ_2$ & $0$ & $\ZZ_2$ & $\ZZ\, (0)$ & $0$  \\
    AII$^{\mathcal{S}_{+}}$ & $4$ & $0$ & $0$ & $2\ZZ\, (4 \ZZ)$ & $\ZZ_2$ & $0$ & $\ZZ\, (2 \ZZ)$ \\
    CII$^{\mathcal{S}_{++}}$ & $5$ & $0$ &$2\ZZ$ & $0$ & $0$ & $2 \ZZ\, (4\ZZ)$ & $\ZZ_2\, (0)$ \\
    C$^{\mathcal{S}_{+}}$ & $6$ & $0$ & $0$ & $\ZZ\, (0)$ & $2\ZZ$ & $0$ & $\ZZ_2$ \\
    CI$^{\mathcal{S}_{++}}$ & $7$ & $0$ & $0$ & $0$ & $0$ & $\ZZ\, (0)$ & $0$ \\
    \hline
    AI$^{\mathcal{CS}_-}$ & $0$ & $1$ & $0$ & $0$ & $0$ & $0$ & $0$ \\
    BDI$^{\mathcal{S}_{+-}}$ & $1$ & $1$ & $0$ & $\ZZ$ & $0$ & $0$ & $2 \ZZ$ \\
    D$^{\mathcal{CS}_{+}}$ & $2$ & $1$ & $\ZZ^2\, (\ZZ)$ & $\ZZ_2$ & $0$ & $\ZZ$ & $0$ \\
    DIII$^{\mathcal{S}_{-+}}$ & $3$ & $1$ & $\ZZ_2^2\, (\ZZ_2)$ & $\ZZ_2$ & ${\ZZ^2\, (\ZZ)}$ & $\ZZ_2$ & $\ZZ\, (0)$ \\
    AII$^{\mathcal{CS}_{-}}$ & $4$ & $1$ & $\ZZ_2^2\, (\ZZ_2)$ & $0$ & ${\ZZ_2^2\, (\ZZ_2)}$ & $\ZZ_2$ & $0$ \\
    CII$^{\mathcal{S}_{+-}}$ & $5$ & $1$ & $0$ & $2 \ZZ$ & ${\ZZ_2^2\, (\ZZ_2)}$ & $0$ & $2\ZZ\, (4\ZZ)$\\
    C$^{\mathcal{CS}_{+}}$ & $6$ & $1$ & $2\ZZ^2\, (2\ZZ)$ & $0$ & $0$ & $2 \ZZ$ & $0$ \\
    CI$^{\mathcal{S}_{-+}}$ & $7$ & $1$ & $0$ & $0$ & ${2\ZZ^2\, (2\ZZ)}$ & $0$ & $\ZZ\, (0)$\\
    \hline 
    AI$^{\mathcal{S}_-}$ & $0$ & $2$ & $0$ & $0$ & $2\ZZ$ & $0$ & $0$ \\
    BDI$^{\mathcal{S}_{--}}$ & $1$ & $2$ & $0$ & $0$ & $0$ & $0$ & $0$  \\
    D$^{\mathcal{S}_{-}}$ & $2$ & $2$ & $0$ & $\ZZ^2\, (\ZZ)$ & $0$ & $0$ & $\ZZ$  \\
    DIII$^{\mathcal{S}_{--}}$ & $3$ & $2$ & $\ZZ\, (2\ZZ)$ & $\ZZ_2^2\, (\ZZ_2)$ & $0$ & $\ZZ^2\, (\ZZ)$ & $\ZZ_2$ \\
    AII$^{\mathcal{S}_{-}}$ & $4$ & $2$ & $\ZZ_2\, (0)$ & $\ZZ_2^2\, (\ZZ_2)$ & $\ZZ\, (2 \ZZ)$ & $\ZZ_2^2\, (\ZZ_2)$ & $\ZZ_2$ \\
    CII$^{\mathcal{S}_{--}}$ & $5$ & $2$ & $\ZZ_2$ & $0$ & $\ZZ_2\, (0)$ & $\ZZ_2^2\, (\ZZ_2)$ & $0$ \\
    C$^{\mathcal{S}_{-}}$ & $6$ & $2$ & $0$ & $2 \ZZ^2\, (2\ZZ)$ & ${\ZZ_2}$ & $0$ & $2\ZZ$ \\
    CI$^{\mathcal{S}_{--}}$ & $7$ & $2$ & $2\ZZ$ & $0$ & $0$ & $2\ZZ^2\, (2\ZZ)$ & $0$ \\
    \hline
    AI$^{\mathcal{CS}_+}$ & $0$ & $3$ & $2\ZZ$ & $0$ & $0$ & $2 \ZZ$ & $0$ \\
    BDI$^{\mathcal{S}_{-+}}$ & $1$ & $3$ & $0$ & $0$ & $2\ZZ$ & $0$ & $0$ \\
    D$^{\mathcal{CS}_{-}}$ & $2$ & $3$ & $\ZZ\, (0)$ & $0$ & $0$ & $0$ & $0$ \\
    DIII$^{\mathcal{S}_{+-}}$ & $3$ & $3$ & $0$ & $\ZZ\, (2 \ZZ)$ & $\ZZ\, (0)$ & $0$ & $\ZZ^2\, (\ZZ)$ \\
    AII$^{\mathcal{CS}_{+}}$ & $4$ & $3$ & $2\ZZ\, (4\ZZ)$ & $\ZZ_2\, (0)$ & $0$ & $\ZZ\, (2 \ZZ)$ & $\ZZ_2^2\, (\ZZ_2)$ \\
    CII$^{\mathcal{S}_{-+}}$ & $5$ & $3$ & $0$ & $\ZZ_2$ & $2\ZZ\, (4\ZZ)$ & $\ZZ_2\, (0)$ & $\ZZ_2^2\, (\ZZ_2)$ \\
    C$^{\mathcal{CS}_{-}}$ & $6$ & $3$ & $\ZZ\, (0)$ & $0$ & $0$ & $\ZZ_2$ & $0$ \\
    CI$^{\mathcal{S}_{+-}}$ & $7$ & $3$ & $0$ & $2 \ZZ$ & $\ZZ\, (0)$ & $0$ & $2\ZZ^2\, (2\ZZ)$ \\\hline\hline
  \end{tabular*}
\caption{Classification of topological crystalline phases with an order-two crystalline symmetry or antisymmetry for the real Altland-Zirnbauer classes, based on Ref. \onlinecite{shiozaki2014}.
  The symbols ${\cal M}$, ${\cal R}$, and ${\cal I}$ refer to mirror reflection
  ($d_{\parallel} = 1$), twofold rotation ($d_{\parallel} = 2$), and inversion
  ($d_{\parallel} = d = 3$), respectively. The entries in brackets give the purely crystalline component $K'^{\RR}(s,t|d_{\parallel},d)$ if different from the full group $K^{\RR}(s,t|d_{\parallel},d)$.
  \label{tab:SS3}}
\end{table}

Some of the topological crystalline phases remain topologically nontrivial if
the crystalline symmetry is broken. These are strong topological insulators or
superconductors, which have gapless states at all boundaries, not only at
boundaries that are invariant under the symmetry operation. The remaining
``purely crystalline'' topological phases, which become trivial if the
crystalline symmetry is broken, are classified by a subgroup of the classifying
groups $K^{\CC}(s,t|d_{\parallel},d)$, $K^{\CC}(s|d_{\parallel},d)$, and
$K^{\RR}(s,t|d_{\parallel},d)$, which we denote
$K'^{\CC}(s,t|d_{\parallel},d)$, $K'^{\CC}(s|d_{\parallel},d)$, and
$K'^{\RR}(s,t|d_{\parallel},d)$, respectively. The quotient groups
$K(s,t|d_\parallel,d)/K'(s,t|d_\parallel,d)$, which are subgroups of the
classifying groups $K(s,d)$ without crystalline symmetries, classify the strong
topological phases that are compatible with the
crystalline symmetry. Tables \ref{tab:SS1}--\ref{tab:SS3} also list the groups
$K'^{\CC}$ and $K'^{\RR}$ between brackets if they are different from the full
classifying groups $K^{\CC}$ and $K^{\RR}$. The ``purely crystalline''
subgroups are evaluated in Sec.~\ref{sec:4b} and App.~\ref{app:3}.

%

The Shiozaki-Sato classification of topological crystalline insulators and
superconductors with an order-two crystalline symmetry,\cite{shiozaki2014} as
well as the preceding complete classifications of mirror-symmetric topological
insulators and superconductors,\cite{chiu2013,morimoto2013} is a ``strong''
classification, in the sense that it addresses topological features that are
robust to a resizing of the unit cell, allowing the addition of perturbations
that break the translation symmetry of the original (smaller) unit cell, while
preserving the crystalline symmetries. Reference \onlinecite{shiozaki2014}
argues that for such a strong classification it is sufficient to classify
Hamiltonians $H_d(\vk)$ with argument $\vk$ defined on a sphere, rather than on
a Brillouin zone of a shape determined by the underlying Bravais lattice. The
construction and classification of second-order topological insulators and
superconductors that we pursue here also follows the paradigm of a strong
classification scheme. Since boundaries play an essential role when considering
second-order topological phases, we will not deform the Brillouin zone into a
sphere, as in Ref.\ \onlinecite{shiozaki2014}, but instead use the freedom
offered by the possibility to resize the unit cell to restrict ourselves to
rectangular Bravais lattice with mirror plane and rotation axes aligned with
the coordinate axes. This is consistent with the mathematical form of the
symmetry operations given in Eqs.\ (\ref{eq:Usym})--(\ref{eq:Aantisym}) above.

\section{Dimensional reduction}
\label{sec:3}

The dimension-raising and lowering isomorphisms devised by Shiozaki and Sato
apply to Hamiltonians with argument $\vk$ defined on a
sphere,~\cite{shiozaki2014} rather than on a torus, which complicates a direct
application to crystals with boundaries and corners. For that reason, we here
make use of an alternative dimension-lowering map, which maps a Shiozaki-Sato
class with index $s$ in $d$ dimensions to a Shiozaki-Sato class with index $s-1$
in $d-1$ dimensions, while preserving the second Shiozaki-Sato index $t$ and the
number of inverted dimensions $d_{\parallel}$. Our dimension-lowering map is a
generalization of a map first proposed by Fulga {\em et al.} for the standard
Altland-Zirnbauer classes,\cite{fulga2012} and recently extended to
mirror-reflection-symmetric models by two of us.\cite{trifunovic2017} Though not
as powerful as the isomorphisms of Ref.\ \onlinecite{shiozaki2014}, which also
relate symmetry classes with different $d_{\parallel}$, this map is sufficient
for the purpose of determining the conditions under which a nontrivial bulk crystalline phase implies the existence of zero-energy corner states (for a two-dimensional crystal) or gapless hinge modes (for a three-dimensional crystal). 

The dimension-lowering procedure starts from the calculation of the reflection
matrix $r_d$ for a $d$-dimensional Hamiltonian $H_d$ embedded in a two-terminal
scattering geometry. Following Ref.\ \onlinecite{fulga2012} the reflection
matrix is reinterpreted as a Hamiltonian $H_{d-1}$ in $d-1$ dimensions, but
with a symmetry class that is different from that of the original Hamiltonian
$H_d$. This reinterpretation is different for Hamiltonians $H_{d}$ with and
without chiral antisymmetry. If $H_d$ has a chiral antisymmetry, one can choose
a basis of scattering states such that $r_d$ is a
hermitian matrix, allowing the definition of a Hamiltonian $H_{d-1}$ {\em
without} chiral antisymmetry as
\begin{equation}
  H_{d-1} = r_d. \label{eq:HdchiralF}
\end{equation}
On the other hand, if $H_d$ has no chiral antisymmetry, Fulga {\em et al.} set
\begin{equation}
  H_{d-1} = 
  \begin{pmatrix}
    0 & r_d \\
    r^\dagger_d & 0
  \end{pmatrix},
  \label{eq:HdnonchiralF}
\end{equation}  
which has a chiral antisymmetry with $U_{\cal C} = \mbox{diag}\,(1,-1)$. A more detailed review of the reflection-matrix based dimensional reduction scheme is given in App.~\ref{app:1}. In the appendix we also show that if $H_d$ has a crystalline symmetry or antisymmetry of Shiozaki-Sato class $(s,t)$ with $d_{\parallel} < d$ then $H_{d-1}$ has a crystalline symmetry of class $(s-1,t)$ with the same value of $d_{\parallel}$. (This was shown in Ref.\ \onlinecite{trifunovic2017} for unitary mirror symmetries and antisymmetries with $d_{\parallel} = 1$.)

\begin{figure}
\includegraphics[width=0.99\columnwidth]{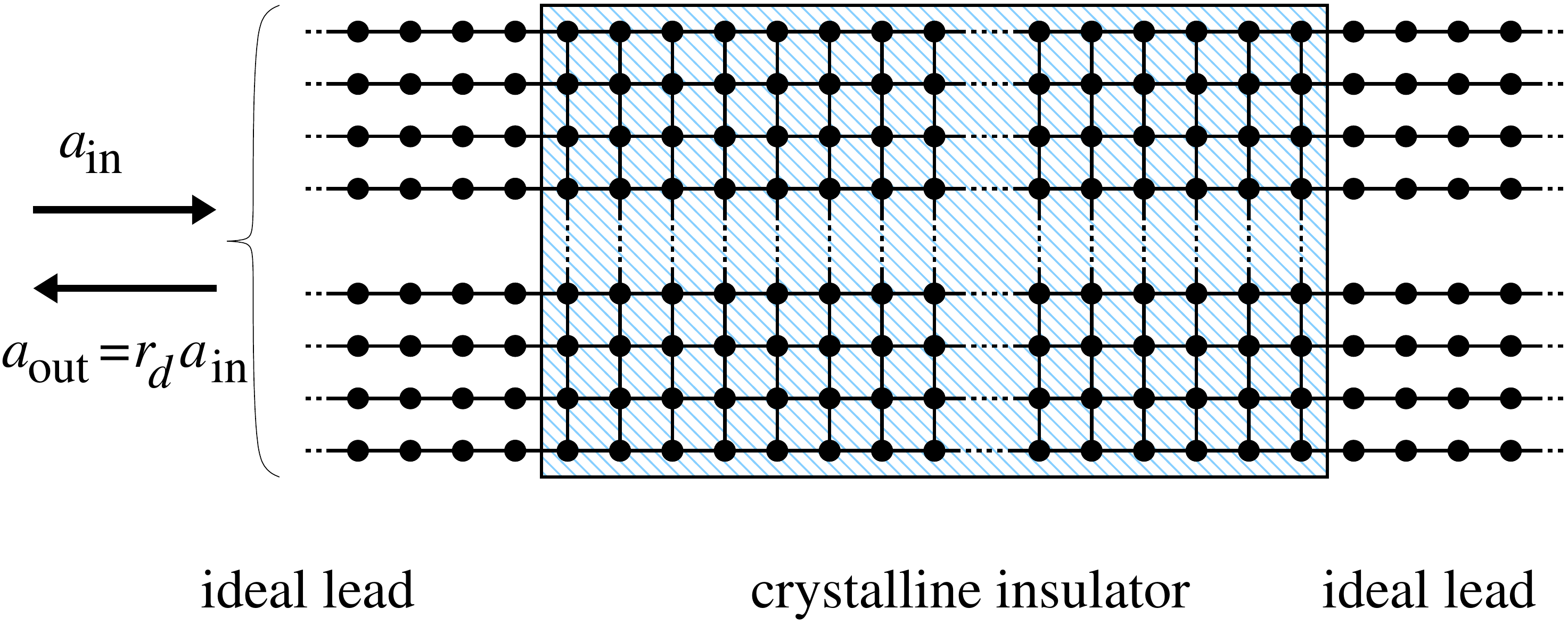}
\caption{\label{fig:setup_finite} Schematic picture of a lattice model for a crystalline insulator in a two-terminal scattering geometry. The reflection matrix $r_d$ relates the amplitudes $a_{\rm in}$ and $a_{\rm out}$ of incident and reflected waves as shown in the figure. The ideal leads are modeled as a grid of parallel one-dimensional chains, which endows the reflection matrix $r_d$ with a real-space structure.}
\end{figure}

Although Refs.~\onlinecite{fulga2012,trifunovic2017} apply the
reflection-matrix-based reduction scheme to Hamiltonians with periodic boundary
conditions, the mapping of Eqs.\ (\ref{eq:HdchiralF}) and
(\ref{eq:HdnonchiralF}) can also be used in a real-space formulation, where it
can be applied to crystals with boundaries. In particular, the mapping of Eqs.\
(\ref{eq:HdchiralF}) and (\ref{eq:HdnonchiralF}) maps $d'$-dimensional protected
boundary modes of $H_{d}$ to $d'-1$-dimensional boundary modes of $H_{d-1}$ for
all $1\le d'<d$, thus not only providing a link between regular first-order
topological insulators and superconductors in different dimensions, but also
between second-order topological insulators and superconductors.

To show how this works explicitly, we consider a $d$-dimensional crystalline
insulator or superconductor, embedded in a two-terminal scattering geometry and
of finite size in the transverse directions, as shown schematically in
Fig.~\ref{fig:setup_finite} for a two-dimensional lattice model. We then
calculate the reflection matrix $r_d(\vr_{\perp},\vr_{\perp}')$ for an ideal
lead consisting of a grid of one-dimensional chains at discrete coordinates
$\vr_{\perp}$ in the transverse direction, see Fig.~\ref{fig:setup_finite}, and
construct a hermitian lattice Hamiltonian $H_{d-1}(\vr_{\perp},\vr_{\perp}')$
using the mapping of Eqs.\ (\ref{eq:HdchiralF}) and (\ref{eq:HdnonchiralF}).
Since it is derived from a reflection matrix $r_d$ for a lead with a finite
$(d-1)$-dimensional cross section and open boundary conditions in the transverse
direction, $H_{d-1}$ also describes a $(d-1)$-dimensional system of finite size
and open boundary conditions. For a crystalline insulator or superconductor of
finite width, the existence of gapless modes along the sample boundary implies
the existence of perfectly transmitted modes along sample boundaries (in case of
a first-order topological insulator or superconductor) or hinges (for a
second-order topological insulator or superconductor). Since the total
scattering matrix, describing reflection {\em and} transmission, is unitary, any
such perfectly transmitted modes correspond to a zero singular value of the
reflection matrix $r_d(\vr_{\perp},\vr_{\perp}')$ and, hence, to a zero-energy
eigenstate of $H_{d-1}$. Since these gapless modes derive from transmitted modes
proceeding along the sample boundary, their eigenvectors have support near the
lead's boundaries (if $H_d$ is a first-order topological insulator) or the
intersection of two of the lead's boundaries (if $H_d$ is a second-order
topological insulator), so that they represent true boundary/corner/hinge modes
of $H_{d-1}$. 

\begin{figure}
\includegraphics[width=0.99\columnwidth]{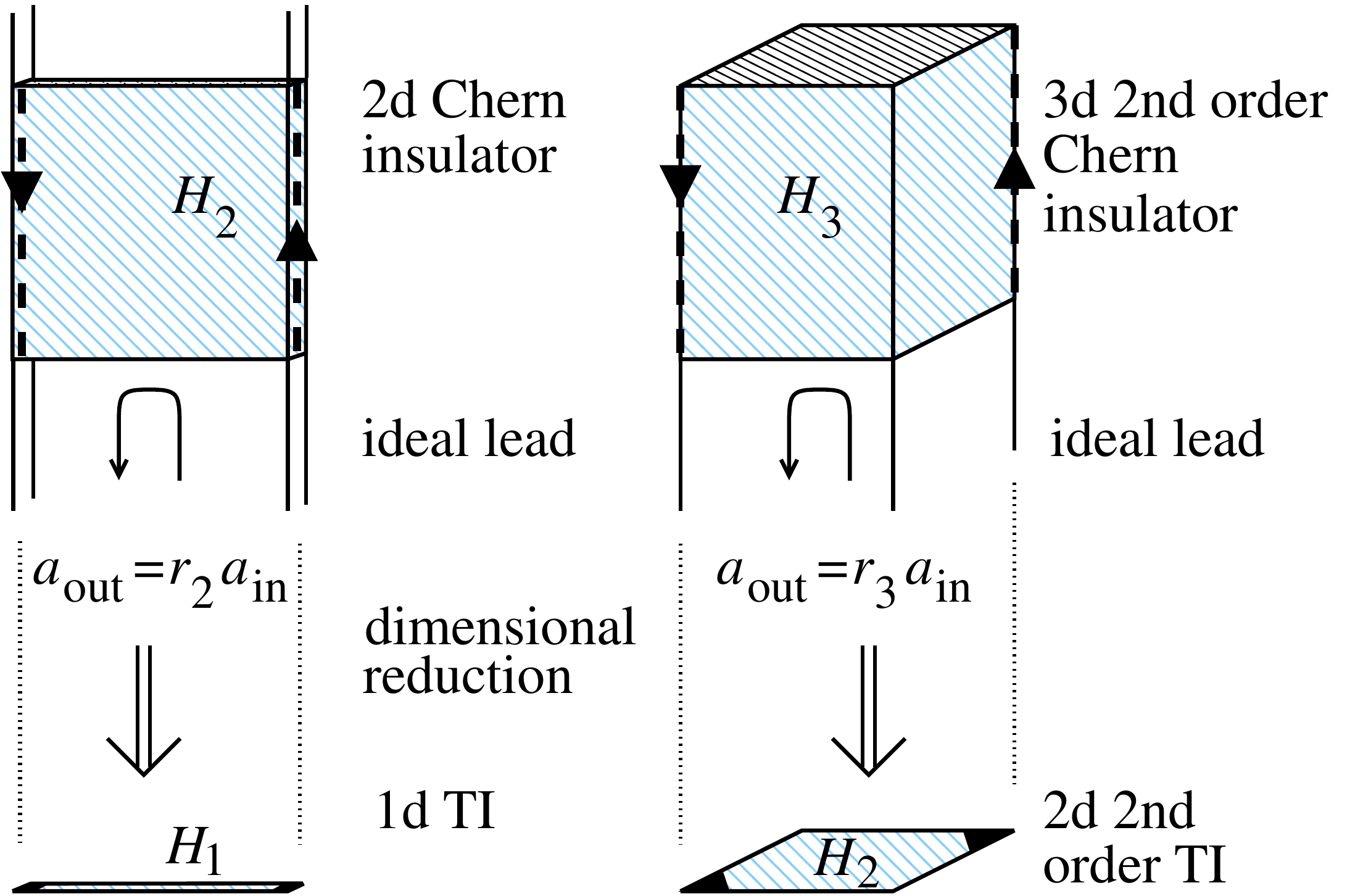}
\caption{\label{fig:chern_lead} Comparison of the reflection-matrix-based dimensional reduction scheme applied to a two-dimensional Chern insulator (left) and a three-dimensional second-order Chern insulator. In each case a lower-dimensional Hamiltonian can be constructed out of the reflection matrix $r_d$ describing scattering from a half-infinite crystal coupled to an ideal lead. Upon constructing the lower-dimensional Hamiltonian $H_{d-1}$, the chiral edge states (left) and hinge states (right) map to protected zero-energy eigenstates localized near ends (left) or corners (right).}
\end{figure}

As an example, we consider a Chern insulator in two dimensions and a
second-order Chern insulator in three dimensions, shown schematically in
Fig.~\ref{fig:chern_lead}. In both cases, the corresponding Altland-Zirnbauer
class is Cartan class A. The two-dimensional Chern insulator has chiral modes
propagating along the sample's edges, see Fig.~\ref{fig:chern_lead} (left). When
the Chern insulator is embedded in a two-terminal scattering geometry, the
presence of the edge modes leads to perfectly transmitted modes or,
equivalently, to zero singular values of the reflection matrix $r_d$. The left
and right eigenvectors corresponding to this zero mode, which build the
corresponding eigenvectors of the Hamiltonian $H_{d-1}$ calculated via Eq.\
(\ref{eq:HdnonchiralF}), are localized near the lead edges. Similarly,
a three-dimensional second-order Chern insulator has chiral hinge modes, as
shown schematically in Fig.~\ref{fig:chern_lead} (right). Again, when embedded
in a scattering geometry, the presence of the hinge modes leads to perfectly
transmitted modes and, hence, zero singular values of the reflection matrix
$r_d$. The support of the corresponding left and right eigenvectors is near the
lead hinges that are connected to the sample hinges carrying the chiral modes.
Correspondingly, the Hamiltonian $H_{d-1}$ obtained from the dimensional
reduction scheme has zero-energy eigenstates at sample corners. Hence, $H_{d-1}$
is a second-order topological insulator. 

\begin{figure}
\includegraphics[width=\columnwidth]{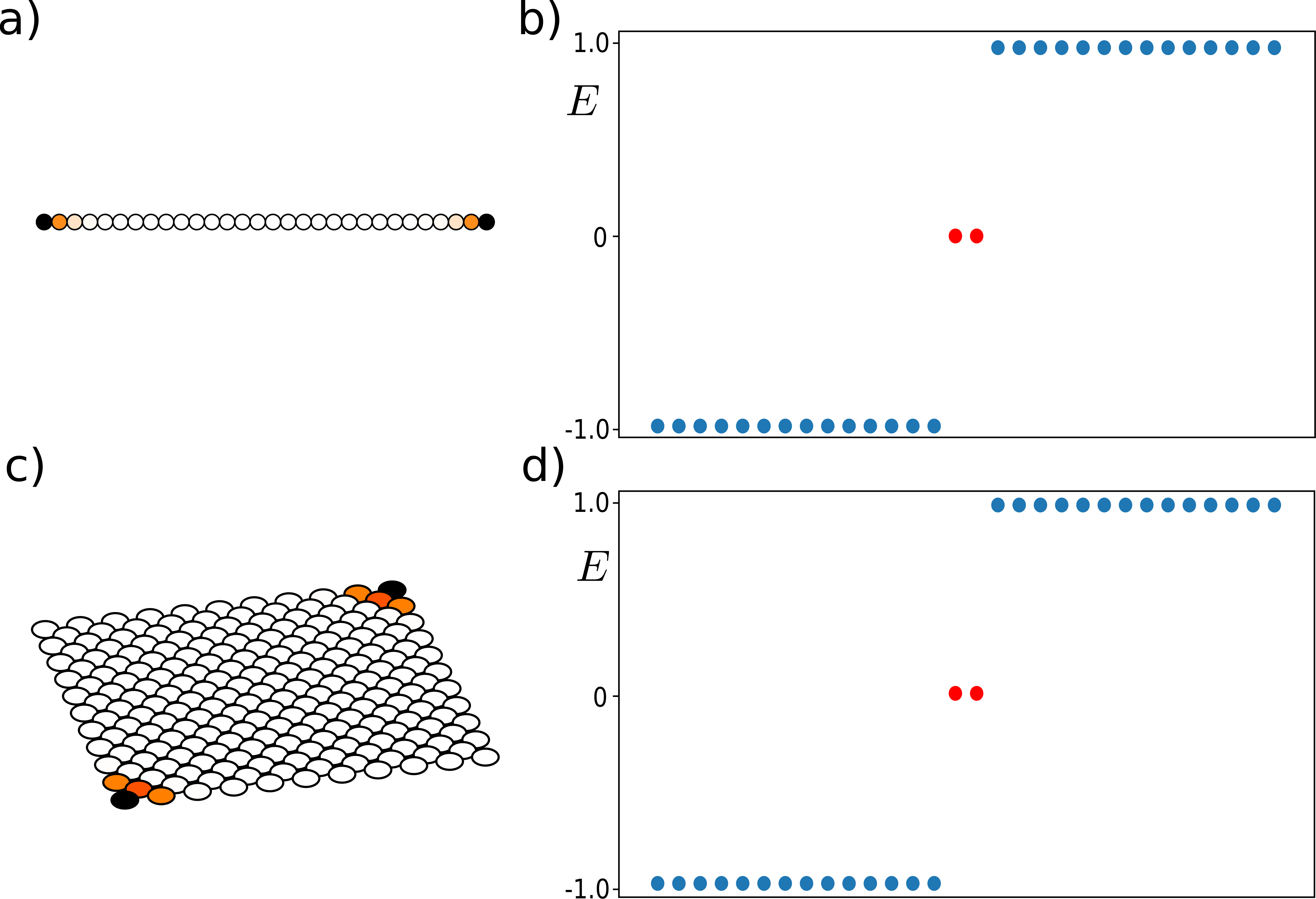}\\
\caption{\label{fig:chern_lead_example} Support of the zero-energy eigenstates
(a) and 30 lowest energies of the spectrum (b) of the mapped Hamiltonian
$H_{1}$ for a two-dimensional Chern insulator with Hamiltonian $H_2$ given in
Eq.~(\ref{eq:HChern}), following the reflection-matrix-based dimensional
reduction scheme. Panels (c) and (d) show the same for the mapped Hamiltonian
$H_{2}$ for the three-dimensional second-order Chern insulator with Hamiltonian
$H_3$ of Eq.~(\ref{eq:HA}) with $b = 0.4$.}
\end{figure}

A numerical simulation of this scenario is shown in Fig.~\ref{fig:chern_lead_example}. The dimensional reduction scheme has been applied to a two-dimensional lattice model with Hamiltonian
\begin{equation}
  H = (m+2-\cos k_1-\cos k_2)\sigma_1+\sin k_1\sigma_2+\sin k_2\sigma_3,
  \label{eq:HChern}
\end{equation}
which describes a two-dimensional Chern insulator for $-2<m<0$, and to a lattice model of a three-dimensional second-order Chern insulator,\cite{langbehn2017,schindler2018} which has Hamiltonian
\begin{align}
  H_3
  =&\, (m+3 - \cos k_1 - \cos k_2 - \cos k_3) \tau_1 \sigma_1
  \label{eq:HA}
  \\ & \, \mbox{}
  + \tau_1 \sigma_3 \sin k_1 + \tau_2 \sin k_2 
  + \tau_3 \sin k_3
  + b \tau_1 \nonumber 
\end{align} 
with $-2 < m < 0$ and $b$ numerically small. In both models, the $\sigma_j$ and $\tau_j$ are Pauli matrices acting on different spinor degrees of freedom. Figure~\ref{fig:chern_lead_example} shows the spectra of the mapped Hamiltonians $H_{d-1}$ (Fig.~\ref{fig:chern_lead_example}b and d), calculated using the Kwant software package,\cite{groth2014} as well as the support of the zero-energy eigenstates (c and d). Consistent with the scenario laid out above, the spectra are gapped up to two zero eigenvalues, which have support at the ends of the mapped one-dimensional chain (Fig.~\ref{fig:chern_lead_example}a) and at mirror-reflection-symmetric corners (Fig.~\ref{fig:chern_lead_example}c).


\section{Mirror reflection-symmetric second-order topological insulators and superconductors}
\label{sec:4}
\subsection{Classification of mirror-symmetric corners and hinges}

\begin{figure}
\includegraphics[width=0.99\columnwidth]{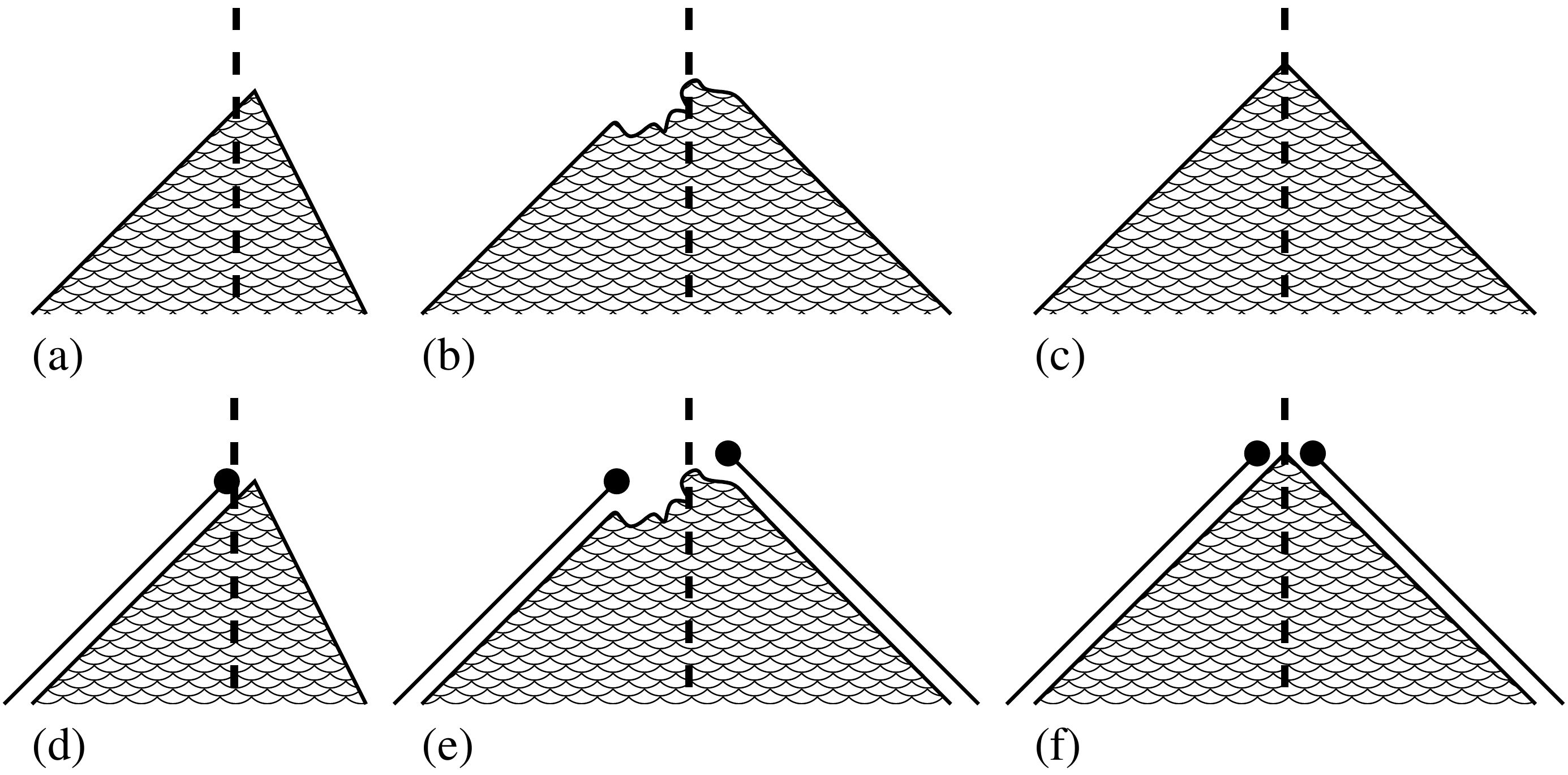}
\caption{\label{fig:rules} Schematic picture of a generic corner (a), a mirror-symmetric corner with locally broken mirror symmetry (b), and a mirror-symmetric corner in a crystal with a bulk mirror symmetry. Panels (d)--(f) represent the possibility to add zero-energy corner states by changing the lattice termination. Effectively, this amounts to the addition of one-dimensional topological insulators or superconductors to the boundaries. At a generic corner, it is possible to change the termination of only one boundary, as shown in panel (d). In a symmetric corner, such a change in termination needs to be applied to both symmetry-related boundaries, shown schematically in panels (e) and (f) for a corner with and without a perturbation that locally breaks the mirror symmetry.}
\end{figure}

We now proceed with the classification of zero-energy states at mirror-symmetric corners and gapless hinge modes at mirror-symmetric hinges of a mirror-symmetric crystalline insulator or superconductor. As explained in the introduction, such a
classification depends on the possible presence of local
mirror-symmetry-breaking perturbations at corners or hinges, and on whether it
is an ``intrinsic'' (termination-independent) classification or an
``extrinsic'' (termination-dependent) one. We recall that we term a
classification intrinsic if it is invariant under a change of lattice
termination, as long as the mirror symmetry of the corner or hinge is
preserved, and extrinsic if it depends on termination. The intrinsic
classification describes properties of the bulk lattice, which is why it is
closely related to the classification of bulk topological crystalline phases,
as we discuss below. Although the extrinsic classification is termination
dependent, it is important to point out that the extrinsic classification
remains valid in the presence of perturbations that do not close the boundary
gap, such as weak disorder. Figure~\ref{fig:rules} schematically shows the four
classification rules that follow from the options discussed above for the case
of a two-dimensional mirror-symmetric crystal, and contrasts these with the
classification of a generic corner discussed in the Introduction.

\begin{table}[t]
\begin{tabular*}{\columnwidth}{l @{\extracolsep{\fill}} ccccc}
\hline\hline 
AZ class & $s$ & $t$ & $d=2$ & $d=3$ \\ \hline
A$^{\cal M}$ & $0$ & $0$ & - & $\ZZ, \ZZ_2\, (\ZZ^2, \ZZ)$ \\
AIII$^{{\cal M}_+}$ & $1$ & $0$ & $\ZZ, \ZZ_2\, (\ZZ^2, \ZZ)$ & - \\ \hline
A$^{\cal CM}$ & $0$ & $1$ & $\ZZ,0\, (\ZZ,0)$ & -  \\
AIII$^{{\cal M}_-}$ & $1$ & $1$ & - & $\ZZ,0\, (\ZZ,0)$ \\ 
\hline \hline
\end{tabular*}
\caption{Classification of mirror-symmetric corners ($d=2$) or hinges ($d=3$) of
  second-order topological insulators and superconductors in the complex
  Altland-Zirnbauer classes with unitary symmetries or antisymmetries. The first
  two entries in the fourth and fifth column give the intrinsic,
  termination-independent, classification without and with perturbations that
  locally break mirror symmetry at the corner or hinge. The entries between
  brackets give the corresponding extrinsic, termination-dependent
  classification. The ordering is $\mathcal{K}_{\rm i}, \mathcal{\bar K}_{\rm
  i} (\mathcal{K}_{\rm e}, \mathcal{\bar K}_{\rm e})$.
  \label{tab:4}}
\end{table}

\begin{table}[t]
\begin{tabular*}{\columnwidth}{l @{\extracolsep{\fill}} cccc}
\hline\hline 
AZ class & $s$ & $d=2$ & $d=3$ \\ \hline
A$^{{\cal T}^+{\cal M}}$ & $0$ & - & - \\
AIII$^{{\cal T}^+{\cal M}_+}$ & $1$ & $\ZZ_2, \ZZ_2\, (\ZZ, \ZZ)$ & - \\
A$^{{\cal P}^+{\cal M}}$ & $2$ & $\ZZ_2, 0\, (\ZZ_2,0)$  & $\ZZ_2, \ZZ_2\, (\ZZ, \ZZ)$ \\
AIII$^{{\cal T}^-{\cal M}_-}$ & $3$ & $0,0\, (\ZZ_2,0)$ &  $\ZZ_2, 0\, (\ZZ_2,0)$ \\
A$^{{\cal T}^-{\cal M}}$ & $4$ & - &  $0,0\, (\ZZ_2,0)$ \\
AIII$^{{\cal T}^-{\cal M}_+}$ & $5$ & $0, 0\, (2 \ZZ, 2 \ZZ)$ & - \\
A$^{{\cal P}^-{\cal M}}$ & $6$ & - & $0, 0\, (2 \ZZ, 2 \ZZ)$  \\
AIII$^{{\cal T}^+{\cal M}_-}$ & $7$ & - & - \\ 
\hline \hline
\end{tabular*}
\caption{Classification of mirror-symmetric corners ($d=2$) or hinges ($d=3$) of second-order topological insulators and superconductors in the complex Altland-Zirnbauer classes with antiunitary symmetries or antisymmetries. The ordering is $\mathcal{K}_{\rm i}, \mathcal{\bar K}_{\rm i} (\mathcal{K}_{\rm e}, \mathcal{\bar K}_{\rm e})$.
  \label{tab:5}}
\end{table}

\begin{table}[t]
\begin{tabular*}{\columnwidth}{l @{\extracolsep{\fill}} ccccc}
\hline\hline 
class & $s$ & $t$ & $d=2$ & $d=3$ \\ \hline
AI$^{{\cal M}_{+}}$ & $0$ & $0$ & - & - \\
BDI$^{{\cal M}_{++}}$ & $1$ & $0$ & $\ZZ,\ZZ_2\, (\ZZ^2,\ZZ)$ & - \\ 
D$^{{\cal M}_{+}}$ & $2$ & $0$ & $\ZZ_2,\ZZ_2\, (\ZZ_2^2,\ZZ_2)$ & $\ZZ,\ZZ_2\, (\ZZ^2,\ZZ)$  \\
DIII$^{{\cal M}_{++}}$ & $3$ & $0$ & $\ZZ_2, \ZZ_2\, (\ZZ_2^2, \ZZ_2)$ & $\ZZ_2,\ZZ_2\, (\ZZ_2^2,\ZZ_2)$ \\
AII$^{{\cal M}_{+}}$ & $4$ & $0$ & - & $\ZZ_2, \ZZ_2\, (\ZZ_2^2, \ZZ_2)$ \\
CII$^{{\cal M}_{++}}$ & $5$ & $0$ & $2 \ZZ, \ZZ_2\, (2 \ZZ^2, 2\ZZ)$ & - \\
C$^{{\cal M}_{+}}$ & $6$ & $0$ & - & $2 \ZZ, \ZZ_2\, (2 \ZZ^2, 2\ZZ)$ \\
CI$^{{\cal M}_{++}}$ & $7$ & $0$ & - & - \\
 \hline
AI$^{{\cal CM}_{-}}$  & $0$ & $1$ & - & -  \\
BDI$^{{\cal M}_{+-}}$ & $1$ & $1$ & - & - \\ 
D$^{{\cal CM}_{+}}$  & $2$ & $1$ & $\ZZ,\ZZ_2\, (\ZZ,\ZZ_2)$ & - \\
DIII$^{{\cal M}_{-+}}$ & $3$ & $1$ & $\ZZ_2, \ZZ_2\, (\ZZ_2, \ZZ_2)$ & $\ZZ,\ZZ_2\, (\ZZ,\ZZ_2)$ \\
AII$^{{\cal CM}_{-}}$  & $4$ & $1$ & $\ZZ_2,0\, (\ZZ_2,0)$ & $\ZZ_2, \ZZ_2\, (\ZZ_2, \ZZ_2)$ \\
CII$^{{\cal M}_{+-}}$ & $5$ & $1$ & - & $\ZZ_2,0\, (\ZZ_2,0)$ \\
C$^{{\cal CM}_{+}}$  & $6$ & $1$ & $2\ZZ,0\, (2\ZZ,0)$ &  - \\
CI$^{{\cal M}_{-+}}$ & $7$ & $1$ & - & $2\ZZ,0\, (2\ZZ,0)$ \\
 \hline
AI$^{{\cal M}_{-}}$ & $0$ & $2$ & - &  $2 \ZZ,0\, (2 \ZZ, 0)$ \\
BDI$^{{\cal M}_{--}}$ & $1$ & $2$ & $0, 0\, (2 \ZZ, 2 \ZZ)$ & - \\ 
D$^{{\cal M}_{-}}$ & $2$ & $2$ & - &  $0, 0\, (2 \ZZ, 2 \ZZ)$ \\
DIII$^{{\cal M}_{--}}$ & $3$ & $2$ & $2\ZZ, \ZZ_2\, (2\ZZ, \ZZ_2)$ & - \\
AII$^{{\cal M}_{-}}$ & $4$ & $2$ & - &  $2\ZZ, \ZZ_2\, (2\ZZ, \ZZ_2)$ \\
CII$^{{\cal M}_{--}}$ & $5$ & $2$ & $\ZZ_2, \ZZ_2\, (2 \ZZ, 2\ZZ)$ & - \\
C$^{{\cal M}_{-}}$ & $6$ & $2$ & - & $\ZZ_2, \ZZ_2\, (2 \ZZ, 2\ZZ)$ \\
CI$^{{\cal M}_{--}}$ & $7$ & $2$ & $2\ZZ,0\, (2 \ZZ, 0)$ & - \\
 \hline
AI$^{{\cal CM}_{+}}$ & $0$ & $3$ & $\ZZ,0\, (\ZZ,0)$ & - \\
BDI$^{{\cal M}_{-+}}$ & $1$ & $3$ & $0, 0\, (\ZZ_2, 0)$ & $\ZZ,0\, (\ZZ,0)$ \\ 
D$^{{\cal CM}_{-}}$ & $2$ & $3$ & $0,0\, (\ZZ_2,0)$ & $0, 0\, (\ZZ_2, 0)$ \\
DIII$^{{\cal M}_{+-}}$ & $3$ & $3$ & - & $0,0\, (\ZZ_2,0)$ \\
AII$^{{\cal CM}_{+}}$ & $4$ & $3$ & $2\ZZ,0\, (2\ZZ,0)$ & - \\
CII$^{{\cal M}_{-+}}$ & $5$ & $3$ & - & $2\ZZ,0\, (2\ZZ,0)$  \\
C$^{{\cal CM}_{-}}$ & $6$ & $3$ & - &  - \\
CI$^{{\cal M}_{+-}}$ & $7$ & $3$ & - & - \\
\hline \hline
\end{tabular*}
\caption{Classification of mirror-symmetric corners ($d=2$) or hinges ($d=3$) of second-order topological insulators and superconductors in the real Altland-Zirnbauer classes. The ordering is $\mathcal{K}_{\rm i}, \mathcal{\bar K}_{\rm i} (\mathcal{K}_{\rm e}, \mathcal{\bar K}_{\rm e})$.\label{tab:6}}
\end{table}

We denote the classifying groups for corners according to the four possible
classification rules  that arise from the above considerations as
$\mathcal{K}_{\rm i}(s,t|d_{\parallel},d)$, $\mathcal{\bar K}_{\rm
i}(s,t|d_{\parallel},d)$, $\mathcal{K}_{\rm e}(s,t|d_{\parallel},d)$, and
$\mathcal{\bar K}_{\rm e}(s,t|d_{\parallel},d)$, where the subscripts ${\rm
i}$, ${\rm e}$ refer to intrinsic (termination-independent)  and extrinsic
(termination-dependent) classification and the bar refers to corners or hinges
with locally broken mirror reflection symmetry. For mirror reflection
$d_{\parallel} = 1$ throughout. (The second argument is omitted for the complex
Altland-Zirnbauer classes with antiunitary symmetries and antisymmetries.)
Tables \ref{tab:4}-\ref{tab:6} contain the complete classification results,
ordered as $\mathcal{K}_{\rm i}$, $\mathcal{\bar K}_{\rm i}$ ($\mathcal{K}_{\rm
e}$, $\mathcal{\bar K}_{\rm e}$).

\begin{figure}
\includegraphics[width=0.59\columnwidth]{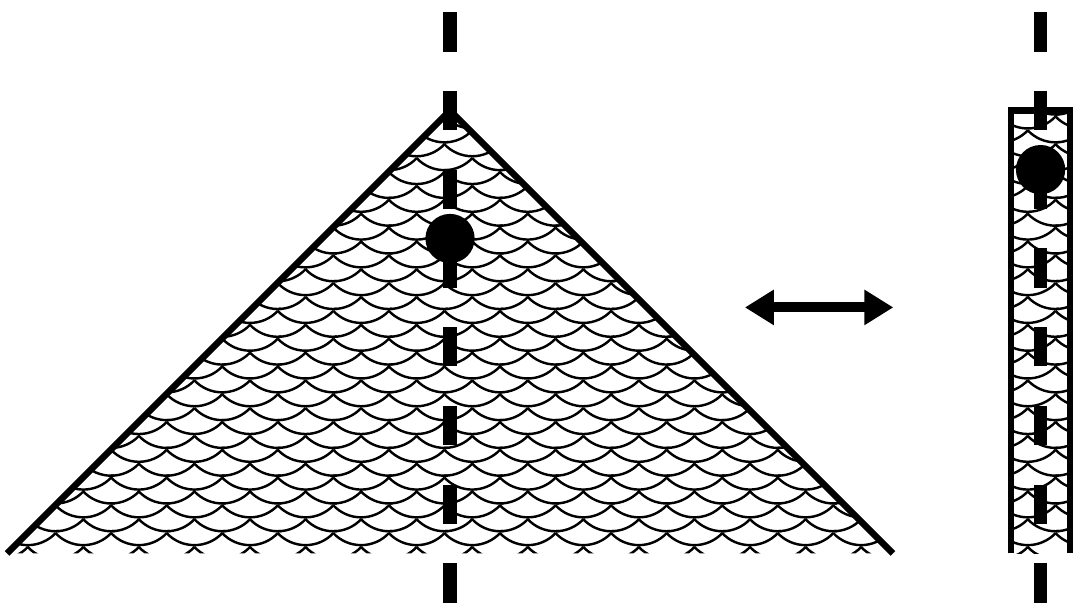}
\caption{\label{fig:iso1} The extrinsic classification of corner states in a mirror-symmetric corner of a two-dimensional crystal is the same as that of end states of a one-dimensional crystal with a transverse mirror symmetry with $d_{\parallel} = 0$. The vertical dashed line is the mirror axis.}
\end{figure}

Although we will explain the derivation of each entry in the table in detail
below and in the appendix, we first outline the general strategy that results
in this classification. Our first observation is that the extrinsic,
termination-dependent, classification of mirror symmetric corners/hinges is
identical to the classification of end states of $(d-1)$-dimensional insulators
and superconductors with a crystalline symmetry with $d_{\parallel}-1=0$
inverted coordinates, see Fig.\ \ref{fig:iso1} for $d=2$. By the bulk-boundary
correspondence, this latter classification is identical to the corresponding
bulk crystalline classification, so that we have
\begin{equation}
  \mathcal{K}_{\rm e}(s,t|d_{\parallel}=1,d) = K(s,t|d_\parallel=0,d-1).
\end{equation}
The classifying groups $K(s,t|d_\parallel=0,d-1)$ are given in Ref.\
\onlinecite{shiozaki2014}. They can also be obtained from Tables
\ref{tab:SS1}--\ref{tab:SS3} using the isomorphisms
(\ref{eq:K1})--(\ref{eq:K3}). Similarly, upon locally breaking the mirror
symmetry, we obtain the equality
\begin{equation}
  \mathcal{\bar K}_{\rm e}(s,t|1,d) = K(s,t|0,d-1)/K'(s,t|0,d-1),
\end{equation}
where $K'(s,t|d_\parallel=0,d-1)$ is the ``purely crystalline'' subgroup of the
classifying group $K(s,t|d_\parallel=0,d-1)$, see the discussion at the end of
Sec.\ \ref{sec:2}.

The intrinsic, termination-independent, classification of mirror-symmetric
corners or hinges can be obtained via the homomorphism
\begin{equation}
	K(s,d-1)\overset{c_t}{\rightarrow} \mathcal{K}_{\rm e}(s,t|d_{\parallel}=1,d),
\end{equation}
which embeds the equivalence class of the Hamiltonian $H(\vk)$ into
corresponding Shiozaki-Sato class of Hamiltonian
\begin{align}
	c_t[H(\bm k)]&=
	\begin{pmatrix}
		H(\bm k) & 0 \\
		0 & \sigma_\mathcal{S} U_\mathcal{S} H(\bm k)U_\mathcal{S}^\dagger
	\end{pmatrix},
	\label{eq:ct}
\end{align}
for $U_\mathcal{S}$ a unitary onsite symmetry ($\sigma_{\mathcal{S}} = 1$) or
antisymmetry ($\sigma_{\mathcal{S}} = -1$) and
\begin{align}
	c_t[H(\bm k)]&=
	\begin{pmatrix}
		H(\bm k) & 0 \\
		0 & \sigma_\mathcal{S} U_\mathcal{S} H^*(-\bm k)U_\mathcal{S}^\dagger
	\end{pmatrix},
	\label{eq:ct}
\end{align}
for $U_\mathcal{S}$ an antiunitary onsite symmetry or antisymmetry. 

For the intrinsic classification corner states or hinge modes that differ by
termination effects are identified. Such corner states or hinge modes are
precisely those in the image $c_t[K(s,d-1)]$, so that we have

\begin{align}
	\mathcal{K}_{\rm i}(s,t|1,d) = \mathcal{K}_{\rm e}(s,t|1,d)/c_t[K(s,d-1)].
  \label{eq:KsKw}
\end{align}
In other words, the elements of the group $\mathcal{K}_{\rm i}(s,t|1,d)$ can be viewed as
topologically non-trivial $d-1$-dimensional crystalline insulators or
superconductors with an onsite twofold symmetry that cannot be obtained by
gluing two corresponding non-crystalline $d-1$-dimensional topological
insulators or superconductors. 


In the next Subsection we demonstrate, by explicit consideration of all
symmetry classes, that the intrinsic (termination-independent) classification
group $\mathcal{K}_{\rm i}(s,t|1,d)$ of corner or hinge states is identical to
the ``pure crystalline'' group $K'(s,t|1,d)$ classifying topological
crystalline bulk phases that are not at the same time strong topological
phases,
\begin{equation}
  \mathcal{K}_{\rm i}(s,t|1,d) =K'(s,t|1,d).
  \label{eq:Ks_TCI}
\end{equation}
Equation (\ref{eq:Ks_TCI}) says that a mirror-symmetric topological crystalline
phase is either a strong topological phase, with gapless modes at all
boundaries, or a topological crystalline phase which can be {\em uniquely}
characterized using protected modes at mirror-symmetric corners (for a
two-dimensional crystal) or hinges (for a three-dimensional crystal). For such
``pure crystalline'' topological crystalline phases Eq.\ (\ref{eq:Ks_TCI}) this
extends the bulk-boundary correspondence to a ``corner-to-bulk correspondence''
or ``hinge-to-bulk correspondence''.


We now discuss the classification table for the complex Altland-Zirnbauer
classes with unitary mirror symmetries and antisymmetries in
detail. The classification of mirror-symmetric corners of 
two-dimensional crystals for the complex Altland-Zirnbauer classes 
with antiunitary mirror symmetries and antisymmetries and of the real 
classes is given in Appendix~\ref{app:2}. The classification of
mirror-symmetric hinges for these classes can be obtained from the 
dimensional reduction scheme of Sec.~\ref{sec:3} and is
not discussed in detail.

{\em Class A$^{\cal M}$, $(s,t) = (0,0)$, $d=2$.---}
This class does not allow protected zero energy states at corners.

{\em Class AIII$^{{\cal M}_{+}}$, $(s,t)=(1,0)$, $d=2$.---}
At a mirror-symmetric corner zero-energy states can be counted according to
their parity under mirror reflection ${\cal M}$ and the chiral operation ${\cal
C}$, since ${\cal M}$ and ${\cal C}$ commute. (Recall that we use the convention that the
mirror operation ${\cal M}$ squares to one.) We denote the number of
corresponding modes with $N_{\sigma_{{\cal C},{\cal M}}}$. Since pairs of zero
modes with opposite $\sigma_{{\cal C}}$ but equal $\sigma_{{\cal M}}$ can be
gapped out by a mirror-symmetric mass term acting locally at the corner, only
$N_{++} - N_{-+}$ and $N_{+-} - N_{--}$ are well defined. This gives the $\ZZ^2$
extrinsic classification listed in Table~\ref{tab:4}.

\begin{figure}
\includegraphics[width=0.49\columnwidth]{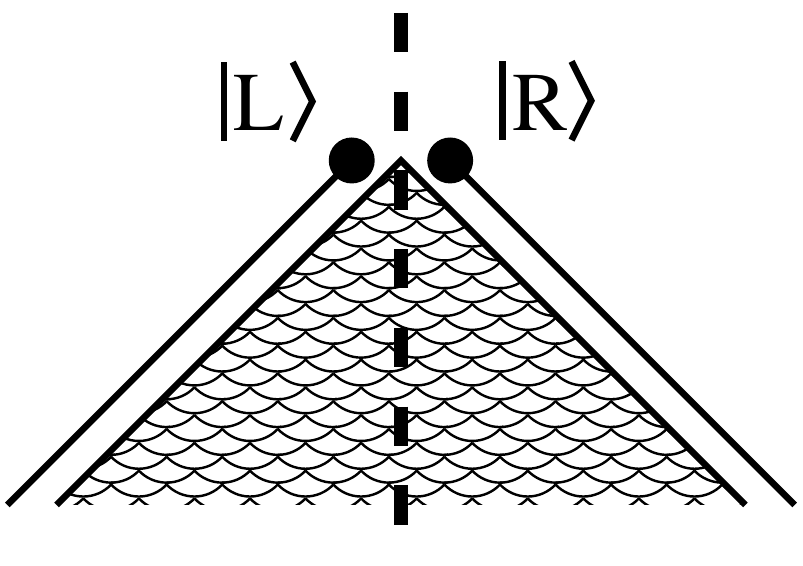}
\caption{\label{fig:decoration} Pairs of corner states may be created by ``glueing'' one-dimensional topologically nontrivial chain to the crystal edges. Mirror symmetry requires that the chains added to mirror-related edges are mirror images of each other.}
\end{figure}

By changing the termination, {\em e.g.} by adding a suitably chosen chain of
atoms on a crystal face, such that the global mirror symmetry is preserved, one can add a pair of zero modes with
the same $\sigma_{\cal C}$, but opposite values of $\sigma_{\cal M}$, see
Fig.~\ref{fig:decoration}. (Note that such a procedure involves closing the boundary gap.) As a result, the difference $N = N_{++} + N_{--} -
N_{+-} - N_{-+}$ is the only remaining invariant, and one finds a $\ZZ$ intrinsic
classification, which is the same classification as the one arising from the
bulk classification of Refs.\
\onlinecite{chiu2013,morimoto2013,shiozaki2014,chiu2016,trifunovic2017}. 

With a mirror-symmetry-breaking local perturbation at the corner, one may only
distinguish corner states by their parity under ${\cal C}$. We use
$N_{\sigma_{\cal C}}$ to denote the number of zero modes with parity
$\sigma_{\cal C}$. Since pairs of zero modes with opposite $\sigma_{{\cal
C}}$ can be gapped out by a mass term acting locally at the corner, only the
difference $N_{+} - N_{-}$ is well defined. This gives a $\ZZ$ extrinsic
classification in the presence of a mirror-symmetry-breaking perturbation.
Moreover, changes of the termination allow one to change $N_{+}$ or $N_{-}$ by
an even number, resulting in a $\ZZ_2$ intrinsic classification in that case.

{\em Class A$^{\cal CM}$, $(s,t)=(0,1)$, $d=2$.---}
This class allows corner modes only if the mirror antisymmetry is not broken
locally at the crystal corner. In that case, corner modes can be counted
according to their parity $\sigma_{\cal CM}$ under the mirror antisymmetry ${\cal
CM}$. (Recall that we use the convention that ${\cal CM}^2 = 1$.) The mirror antisymmetry
protects zero modes at the same value of $\sigma_{\cal CM}$, but allows pairs of
zero modes at opposite mirror parity $\sigma_{\cal CM}$ to gap out. We conclude
that the difference $N = N_{+} - N_{-}$ is the corresponding topological invariant,
giving the $\ZZ$ classification listed in Table~\ref{tab:4}. There is no
difference between an ``extrinsic'' and an ``intrinsic'' classification because the
Altland-Zirnbauer class A is trivial for $d=1$, so that no protected zero modes
can be added by changing the lattice termination. This phase is trivial if the mirror antisymmetry is broken locally at the corner.

{\em Class AIII$^{{\cal M}_-}$, $(s,t) = (1,1)$, $d=2$.---}
The bulk crystalline phase in this class is trivial. However, since the
Altland-Zirnbauer class AIII is nontrivial in one dimension, one should
consider the possibility that corner states can arise by suitable decoration at
the crystal edges, see Fig.~\ref{fig:decoration}. Hereto, consider the addition
of two one-dimensional chains with zero-energy end states, labeled $\ket{\rm
L}$ and $\ket{\rm R}$. 
The chains are placed symmetrically, so that $\ket{\rm L} = {\cal
M} \ket{\rm R}$. Since ${\cal M}$ anticommutes with ${\cal C}$, the end states
$\ket{\rm L}$ and $\ket{\rm R}$ have opposite parity under ${\cal C}$. Upon coupling the chains to each other, a term $\ket{\rm L} \bra{\rm R} +
\ket{\rm R} \bra{\rm L}$ that gaps the two zero modes out is allowed under
${\cal C}$ antisymmetry and mirror reflection symmetry. Hence, we conclude that no
stable corner states can be created by changing the lattice termination.
(Alternatively, one may note that a mirror reflection operation that
anticommutes with ${\cal C}$ can be viewed as a valid term in the Hamiltonian,
which gaps out zero-energy states on the left and right of the corner.)

We point out that whereas in this symmetry class a mirror-symmetric corner does
not allow for protected zero-energy states, a generic corner still does. The
reason is that in a generic corner one may separately choose lattice
terminations at both edges that meet at that corner,
whereas in a mirror-symmetric corner the lattice terminations at the edges
meeting in that corner are symmetry-related. 

{\em Class A$^{{\cal M}}$, $(s,t) = (0,0)$, $d=3$.---} We use $y$ to denote the coordinate running along the hinge. Hinge modes can be characterized by their mirror parity $\sigma_{\cal M}$ and by their propagation direction in the $y$ direction. Whereas counterpropagating modes with the same mirror parity can mutually gap out, counterpropagating hinge modes constructed with opposite $\sigma_{\cal M}$ are protected by mirror symmetry. Using $N_{\sigma_{\cal M}\pm}$ to denote the number of hinge modes of mirror parity $\sigma_{\cal M}$ propagating in the $\pm y$ direction, the differences $N_{++} - N_{+-}$ and $N_{-+} - N_{--}$ are two well-defined integer extrinsic topological invariants, consistent with the $\ZZ^2$ extrinsic classification of gapless hinge states. 

By adding, {\em e.g.}, integer quantum Hall insulators on the mirror-related faces adjacent to the hinge, two co-propagating hinge modes with opposite mirror parity can be created, leaving $N_{++} - N_{+-} - N_{-+} + N_{--}$ as the only remaining intrinsic integer topological invariant. If mirror symmetry is broken locally at the hinge, all counterpropagating modes can in principle be gapped out, giving rise to $\ZZ$ and $\ZZ_2$ extrinsic and intrinsic topological invariants, respectively. 

{\em Class AIII$^{{\cal M}_{+}}$, $(s,t) = (1,0)$, $d=3$.---} This class does not allow for topologically protected hinge modes.

{\em Class A$^{{\cal CM}}$, $(s,t) = (0,1)$, $d=3$.---} The mirror antisymmetry rules out the existence of protected hinge modes for this class --- recall that for a mirror-symmetric hinge the mirror antisymmetry ${\cal CM}$ is effectively a {\em local} operation. Whereas a single dispersing hinge mode can not be an eigenmode of the antisymmetry ${\cal CM}$, two modes $\ket{\rm L}$ and $\ket{\rm R} = {\cal CM}\ket{R}$ can be gapped out by the mirror-antisymmetric perturbation $i(\ket{\rm L} \bra{\rm R} - \ket{\rm R} \bra{\rm L})$. Note that for class A$^{{\cal CM}}$ a generic hinge may still carry a protected hinge mode. (Compare with the discussion of class AIII$^{{\cal M}_-}$ for $d=2$.)

{\em Class AIII$^{{\cal M}_{-}}$, $(s,t) = (1,1)$, $d=3$.---}
The hinge modes can be chosen to have a well-defined mirror parity $\sigma_{\cal M}$. Since ${\cal M}$ anticommutes with ${\cal C}$, they occur as doublets with opposite $\sigma_{\cal M}$ and opposite propagation direction. For each doublet the ``mixed parity'' $\sigma$, the product of propagation direction and mirror parity $\sigma_{\cal M}$, is well-defined. The corresponding integer invariant $N$ counts the difference of the number of such doublets with positive and negative $\sigma$. Since Altland-Zirnbauer class AIII is trivial in two dimensions, there is no difference between an extrinsic and intrinsic classifications for this class. Breaking the mirror symmetry locally at the hinge removes the protection of the hinge modes.

\subsection{From bulk crystalline phase to second-order phase}
\label{sec:4b}

\begin{figure}
\includegraphics[width=0.99\columnwidth]{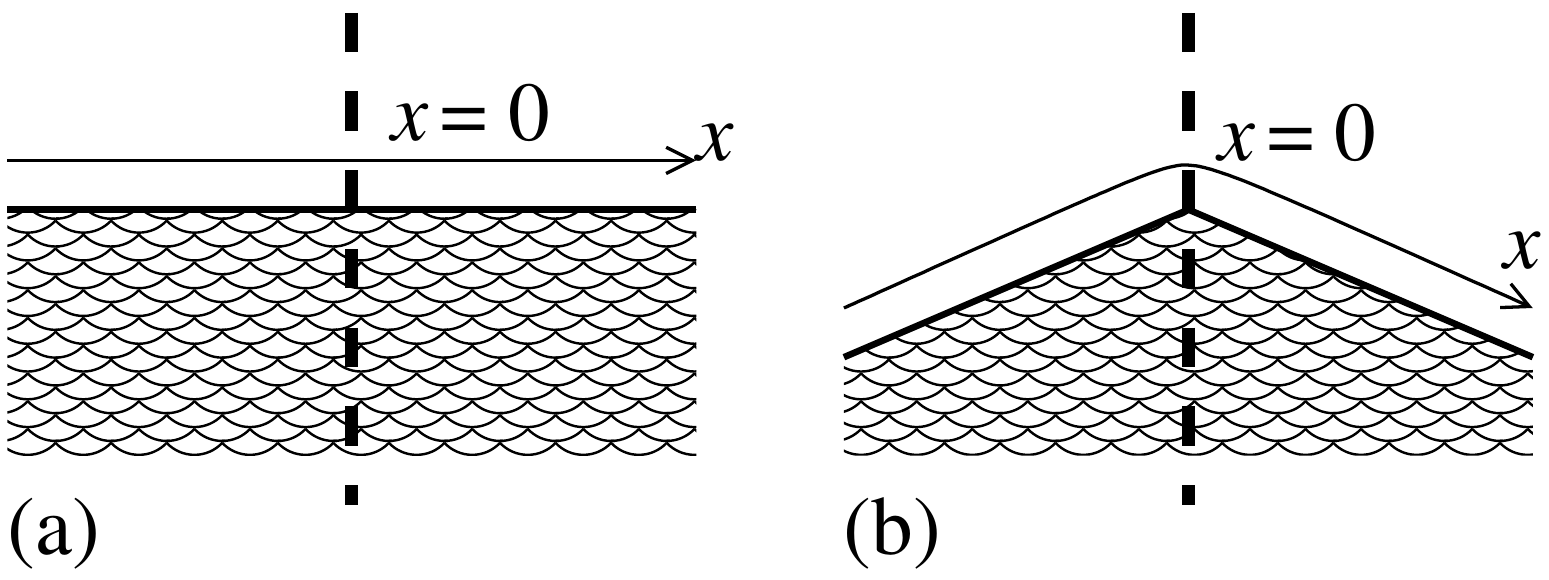}
\caption{\label{fig:edge_deformation} A mirror-symmetric edge, with coordinate $x$ running along the edge (a) can be deformed into a corner joining two mirror-related edges (b). The situation shown in (a) has mirror symmetry acting everywhere along the edge; in (b) mirror symmetry exists only for a mirror reflection axis going through the corner at $x=0$.}
\end{figure}

The above classification of tables~\ref{tab:4}--\ref{tab:6} is based on a
classification of zero-energy states localized at corners and gapless modes at
hinges only. To make a connection with the bulk topology we use
the bulk-boundary correspondence for mirror-symmetric topological crystalline
insulators, which uniquely connects the bulk crystalline phase with the existence of gapless boundary
modes at boundaries that are invariant under the mirror reflection operation. 

In a two-dimensional crystal the edge is one dimensional and we can introduce a
coordinate $x$ running along the edge. If the boundary is tilted slightly away
from the invariant direction, such that a corner connecting to mirror-related
edges emerges at $x=0$, as shown schematically in
Fig.~\ref{fig:edge_deformation}, generically a mass term is generated, which is
{\em odd} under the mirror reflection operation ${\cal M}$. Such a mass term
gaps out the edge states, but the fact that it is odd under mirror reflection
implies the existence of a domain wall and an associated zero-energy state at
the corner at $x=0$. There is a one-to-one relationship between the number of
topologically protected edge modes and the number of zero modes obtained in this
way --- with the caveats that such zero modes may be annihilated by local
mirror-symmetry breaking perturbations at the corner and that additional zero
modes may be generated by a modification of the lattice termination. In a
three-dimensional crystal in principle the same arguments apply, with the only
modification that in this case the invariant boundary is a surface.

Reference \onlinecite{langbehn2017} has implemented this construction
for all Shiozaki-Sato classes that have unitary mirror symmetries, and for which
the mass term is unique. A unique mass term guarantees that a single corner or
hinge mode cannot be gapped out by a perturbation that breaks the mirror
symmetry locally at the corner. To complete the discussion of the complex
Altland-Zirnbauer classes with a unitary mirror symmetry, we here discus how
the presence of gapless states at a mirror-symmetric edge or surface gives rise to zero-energy corner states at mirror-symmetric corners or gapless hinge modes at mirror-symmetric hinges. Comparing to the analysis of the previous Subsection, we thus verify that we precisely recover the zero-energy corner state found by inspection of the corner alone. Appendix~\ref{app:3} carries out the same program for the remaining Shiozaki-Sato classes.

{\em Class AIII$^{{\cal M}_{+}}$, $(s,t)=(1,0)$, $d=2$.---}
For concreteness, we use $U_{\cal C} = \sigma_3$ and $U_{\cal M} = \sigma_3 \tau_3$ to represent the commuting operations ${\cal C}$ and ${\cal M}$. The bulk phase has a $\ZZ$ classification\cite{chiu2013,morimoto2013,shiozaki2014} with an integer topological invariant $N$, which counts the difference of counterpropagating pairs of edge modes with positive and negative mixed parity $\sigma_{{\cal MC}}$ at zero energy. (Although the product ${\cal MC}$ is an antisymmetry of the Hamiltonian, not a symmetry, edge modes can be chosen to be eigenmodes of ${\cal MC}$ at zero energy. Pairs of counterpropagating edge modes can not mutually gap out if they have the same eigenvalue $\sigma_{{\cal M C}}$.) After a suitable basis transformation and rescaling, the Hamiltonian of a ``minimal'' edge, in which all gapless modes have the same mixed parity $\sigma_{{\cal MC}}$, may be written as
\begin{equation}
  H_{\rm edge} = - i v \sigma_1 \partial_x \openone_N,
\end{equation}
where $x$ is the coordinate along the edge, see Fig.~\ref{fig:edge_deformation}a, $\openone_N$ the $N \times N$ unit matrix, and $v$ a constant with the dimension of velocity.
A corner between two mirror-related edges meeting
at $x=0$, as shown in Fig.~\ref{fig:edge_deformation}b, is represented by a mass
term $m(x) \sigma_2$ with $m(x) = -m(-x)$ a $N \times N$ hermitian matrix. The
eigenvalues of $m(x)$ have ``domain walls'' at $x=0$, allowing for $N$ zero
modes localized around $x=0$. The bulk theory does not specify the sign of the
limiting values of the eigenvalues of the mass term $m(x)$ at a large distance from the corner. The two
choices for this sign give corner states with different parity eigenvalues
$\sigma_{\cal C}$ and $\sigma_{\cal M}$, but the same value of $\sigma_{{\cal MC}}
= \sigma_{\cal C} \sigma_{\cal M}$: A domain wall with $m(x) > 0$ for $x \gg 0$
gives a solution with $\sigma_{{\cal C}} = \sigma_{{\cal MC}} \sigma_{{\cal M}} =
+$, whereas a domain wall with $m(x) < 0$ for $x \gg 0$ gives a solution with
$\sigma_{{\cal C}} = \sigma_{{\cal MC}} \sigma_{{\cal M}} = -$. One verifies that
if mirror symmetry is present locally around $x=0$, neither perturbations coupling
such zero-energy states with the same value of $\sigma_{{\cal MC}} = \sigma_{\cal C} \sigma_{\cal M}$, nor perturbations coupling zero-energy states with different values of $\sigma_{\cal M}$ are allowed. 

The analysis of corner states of the previous Subsection counted their numbers
$N_{\sigma_{\cal C},\sigma_{\cal M}}$ with parities $\sigma_{\cal C}$ and
$\sigma_{\cal M}$ (at zero energy) and found that the differences $N_{++} -
N_{-+}$ and $N_{+-} - N_{--}$ are the extrinsic topological invariants, whereas
$N = N_{++} + N_{--} - N_{+-} - N_{-+}$ is the intrinsic topological invariant.
The above analysis provides a confirmation of the differences $N_{++} - N_{-+}$
and $N_{+-} - N_{--}$ as extrinsic, termination-dependent invariants, and
identifies the intrinsic invariant $N$ describing the corner states with the
bulk topological invariant $N$.

{\em Class A$^{\cal CM}$, $(s,t)=(0,1)$, $d=2$.---}
This phase has a $\ZZ^2$ bulk classification,\cite{shiozaki2014,trifunovic2017} with a purely crystalline classifying group $K' = \ZZ$. The first-order (strong) topological phase has chiral edge modes. For a second-order topological phase we restrict ourselves to purely crystalline topological phases with equal numbers of counterpropagating modes. The corresponding integer index $N$ counts the difference of the numbers of pairs of counterpropagating edge modes with positive and negative parity $\sigma_{\cal CM}$ at zero energy. (One verifies that a pair of counterpropagating modes can not mutually gap out if both modes have the same parity under ${\cal CM}$, {\em i.e.}, the same eigenvalue of $U_{\cal CM}$.) For a minimal edge, in which all edge modes have the same parity $\sigma_{\cal CM}$, we may represent the mirror antisymmetry with the unit matrix, $U_{\cal CM} = 1$. After a suitable rescaling and basis transformation, the edge Hamiltonian may then be written as
\begin{equation}
  H_{\rm edge} = -i v \sigma_3 \partial_x \openone_N,
\end{equation}
where $x$ is the coordinate along the crystal edge, $\openone_N$ is the $N
\times N$ unit matrix, and $\sigma_3$ a Pauli matrix acting on pairs of
counterpropagating modes. Although mirror antisymmetry does
not allow a uniform mass term, a mass term $m_1(x) \sigma_1 + m_2(x) \sigma_2$ in
which $m_1(x)$ and $m_2(x)$ are hermitian $N \times N$ matrix-valued
antisymmetric functions of $x$ is allowed if the edge is deformed into two
mirror-related edges meeting in a corner at $x=0$. Such a mass term allows for
$N$ zero-energy states localized near $x=0$. No further topology or symmetry related numbers can be associated with the zero-energy states, consistent with the integer classification obtained by inspection of corner states given in the previous Subsection.

\begin{figure}
\includegraphics[width=0.89\columnwidth]{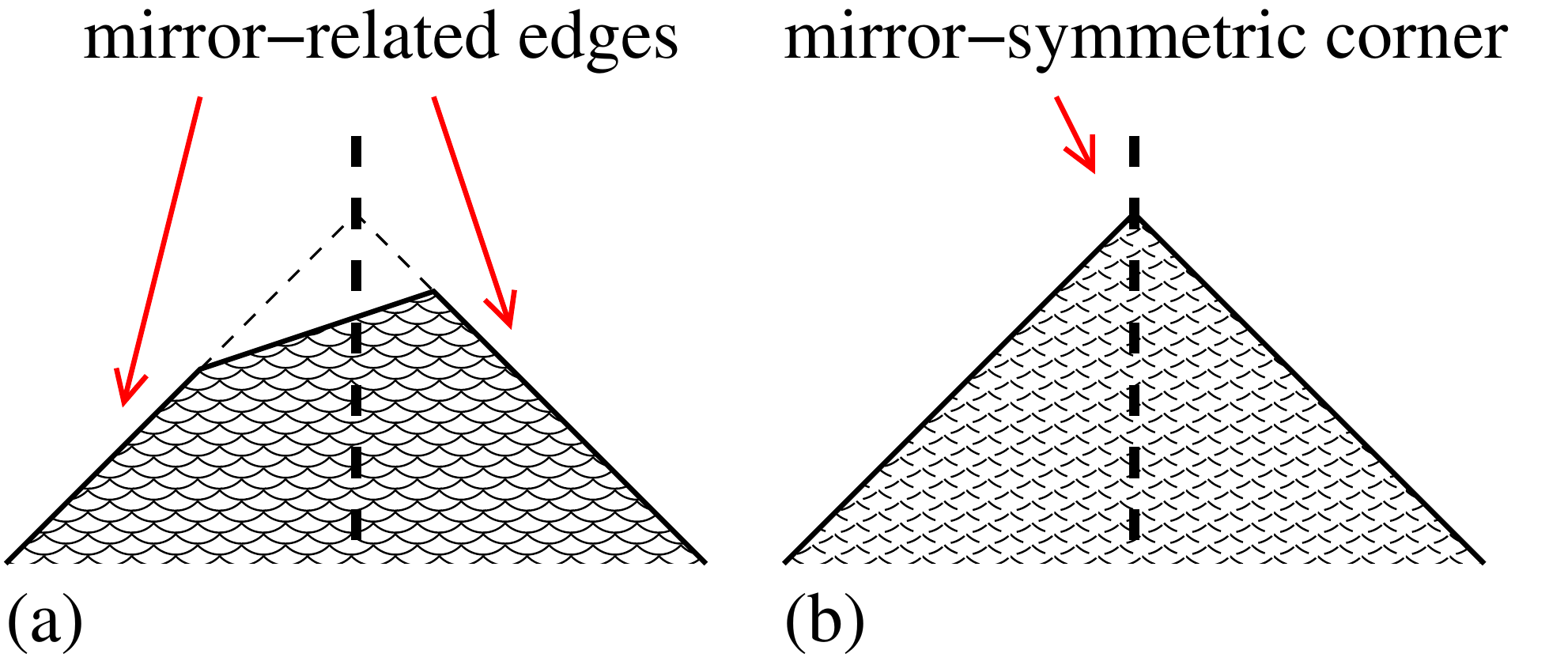}
\caption{\label{fig:nocorner} (a) A two-dimensional crystal with a pair of mirror-related edges, but without a mirror-symmetric corner. (b) The crystal may be smoothly deformed into a crystal with a mirror-symmetric corner. The parity of the number of zero-energy states between the two mirror-related edges in (a) is the same as the parity of the number of zero-energy states at the mirror-symmetric corner in (b).}
\end{figure}

{\em Class $A^{{\cal M}}$, $(s,t) = (0,0)$, $d=3$.---} We use $U_{\cal M} = \sigma_2$ to represent mirror reflection. This class admits surface states with dispersion $-i v (\sigma_1 \partial_x \pm \sigma_2 \partial_y)$, where the sign $\pm$ defines the ``mirror chirality'' and $x$ and $y$ are coordinates along the surface, such that the mirror reflection sends $x \to -x$. The bulk crystalline phase has a $\ZZ$ topological classification,\cite{chiu2013,morimoto2013,shiozaki2014,trifunovic2017} with an integer topological invariant $N$ equal to the difference of surface states with positive and negative mirror chirality.\cite{teo2008} For a minimal surface, all surface states have the same mirror chirality. With a suitable choice of basis and after rescaling the corresponding surface Hamiltonian reads
\begin{equation}
  \label{eq:HsurfaceAM}
  H_{\rm surface} = -i v (\sigma_1 \partial_x + \sigma_2 \partial_y) \openone_N,\end{equation}
with $\openone_N$ the $N \times N$ identity matrix and $x$ and $y$ coordinates at the invariant surface. The unique mass term $m(x,y) \sigma_3$ with $m(x,y) = - m(-x,y)$ an $N \times N$ hermitian matrix gaps out the surface states. The fact that the mass term is odd under mirror reflection guarantees the existence of gapless hinge modes at mirror-symmetric hinges. 

Considering the surface Hamiltonian (\ref{eq:HsurfaceAM}) with a mass term $m(x) \sigma_3$ with $m(x) = -m(-x)$, the propagation direction of the hinge states and their mirror parity $\sigma_{\cal M}$ are determined by the signs of the eigenvalues of $m(x)$ for $x \gg 0$, such that a positive eigenvalue corresponds to a hinge state with positive $\sigma_{\cal M}$, moving in the positive $y$ direction, whereas a negative eigenvalue corresponds to a hinge state with negative $\sigma_{\cal M}$, moving in the negative $y$ direction. (The mirror parity $\sigma_{\cal M}$ and the propagation direction are opposite if we would have started from a surface Hamiltonian describing surface states with negative mirror chirality.) Counterpropagating hinge modes constructed this way have opposite $\sigma_{\cal M}$ and are, hence, protected by mirror symmetry. Since the sign of $m$ depends on the details of the surface termination, changing the surface termination allows to simultaneously switch the propagation direction $\pm$ and the mirror parity $\sigma_{\cal M}$ of the hinge states, consistent with the intrinsic topological invariant $N_{++} - N_{+-} - N_{-+} + N_{--}$. 

{\em Class AIII$^{{\cal M}_-}$, $(s,t) = (1,1)$, $d=3$.---} 
We choose $U_{\cal C} = \sigma_3$ and $U_{\cal M} = \sigma_2 $ to represent ${\cal C}$ and ${\cal M}$, respectively. This class supports gapless surface states with dispersion $\propto -i v (\sigma_1 \partial_x \pm \sigma_2 \partial_y)$, which defines the chirality $\pm$. The crystalline bulk has a $\ZZ^2$ classification,\cite{chiu2013,morimoto2013,shiozaki2014,trifunovic2017} with purely crystalline classifying group $K' = \ZZ$, see Table \ref{tab:SS1}. The strong integer index counts the number of such surface Dirac cones, weighted by chirality. For a second-order topological phase we are interested in the purely crystalline topological phases, in which the surface carries multiple pairs of Dirac cones of opposite chirality. Their dispersion is $-i v (\sigma_1 \tau_3 \partial_x \pm \sigma_2 \tau_0 \partial_y)$, where the sign $\pm$ defines the mirror chirality and the $\tau_j$, $j=0,1,2,3$, are Pauli matrices acting on a different spinor degree of freedom than the matrices $\sigma_j$, $j=0,1,2,3$. The corresponding (second, purely crystalline) integer topological invariant $N$ counts the number of such pairs of Dirac cones, weighted by mirror chirality. A minimal surface with $N \ge 0$ has surface Hamiltonian
\begin{equation}
  H_{\rm surface} = -i v (\sigma_1 \tau_3 \partial_x + \sigma_2 \tau_0 \partial_y) \openone_N, \label{eq:HsurfaceAIII}
\end{equation}
where $\openone_N$ is the $N \times N$ unit matrix. The mass terms allowed by chiral symmetry and mirror reflection symmetry are $m_1(x,y) \sigma_1 \tau_1 + m_2(x,y) \sigma_1 \tau_2$ with $m_{1,2}(x,y) = -m_{1,2}(-x,y)$ $N \times N$ hermitian matrices. This ensures the presence of gapless hinge modes at mirror-symmetric hinges. One verifies that the surface Hamiltonian (\ref{eq:HsurfaceAIII}) gives $N$ doublets for which the mixed parity, the product of mirror parity $\sigma_{\cal M}$ and the propagation direction, is positive. Similarly, surface Dirac cones with negative mirror chirality give hinge doublets of negative mixed parity, thus allowing one to identify the topological invariants derived from counting gapless hinge states and the (purely crystalline) topological invariant $N$ describing the bulk crystalline topology. 

\subsection{Mirror-symmetric crystals without mirror-symmetric corners}

In principle, a mirror-symmetric crystal need not have mirror-symmetric corners. However, as long as the crystal has at least a pair of mirror-related edges (for a two-dimensional crystal) or a pair of mirror-related faces (for a three-dimensional crystal), the bulk topology determines the {\em parity} of the number of corner or hinge states {\em between} the two mirror-related edges or surfaces. An example of such a situation is shown in Fig.\ \ref{fig:nocorner}. Since such a crystal without mirror-symmetric corners or hinges (but with two mirror-related edges or surfaces) may be smoothly deformed into a crystal with a mirror-symmetric corner without closing the bulk gap, and since corner states and hinge modes can only be generated or annihilated pairwise in such a deformation, one immediately finds that the parity of corner states or hinge modes is the same as the parity of corner states or hinge modes at a mirror-symmetric corner. The corresponding entry in Tables \ref{tab:4}--\ref{tab:6} is the classifying group $\mathcal{\bar K}_{\rm i}$.

\section{Classification of second-order topological insulators and superconductors with twofold rotation and inversion symmetry}
\label{sec:5}

\subsection{Twofold rotation symmetry for $d=3$} \label{sec:5b}

\begin{figure}
\includegraphics[width=0.99\columnwidth]{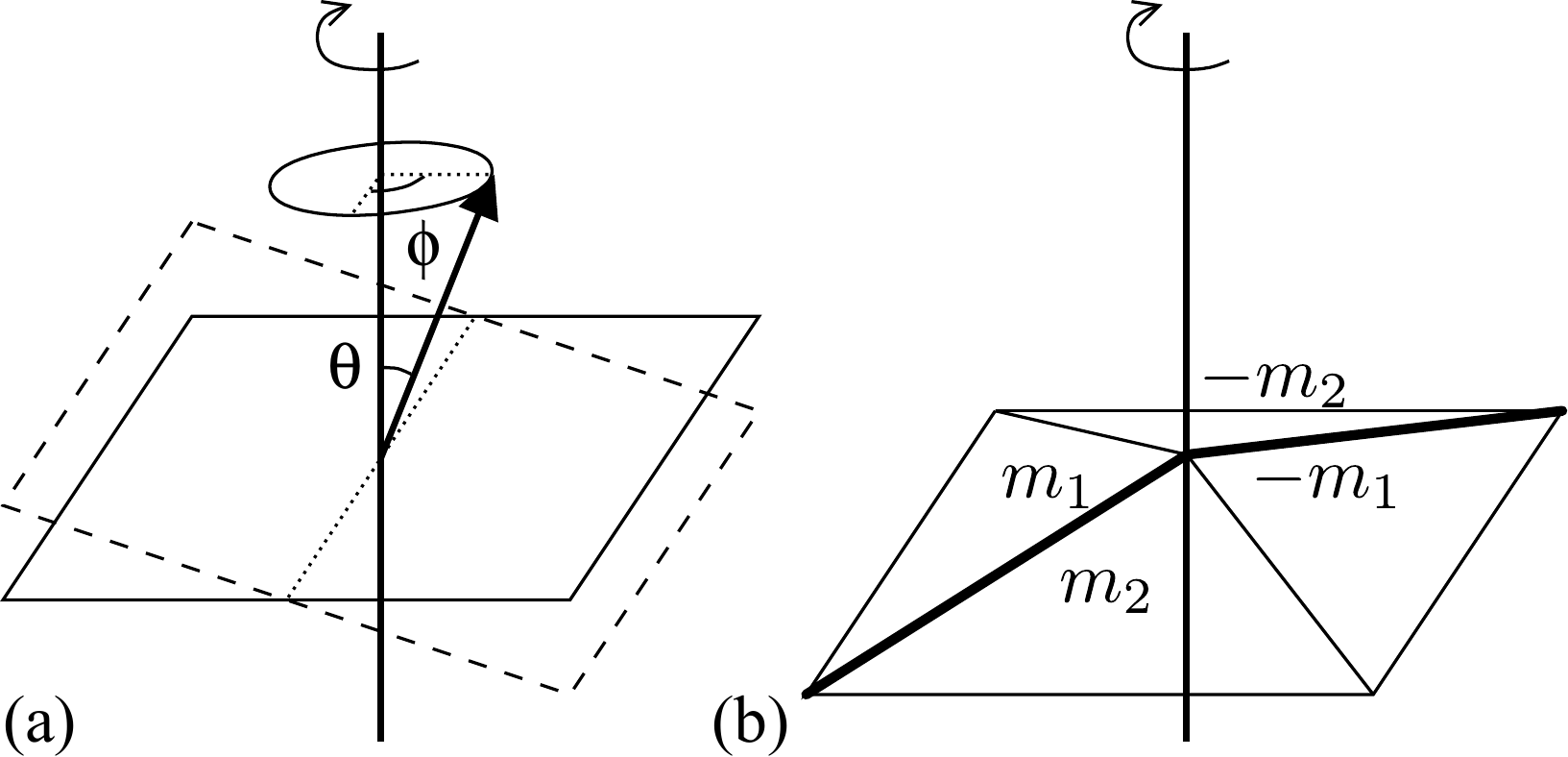}
\caption{\label{fig:tilted_surface} (a) A surface perpendicular to the twofold rotation
axis hosts a gapless mode in a nontrivial topological crystalline phase. The
surface Hamiltonian acquires a mass term $m(\theta,\phi)$ upon tilting the
surface away from the normal direction, which depends on the tilt angle
$\theta$ and the azimuthal angle $\phi$. The mass term is odd under the twofold
rotation operation, $m(\theta,\phi) = -m(\theta,\phi + \pi)$ (b) A generic
rotation symmetric surface. Surfaces related by twofold rotation have opposite
mass terms.  As a result, a protected gapless hinge mode (thick black line)
forms at the intersection of surfaces with masses of different sign. The situation shown in the figure corresponds to $\mathrm{sign}(m_1) =-\mathrm{sign}(m_2)$.}
\end{figure}

The construction of Sec.\ \ref{sec:4b}, in which the existence of a protected
corner state or hinge mode is derived from a nontrivial bulk crystalline topology,
can be directly extended to the case of a three-dimensional insulator or
superconductor with a twofold rotation symmetry, provided a (generic) hinge allows
for the existence of a protected hinge mode, see Table \ref{tab:AZ}.
In
that case, the argument starts from the existence of a gapless mode on a
surface that is invariant under the twofold rotation operation. We first
consider the case that the number of gapless modes is ``minimal'', {\em i.e.},
we consider a generator of the topological crystalline phase. Following the
construction of Sec.\ \ref{sec:4b}, one then argues that a unique mass term is
generated upon tilting this surface away from the invariant direction. 
The mass term $m$ depends on the
tilt angle $\theta$ and the azimuthal angle $\phi$ of the tilted surface, see Fig.\
\ref{fig:tilted_surface}(a), and is odd under the twofold rotation operation,
$m(\theta,\phi) = -m(\theta,\phi+\pi)$, since the twofold rotation symmetry
forbids a mass term for the rotation-invariant surface. As a consequence, a
protected gapless hinge mode forms at the intersection of surfaces with masses
of different sign, see Fig.\ \ref{fig:tilted_surface}(b). Since the number of
sign changes of the mass term for $0  \le \phi < 2 \pi$ must be an odd multiple
of two, the number of such hinge modes intersecting a generic cross section of
the crystal is an odd multiple of two.

The above argument guarantees the existence of hinge modes globally, as long as the lattice termination is consistent with the twofold rotation symmetry, but it does not address the existence of a hinge mode at a given hinge. Indeed, generically, single hinges are not mapped to themselves under the twofold rotation operation; in this sense, {\em all} hinges are ``generic'' in a crystal with twofold rotation symmetry. This is a difference with the mirror-reflection symmetric case, for which a nontrivial mirror-symmetric topological crystalline phase can guarantee the existence of hinge modes at mirror-symmetric hinges. 

All hinges being generic, hinges modes at a given hinge can also be induced by a suitable manipulation of the lattice termination. 
However, a change of lattice termination that is compatible with the twofold rotation symmetry always changes the total number of hinge modes passing a generic cross section of the crystal by a multiple of {\em four}. Since, as seen above, nontrivial bulk crystalline topology can also induce a number of hinge modes that is an odd multiple of two, we conclude that second-order topological phases protected by a twofold rotation symmetry have a $\ZZ_2$ invariant, which is nontrivial if the number of hinge modes is an odd multiple of two. Generators of the topological crystalline classes have a nontrivial $\ZZ_2$ index; if the bulk topological crystalline phase has an integer classification, only the odd topological numbers map to a nontrivial second-order phase. 

It is interesting to point out that for a nontrivial bulk crystalline phase in a symmetry class that does not allow for protected hinge modes, {\em i.e.}, for which the corresponding Altland-Zirnbauer class in $d=2$ dimensions is trivial, the mass term obtained by tilting the surface away from the invariant direction is not unique. With two or more masses $m_1(\theta,\phi)$ and $m_2(\theta,\phi)$, the antisymmetry relation $m_{1,2}(\theta,\phi) = -m_{1,2}(\theta,\phi+\pi)$ no longer forces the mass to be zero for certain values of the azimuthal angle $\phi$, so that no stable gapless modes exist at hinges. This is a key difference with the case of mirror reflection-symmetric crystalline insulators, where protected modes are guaranteed at mirror-symmetric corners or hinges even in the presence of multiple mass terms.

The resulting classification is shown in the Tables \ref{tab:R1}--\ref{tab:R3}. The nontrivial entries in these tables are those Shiozaki-Sato symmetry classes, for which both the purely crystalline classification groups $K'$ of Tables \ref{tab:SS1}--\ref{tab:SS3} and the corresponding entry Table \ref{tab:AZ} are both nonzero. Below we give detailed considerations making this construction explicit for the nontrivial complex Altland-Zirnbauer classes with unitary twofold rotation symmetry or antisymmetry. The complex Altland-Zirnbauer classes with antiunitary twofold rotation symmetry or antisymmetry and the real Altland-Zirnbauer classes are discussed in App.\ \ref{app:4}.

\begin{table}
\begin{tabular*}{\columnwidth}{l @{\extracolsep{\fill}} cccccc}
\hline\hline 
class & $s$ & $t$ & \begin{tabular}{c} $d=2$\\ ${\cal R}$ \end{tabular} & \begin{tabular}{c} $d=3$\\ ${\cal R}$ \end{tabular} &  \begin{tabular}{c} $d=3$\\ ${\cal I}$ \end{tabular} \\ \hline
A$^{\cal S}$ & $0$ & $0$ & $0$ & $0$ & $\ZZ_2$ \\
AIII$^{\cal S_+}$ & $1$ & $0$ & $0$ & $0$ & $0$ \\ \hline
A$^{\cal CS}$ & $0$ & $1$ & $0$ & $\ZZ_2$ & $0$ \\
AIII$^{\cal S_-}$ & $1$ & $1$ & $\ZZ_2$ & $0$ & $0$ \\ \hline\hline
\end{tabular*}
\caption{Classification of topological crystalline phases with an order-two rotation symmetry or an inversion symmetry for the complex Altland-Zirnbauer classes. The symbols ${\cal R}$ and ${\cal I}$ refer to twofold rotation ($d_{\parallel} = 2$) and inversion ($d_{\parallel} = d = 3$), respectively.\label{tab:R1}}
\end{table}

\begin{table}
\begin{tabular*}{\columnwidth}{l @{\extracolsep{\fill}} ccccc}
\hline\hline 
class & $s$ & \begin{tabular}{c} $d=2$\\ ${\cal R}$ \end{tabular} & \begin{tabular}{c} $d=3$\\ ${\cal R}$ \end{tabular} &  \begin{tabular}{c} $d=3$\\ ${\cal I}$ \end{tabular} \\ \hline
A$^{{\cal T}^+{\cal S}}$ & $0$ & $0$ & $\ZZ_2$ & $0$ \\
AIII$^{{\cal T}^+{\cal S}_+}$ & $1$ & $0$ & $0$ & $0$ \\
A$^{{\cal P}^+{\cal S}}$ & $2$ & $0$ & $0$ & $0$  \\
AIII$^{{\cal T}^-{\cal S}_-}$ & $3$ &  $0$ & $0$ & $0$ \\
A$^{{\cal T}^-{\cal S}}$ & $4$ & $0$ & $0$ & $0$ \\
AIII$^{{\cal T}^-{\cal S}_+}$ & $5$ & $0$ & $0$ & $0$ \\
A$^{{\cal P}^-{\cal S}}$ & $6$ & $0$ & $0$ & $\ZZ_2$ \\
AIII$^{{\cal T}^+{\cal S}_-}$ & $7$ & $\ZZ_2$ & $0$ & $0$ \\\hline\hline
\end{tabular*}
\caption{Same as table \ref{tab:R1}, but for antiunitary symmetries and antisymmetries.\label{tab:R2}}
\end{table}

\begin{table}
\begin{tabular*}{\columnwidth}{l @{\extracolsep{\fill}} cccccc}
\hline\hline 
class & $s$ & $t$ & \begin{tabular}{c} $d=2$\\ ${\cal R}$ \end{tabular} & \begin{tabular}{c} $d=3$\\ ${\cal R}$ \end{tabular} &  \begin{tabular}{c} $d=3$\\ ${\cal I}$ \end{tabular} \\ \hline
    AI$^{\mathcal{S}_+}$ & $0$ & $0$ & $0$ & $0$ & $0$ \\
    BDI$^{\mathcal{S}_{++}}$ & $1$ & $0$ & $0$ & $0$ & $0$ \\
    D$^{\mathcal{S}_{+}}$ & $2$ & $0$ & $0$ & $0$ & $0$  \\
    DIII$^{\mathcal{S}_{++}}$ & $3$ & $0$ & $0$ & $0$ & $0$  \\
    AII$^{\mathcal{S}_{+}}$ & $4$ & $0$ & $0$ & $0$ & $\ZZ_2$ \\
    CII$^{\mathcal{S}_{++}}$ & $5$ & $0$ & $0$ & $0$ & $0$ \\
    C$^{\mathcal{S}_{+}}$ & $6$ & $0$ & $0$ & $0$ & $\ZZ_2$ \\
    CI$^{\mathcal{S}_{++}}$ & $7$ & $0$ & $0$ & $0$ & $0$ \\
    \hline
    AI$^{\mathcal{CS}_-}$ & $0$ & $1$ & $0$ & $0$ & $0$ \\
    BDI$^{\mathcal{S}_{+-}}$ & $1$ & $1$ & $\ZZ_2$ & $0$ & $0$ \\
    D$^{\mathcal{CS}_{+}}$ & $2$ & $1$ & $\ZZ_2$ & $\ZZ_2$ & $0$ \\
    DIII$^{\mathcal{S}_{-+}}$ & $3$ & $1$ & $\ZZ_2$ & $\ZZ_2$ & $0$ \\
    AII$^{\mathcal{CS}_{-}}$ & $4$ & $1$ & $0$ & $\ZZ_2$ & $0$ \\
    CII$^{\mathcal{S}_{+-}}$ & $5$ & $1$ & $\ZZ_2$ & $0$ & $0$\\
    C$^{\mathcal{CS}_{+}}$ & $6$ & $1$ & $0$ & $\ZZ_2$ & $0$ \\
    CI$^{\mathcal{S}_{-+}}$ & $7$ & $1$ & $0$ & $0$ & $0$ \\
    \hline 
    AI$^{\mathcal{S}_-}$ & $0$ & $2$ & $0$ & $0$ & $0$ \\
    BDI$^{\mathcal{S}_{--}}$ & $1$ & $2$ & $0$ & $0$ & $0$  \\
    D$^{\mathcal{S}_{-}}$ & $2$ & $2$ & $\ZZ_2$ & $0$ & $\ZZ_2$  \\
    DIII$^{\mathcal{S}_{--}}$ & $3$ & $2$ & $\ZZ_2$ & $\ZZ_2$ & $\ZZ_2$ \\
    AII$^{\mathcal{S}_{-}}$ & $4$ & $2$ & $0$ & $\ZZ_2$ & $\ZZ_2$ \\
    CII$^{\mathcal{S}_{--}}$ & $5$ & $2$ & $0$ & $0$ & $0$ \\
    C$^{\mathcal{S}_{-}}$ & $6$ & $2$ & $0$ &$0$ & $\ZZ_2$ \\
    CI$^{\mathcal{S}_{--}}$ & $7$ & $2$ & $0$ & $0$ & $0$ \\
    \hline
    AI$^{\mathcal{CS}_+}$ & $0$ & $3$ & $0$ & $0$ & $0$ \\
    BDI$^{\mathcal{S}_{-+}}$ & $1$ & $3$ & $0$ & $0$ & $0$ \\
    D$^{\mathcal{CS}_{-}}$ & $2$ & $3$ & $0$ & $0$ & $0$ \\
    DIII$^{\mathcal{S}_{+-}}$ & $3$ & $3$ & $\ZZ_2$ & $0$ & $\ZZ_2$ \\
    AII$^{\mathcal{CS}_{+}}$ & $4$ & $3$ & $0$ & $\ZZ_2$ & $\ZZ_2$ \\
    CII$^{\mathcal{S}_{-+}}$ & $5$ & $3$ & $\ZZ_2$ & $0$ & $0$ \\
    C$^{\mathcal{CS}_{-}}$ & $6$ & $3$ & $0$ & $\ZZ_2$ & $0$ \\
    CI$^{\mathcal{S}_{+-}}$ & $7$ & $3$ & $0$ & $0$ & $0$ \\\hline\hline
  \end{tabular*}
\caption{Classification of topological crystalline phases with an order-two
  crystalline symmetry or antisymmetry for the real Altland-Zirnbauer classes.
  The symbols ${\cal R}$ and ${\cal I}$ refer to twofold rotation ($d_{\parallel} = 2$), and inversion
  ($d_{\parallel} = d = 3$), respectively. 
  \label{tab:R3}}
\end{table}

{\em Class AIII$^{{\cal R}_+}$, $(s,t) = (1,0)$.---} 
The presence of the chiral antisymmetry with $U_{\cal C} = \sigma_3$ allows one to assign a chirality $\pm$ to surface modes with Dirac-like dispersion $\propto -i \sigma_1 \partial_x \pm i \sigma_2 \partial_y$, where $x$ and $y$ are the Cartesian coordinates parameterizing the surface and the twofold rotation operation sends $x \to -x$ and $y \to -y$. The crystalline bulk has a $\ZZ^2$ classification,\cite{shiozaki2014} with purely crystalline classifying group $K' = \ZZ$, see Table \ref{tab:SS1}. For a second-order topological phase we restrict ourselves to the purely crystalline topological phases, which have equal numbers of Dirac cones of both chiralities. Such Dirac cones can not mutually gap out for a rotation-invariant surface if they have the same parity under ${\cal R C}$. At a minimal surface, in which all surface modes have the same parity under ${\cal R C}$, the twofold rotation symmetry may be represented by $U_{\cal R} = U_{\cal C} = \sigma_3$.

With a suitable choice of basis and after rescaling, the surface Hamiltonian of a minimal surface may be written as
\begin{equation}
  H_{\rm surface} = -i v(\sigma_1 \tau_3 \partial_x + \sigma_2 \partial_y) \openone_N,
\end{equation}
where $\openone_N$ is the $N \times N$ unit matrix. The mass terms allowed by chiral symmetry and rotation symmetry are $m_1(x,y) \sigma_1\tau_1 + m_2(x,y) \sigma_1\tau_2$ with $m_{1,2}(x,y) = -m_{1,2}(-x,-y)$ $N \times N$ hermitian matrices. Although surfaces related by the twofold rotation operation have opposite masses, the existence of two mass terms allows the crystal faces to avoid domain walls and the associated protected hinge modes.

{\em Class $A^{{\cal CR}}$, $(s,t) = (0,1)$.---} The bulk has a $\ZZ$ topological classification, with an integer topological invariant $N$ equal to the difference of surface states with positive and negative parity $\sigma_{\cal CR}$ at zero energy. For a minimal surface, all surface states have the same value of $\sigma_{\cal CR}$ and one may effectively represent ${\cal CR}$ using $U_{\cal CR} = 1$. With a suitable choice of basis and after rescaling the corresponding surface Hamiltonian reads
\begin{equation}
  H_{\rm surface} = -i v (\sigma_1 \partial_x + \sigma_2 \partial_y) \openone_N,\end{equation}
with $\openone_N$ the $N \times N$ identity matrix and $x$ and $y$ are coordinates at the invariant surface. The unique mass term $m(x,y) \sigma_3$ with $m(x,y) = - m(-x,-y)$ an $N \times N$ hermitian matrix gaps out the surface states. If $N$ is odd the existence of hinge modes at the intersection of surfaces with opposing signs of $\det m(x,y)$ is guaranteed by the rotation antisymmetry. If $N$ is even one can still construct a mass term which is nonzero everywhere (except at the origin), corresponding to a state without hinge modes.

\subsection{Twofold rotation symmetry for $d=2$ and inversion symmetry}

\begin{figure}
\includegraphics[width=0.99\columnwidth]{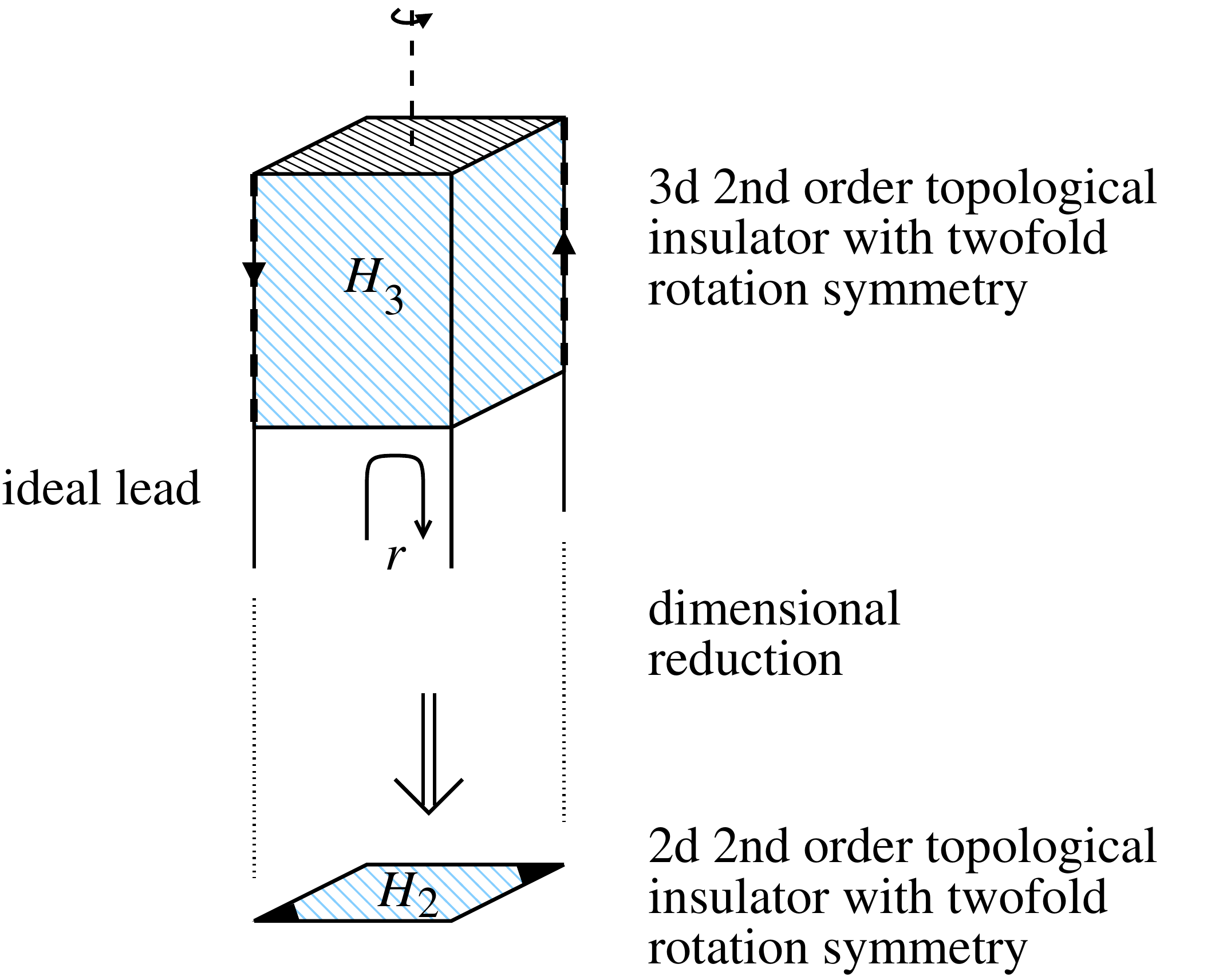}
\caption{\label{fig:rot} Dimensional reduction scheme from a three-dimensional second-order topological insulator with twofold rotation symmetry to a two-dimensional second-order topological insulator with inversion symmetry. Upon dimensional reduction, the Altland-Zirnbauer class changes from $s$ to $s-1$ (modulo $2$ for the complex classes, modulo $8$ for the real classes), see the discussion in the main text.}
\end{figure}

The above construction can not be applied to two-dimensional crystals with
twofold rotation symmetry and to three-dimensional crystals with inversion
symmetry, because these do not have symmetry-invariant boundaries. Instead, we
argue for the existence of a second-order topological phase in this case using
the reflection-matrix based dimensional reduction scheme outlined in
Sec.~\ref{sec:3}. Starting from a second-order topological phase in $d+1$
dimensions in Shiozaki-Sato symmetry class $(s+1,t)$ (class $s+1$ for complex
Hamiltonians with antiunitary symmetries) and $d_{\parallel}<d+1$ inverted
coordinates, the dimensional reduction scheme allows one to 
construct a second-order topological insulator or
superconductor in Shiozaki-Sato symmetry class $(s,t)$ (class $s$ for complex
Hamiltonians with antiunitary symmetries) in $d$ dimensions, with the same number $d_{\parallel}$ of inverted dimensions. The real-space version of the
reflection-matrix based dimensional reduction scheme directly maps hinge states
in a three-dimensional second-order topological insulator or superconductor
with twofold rotation symmetry to corner states in a two-dimensional
topological insulator or superconductor with twofold rotation symmetry, see
Fig.~\ref{fig:rot}. Similarly, it maps generalized hinge states of a
four-dimensional second-order topological insulator or superconductor with an
order-two inversion with $d_{\parallel} = 3$ to hinge states of a
three-dimensional second-order topological insulator or superconductor with
inversion symmetry. The resulting $\ZZ_2$ classification is given in Tables
\ref{tab:R1}--\ref{tab:R3}.


\section{Examples}
\label{sec:6}

In this section we give various tight-binding model realizations of the second-order topological insulators. The models we consider all follow the same pattern. We first describe their general structure and then turn to a description of specific Shiozaki-Sato symmetry classes. The model Hamiltonian we consider is of the general form $H(\vk) = H_0(\vk) + H_1$, with
\begin{align}
  H_0(\vk)&= \sum_{j=0}^{d} d_j(\vk) \Gamma_j, \ \
  H_1 = \sum_{j=1}^{d} b_{j} B_{j},
	\label{eq:HdG}
\end{align}
where the $\Gamma_j$ and the $B_j$, $j=1,\ldots,d$, are matrices that depend on the specific Shiozaki-Sato class and that satisfy $\Gamma_j^2 = B_j^2 = 1$, the $b_j$ are real numbers typically chosen to be numerically small, and
\begin{align}
  d_0(\vk) =&\, m + \sum_{j'=1}^{d} (1 - \cos k_{j'}),\nonumber \\
  d_j(\vk) =&\, \sin k_j, \ \ j=1,\ldots,d.
  \label{eq:d(k)}
\end{align}
The matrices $\Gamma_0$ and $\Gamma_j$, $j=1,\ldots,d$, anticommute mutually, which ensures that for small numbers $b_j$, the Hamiltonian~(\ref{eq:HdG}) is in a nontrivial topological crystalline phase for $-2<m<0$. We further choose the matrix $B_1$ such that it commutes with $\Gamma_1$ and $\Gamma_0$ and anticommutes with $\Gamma_j$ with $j \ge 1$. For the remaining matrices $B_j$ with $j > 1$ we set
\begin{equation}
  B_j = \Gamma_j \Gamma_1 B_1
  ,\ \ j=2,\ldots,d,
  \label{eq:BGamma}
\end{equation}
which ensures that $B_i$ commutes with $\Gamma_i$ and $\Gamma_0$ and anticommutes with $\Gamma_j$ for $j \neq i$. Mirror symmetry with $k_1 \to -k_1$ requires $b_2 = b_3 = 0$; twofold rotation symmetry with rotation around the $x_3$ axis requires $b_3 = 0$. The role of the perturbation $H_1$ is to reduce the symmetry of the Hamiltonian, while preserving the crystalline symmetry of interest. Further, as we will show below, each $B_j$ term gaps the surface that is perpendicular to the $x_j$ direction. When appropriate, we will simplify our notation by writing the matrices $\Gamma_j$ and $B_j$ and the numbers $b_j$ as vectors, $\bm \Gamma = (\Gamma_1,\ldots,\Gamma_d)$, $\vB = (B_1,\ldots,B_d)$, and $\vb = (b_1, \ldots, b_d)$. 

For all of the examples that we discuss below we verified the existence of Majorana corner modes or gapless hinge modes by numerical diagonalization of a finite cluster. (All numerical calculations in this Section were performed using the Kwant software package.\cite{groth2014}) Alternatively, for a Hamiltonian of the form (\ref{eq:HdG}), with the constraints as described above, the existence of zero-energy corner modes or gapless hinge modes can also be concluded from an explicit solution of the low-energy theory, modeling the crystal boundaries as interfaces between regions with negative and positive $m$, with negative $m$ corresponding to the interior of the crystal. The low-energy limit of $H_0$ near a sample boundary has the form
\begin{equation}
  H_0 = m(x_{\perp}) \Gamma_0 -i \hbar \vGamma \cdot \partial_{\vr},
  \label{eq:H0boundary}
\end{equation}
where $x_{\perp} = \vn \cdot \vr$ is the coordinate transverse to a boundary with outward-pointing normal $\vn$. We require $m(x_{\perp}) > 0$ for $x_{\perp} > 0$ and $m(x_{\perp}) < 0$ for $x_{\perp} < 0$, so that the sample interior corresponds to negative $x_{\perp}$. The Hamiltonian (\ref{eq:H0boundary}) admits a zero-energy boundary mode with spinor wavefunction $\psi(x_{\perp})$ satisfying
\begin{equation}
  \partial_{x_{\perp}} \psi(x_{\perp}) = - \frac{i}{\hbar} m(x_{\perp}) (\vn \cdot \vGamma) \Gamma_0 \psi(x_{\perp}).
\end{equation}
For $2 b$-dimensional spinors, this equation has $b$ bounded solutions with an $x_{\perp}$-independent spinor structure. The projection operator to the $b$-dimensional space of allowed spinors is
\begin{equation}
  P(\vn) = \frac{1}{2}[i (\vn \cdot \vGamma) \Gamma_0 + 1].
\end{equation}

The effective $b$-band surface Hamiltonian is obtained using the projection operator $P(\vn)$. To illustrate this procedure, we consider a family of surfaces with surface normal $\vn = (\cos \phi,\sin \phi)$ for $d=2$ or $\vn = (\cos \phi, \sin \phi,0)$ for $d=3$. In this case we write the projection operator as 
\begin{align}
  P(\phi) &= \frac{1}{2}(i \Gamma_1 \Gamma_0 \cos \phi + i \Gamma_2 \Gamma_0 \sin \phi + 1)
  \nonumber \\
  &= e^{\phi \Gamma_2 \Gamma_1/2} P(0) e^{-\phi \Gamma_2 \Gamma_1/2}.
\end{align}
The projected Hamiltonian then reads, 
\begin{align}
  P(\vn) H P(\vn) =&\,
  e^{\phi \Gamma_2 \Gamma_1/2} P(0) \nonumber \\ &\, \mbox{} \times
  [-i \hbar (\Gamma_2 \partial_{x_{\parallel}} + \Gamma_3 \partial_{x_3}) + m(\phi) B_1] \nonumber \\ &\, \mbox{} \times
  P(0) e^{-\phi \Gamma_2 \Gamma_1/2},
  \label{eq:HP}
\end{align}
where $m(\phi) = b_1 \cos \phi + b_2 \sin \phi$ and $\partial_{x_{\parallel}} = \cos \phi\, \partial_{x_2} - \sin \phi\, \partial_{x_1}$ is the derivative with respect to a coordinate along the surface. (For $d=2$ the terms proportional to $\partial_{x_3}$ should be omitted from Eq.\ (\ref{eq:HP}) and from Eq.\ (\ref{eq:HboundaryGamma}) below.) From Eq.\ (\ref{eq:HP}) we derive the effective boundary Hamiltonian
\begin{equation}
  H_{\rm boundary} = -i \hbar (\Gamma_2' \partial_{x_{\parallel}} + \Gamma_3' \partial_{x_3}) + m(\phi) B_1', \label{eq:HboundaryGamma}
\end{equation}
where $\Gamma_2' = P(0) \Gamma_2 P(0)$, $\Gamma_3' = P(0) \Gamma_3 P(0)$, and $B_1' = P(0) B_1 P(0)$ are effectively $b \times b$ matrices because of the projection operator $P(0)$. (Note that $\Gamma_2$, $\Gamma_3$, and $B_1$ commute with $P(0)$.) The boundary Hamiltonian (\ref{eq:HboundaryGamma}) supports boundary modes with a gap $|m(\phi)|$. 
For $d=2$ zero-energy corner states appear between crystal edges with opposite sign of $m(\phi)$; for $d=3$ gapless hinge modes appear between crystal faces with opposite sign of $m(\phi)$.

\subsection{Examples in two dimensions}

\subsubsection{Class D with $t=d_\parallel$}

This example applies to symmetry class D$^{{\cal CM}_+}$, $(s,t) = (2,1)$ and to symmetry class $D^{{\cal R}_-}$, $(s,t) = (2,2)$. We represent the symmetry operations using $U_{\cal P} = \sigma_1$, $U_{\mathcal{CM}} = \sigma_1$, and $U_{\mathcal{R}} = \sigma_3$. The mirror operation sends $k_1 \to -k_1$. For the matrices $\Gamma_j$ and $B_j$ in the tight-binding Hamiltonian (\ref{eq:HdG}) we choose
\begin{equation}
	\Gamma_0 = \sigma_3,\ \ \bm \Gamma =(\tau_3\sigma_1,\sigma_2),\ \
	\vB=(\tau_2\sigma_3,-\tau_1).
	\label{eq:D0}
\end{equation}
For class D$^{{\cal CM}_+}$, the mirror antisymmetry imposes that $b_2 = 0$; for class $D^{{\cal R}_-}$ nonzero $b_1$ and $b_2$ are allowed. We note that for $b_1 = 0$ this example also possesses a mirror symmetry for mirror reflection $k_2 \to -k_2$, which is represented by $\sigma_2 \tau_3$. The mirror-symmetric case hosts Majorana zero modes at corners that are bisected by the mirror axis. The rotation-symmetric case also hosts Majorana modes at corners, but these corners are determined by the orientation of the vector $\vb$ (numerical data not shown).

\subsubsection{Class D with $t=d_\parallel+3 \mod 4$}

This example applies to symmetry class D$^{{\cal M}_+}$, $(s,t) = (2,0)$ and to symmetry class $D^{{\cal CR}_+}$, $(s,t) = (2,1)$. We represent the symmetry operations using $U_{\cal P} = 1$, $U_{\mathcal{M}} = \sigma_1$, and $U_{\mathcal{CR}} = \tau_3\sigma_1$. For the matrices $\Gamma_j$ and $B_j$ in the tight-binding Hamiltonian (\ref{eq:HdG}) we choose
\begin{equation}
	\Gamma_0 = \tau_2,\ \ \bm \Gamma=(\tau_1\sigma_3,\tau_3),\ \
	\vB=(\tau_2\sigma_1,-\sigma_2).
	\label{eq:D3}
\end{equation}
Again the mirror symmetry imposes that $b_2 = 0$; for class $D^{{\cal CR}_-}$ nonzero $b_1$ and $b_2$ are allowed. As in the previous example, the mirror-symmetric case hosts Majorana zero modes at corners that are bisected by the mirror axis.\cite{langbehn2017} The rotation-symmetric case also hosts Majorana zero modes at corners that are determined by the orientation of the vector $\vb$ (numerical date not shown).

\subsubsection{Class DIII with $t=d_\parallel-1 \mod 4$}

\begin{figure}
\includegraphics[width=\columnwidth]{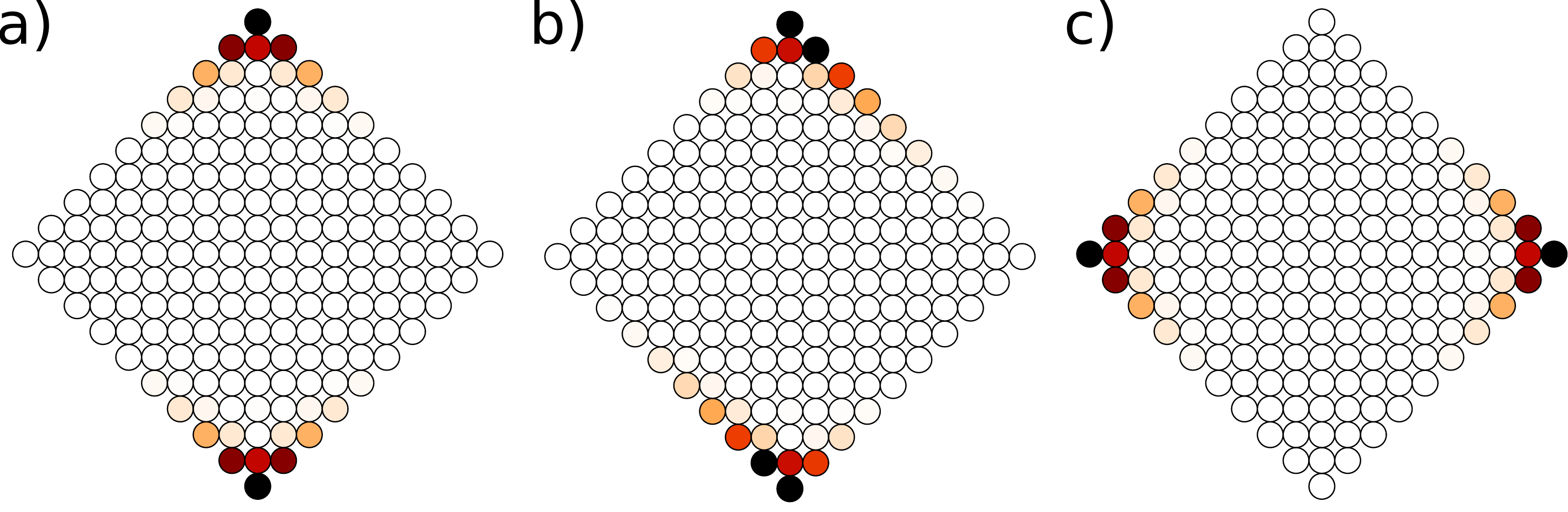}
\caption{Support of the zero-energy corner state obtained from exact diagonalization of the two-dimensional time-reversal invariant superconductor in class DIII with Hamiltonian~(\ref{eq:HdG}) with $m = -1$ and $\bm\Gamma$ and $\vB$ given by Eq.~(\ref{eq:DIII2G}) with $\vb = (0.3,0)$ (a), $\vb = (0.3,0.1)$ (b), and $\vb = (0,0.3)$ (c). \label{fig:WF_2d_DIII_Deformation}}
\end{figure}

This example applies to symmetry classes DIII$^{\mathcal{M}_{++}}$ and
DIII$^{\mathcal{R}_{-+}}$, which both have a $\ZZ_2$ classification. We consider
an eight-band model, for which we represent the symmetry operations using
$U_{\cal T} = \sigma_2$, $U_{\cal P} = \tau_1$, $U_{\cal M} = \rho_3$, and $U_{\cal
R} = \sigma_3$, where the $\rho_j$, $\sigma_j$, and $\tau_j$ are Pauli matrices
acting on different spinor degrees of freedom. For the matrices $\Gamma_j$ and
$B_j$ in the tight-binding Hamiltonian (\ref{eq:HdG}) we choose
\begin{align}
  \Gamma_0 &= \tau_3, \ \
  \bm\Gamma = (\rho_1 \tau_1 \sigma_1, \rho_3 \tau_1 \sigma_1),\nonumber \\
  \vB &= (\rho_3 \tau_3, -\rho_1 \tau_3).
  \label{eq:DIII2G}
\end{align}
Mirror symmetry imposes that $b_2 = 0$. The perturbation $b_1 B_1$ preserves both mirror and rotation symmetries, but breaks a mirror symmetry with $x_2 \to -x_2$, represented by $\rho_1$. As shown in figure \ref{fig:WF_2d_DIII_Deformation}a, the mirror-symmetric model with nonzero $b_1$ hosts Majorana Kramers pairs at its symmetry-invariant corners. The corner states persist if the mirror-symmetry-breaking perturbation $b_2 B_2$ is switched on, see Fig.\ \ref{fig:WF_2d_DIII_Deformation}b. In this case, the ratio of $b_1$ and $b_2$ determines the corner at which the Majorana Kramers pairs reside, such that they move to the other corners if $b_1 = 0$, see Fig.\ \ref{fig:WF_2d_DIII_Deformation}c.

\subsubsection{Class AII with $t=d_\parallel+2 \mod 4$}
\label{sec:AII2d}

This example applies to symmetry classes AII$^{\mathcal{CM}_+}$ and AII$^{\mathcal{R}_+}$. We represent the symmetry operations using $U_{\cal T} = \sigma_2$, $U_{\mathcal{CM}} = \tau_2 \sigma_3$, and $U_{\mathcal{R}} = \tau_2\sigma_1$. This symmetry class allows a perturbation $H_1$ of the form
\begin{equation}
  H_1 = \sum_{j=1}^{d} b_j B_j + \sum_{j=1}^{d} c_j C_j,
\end{equation}
where the matrices $C_j$ anticommute with the matrices $B_j$ and otherwise satisfy the same properties, see Eq.\ (\ref{eq:BGamma}) and the discussion preceding that equation. For the matrices $\Gamma_j$, $B_j$, and $C_j$ we choose
\begin{align}
	\label{eq:AII2G}
  & \Gamma_0 = \tau_2 \sigma_1,\ \ \bm\Gamma =(\sigma_3,\sigma_2),\ \nonumber \\
  &  \vB =(\mu_2\tau_3\sigma_3,\mu_2\tau_3\sigma_2),\ \
  \vC = (\mu_2 \tau_1 \sigma_3,\mu_2 \tau_1 \sigma_2),
\end{align}
where the $\mu_j$, $\sigma_j$, and $\tau_j$ are Pauli matrices acting on
different spinor degrees of freedom. As in the previous examples the mirror
antisymmetry imposes that $b_2 = c_2 = 0$; for class AII$^{{\cal R}_+}$ nonzero $b_{1,2}$ and $c_{1,2}$ are allowed. The mirror antisymmetry can protect a zero-energy Kramers pair at mirror-symmetric corners. However, if the mirror antisymmetry is broken, the twofold rotation symmetry alone cannot protect a topologically protected zero-energy state if both $\vb$ and $\vc$ are nonzero and linearly independent. (If $\vb$ and $\vc$ are both nonzero and linearly dependent, the model specified by Eq.\ (\ref{eq:AII2G}) obeys an accidental chiral antisymmetry, effectively placing it in the Shiozaki-Sato symmetry classes CII$^{\mathcal M_{--}}$ and CII$^{\mathcal R_{+-}}$, which stabilizes a zero-energy corner mode even if mirror symmetry is broken.)

\begin{figure}
\includegraphics[width=\columnwidth]{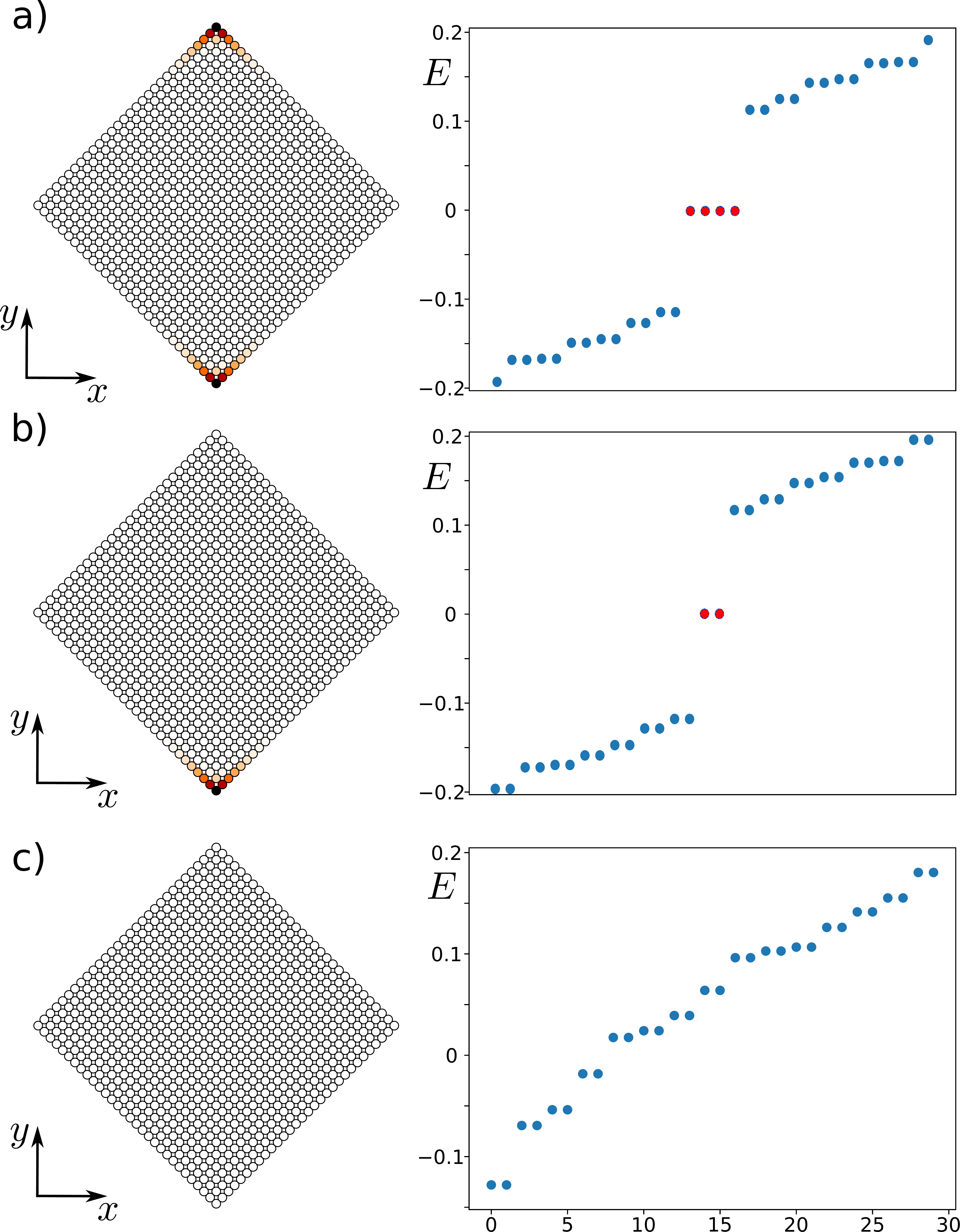}\\
\caption{\label{fig:AII_t2} Support of the zero-energy eigenstates (if present, left) and the lowest 30 eigenvalues (right a, b and c) of the model discussed in Sec.\ \ref{sec:AII2d}. Panel (a) is for the case that mirror antisymmetry is present, $\vb = (0.4,0)$ and $\vc = (0,0)$, which has a Kramers pair of zero-energy states localized at the mirror-symmetric top and bottom corners. Breaking the mirror antisymmetry locally at the top corner removes one zero-energy Kramers pair, as shown in panel (b). No zero-energy Kramers pairs remain after removing the mirror antisymmetry at both the top and the bottom corner, as shown in panel (c).}
\end{figure}

Figure \ref{fig:AII_t2} shows the result of the exact diagonalization of this
model on a finite-sized lattice. Panel (a) shows the support of the Kramers
pairs for a system with $b_2 = c_1 = c_2=0$ as well as the spectrum near zero energy. Upon
adding the mirror-antisymmetry-breaking perturbation $c_2 C_2$
locally near the top corner, the Kramers pair located there acquires a finite
energy, see panel (b). 
Both Kramers pairs disappear if the mirror-symmetry-breaking
perturbation is added to both top and bottom corners, see
Fig.~\ref{fig:AII_t2}c.

\subsection{Examples in three dimensions}

\subsubsection{Class A with $t=d_\parallel+1 \mod 4$}

\begin{figure}
\includegraphics[width=0.9\columnwidth]{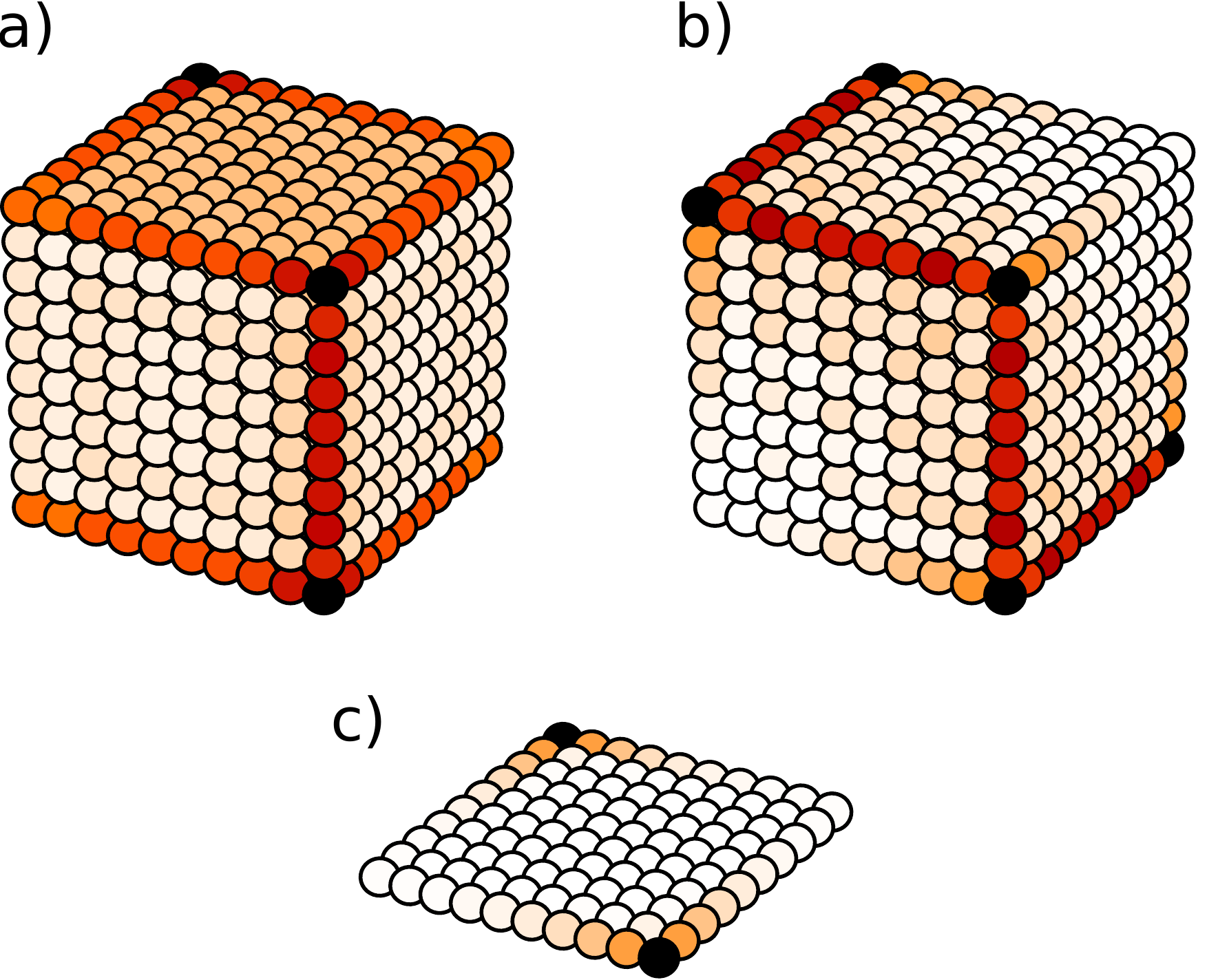}\\
\caption{\label{fig:Aexample3d} Support of the zero-energy hinge modes for a
three-dimensional crystal with tight-binding Hamiltonian specified by Eqs.\
(\ref{eq:HdG}) and (\ref{eq:At1G}) for $\vb = (0.8,0.8,0.8)/\sqrt{3}$ (a) and
$\vb=(0.8,0.8,0)/\sqrt{2}$ (b). The example shown in panel (a) has
mirror-reflection symmetry, twofold rotation symmetry, and inversion symmetry;
the example in panel (b) has inversion symmetry only. Panel (c) shows the
support of the zero-energy corner modes obtained for the two-dimensional
tight-binding model specified by Eqs.\ (\ref{eq:HdG}) and (\ref{eq:At1Gd2})
with $\vb = (0.4,0)$.}
\end{figure}

Langbehn \textit{et al.}\cite{langbehn2017} considered this class for the case of a mirror symmetry with $k_1 \to -k_1$ represented by $U_{\mathcal{M}}=\sigma_1$. Here we give an example that also has twofold rotation antisymmetry, represented by  $U_{\mathcal{CR}}=\tau_2\sigma_1$, and inversion symmetry, represented by $U_{\mathcal{I}}=\tau_1\sigma_1$. For the matrices $\Gamma_j$ and $B_j$ in the tight-binding Hamiltonian (\ref{eq:HdG}) we choose
\begin{equation}
	\label{eq:At1G}
  \Gamma_0 = \tau_1 \sigma_1,\ \
	\bm\Gamma=(\tau_1\sigma_3,\tau_2,\tau_3),\ \
	\vB =(\tau_1,\tau_2\sigma_3,\tau_3\sigma_3).
\end{equation}
Mirror symmetry imposes that $b_2 = b_3 = 0$; twofold rotation antisymmetry
imposes that $b_3 = 0$. The mirror-symmetric model with $b_2=b_3=0$ was already
considered in Sec.\ \ref{sec:3}. Additionally, the system has a mirror symmetry
sending $k_2 \to -k_2$ ($k_3 \to -k_3$) represented by
$U_{\mathcal{M}}=\tau_3\sigma_2$ ($U_{\mathcal{M}}=\tau_2\sigma_2$) and a twofold
rotation antisymmetry around $x_1$-axis ($x_2$-axis) represented by
$U_{\mathcal{CR}}=\sigma_2$ ($U_{\mathcal{CR}}=\tau_3\sigma_1$). The
mirror-symmetric case A$^{\cal M}$ in which only the perturbation $b_1 B_1$ is
present has a single chiral mode wrapping around the sample
hinges.~\cite{langbehn2017} These modes persist when all three perturbations
$b_j B_j$ are switched on, where the orientation of the vector $\vb$ determines
which hinges support the chiral hinge modes. As an example, Figure
\ref{fig:Aexample3d} shows the support of the chiral hinge modes for two
different choices of $\vb$.

Upon performing the reflection-matrix dimensional reduction scheme of Sec.\
\ref{sec:3} the model defined by the choice (\ref{eq:At1G}) can be used to
generate an eight-band
two-dimensional Hamiltonian in classes AIII$^{\mathcal{M}_+}$ and
AIII$^{\mathcal{R}_-}$ with $U_{\cal C} = \mu_3$, $U_{\cal M} = \sigma_1$, and
$U_{\cal R} = \mu_1\tau_2 \sigma_1$. Figure~\ref{fig:chern_lead_example} shows
the support of the zero-energy corner states of the two-dimensional Hamiltonian
that is obtained this way. For comparison, we may consider a four-band model for a two-dimensional tight-binding Hamiltonian, with $U_{\cal C} = \tau_3$, $U_{\cal M} = \sigma_1$, and $U_{\cal R} = \tau_1 \sigma_1$ and Hamiltonian specified by
\begin{equation}
  \Gamma_0 = \tau_1\sigma_1,\ \ \bm \Gamma = (\tau_1\sigma_3,\tau_2),\ \ \vB = (\tau_1,\tau_2\sigma_3).
  \label{eq:At1Gd2}
\end{equation}
The above model has a mirror symmetry
for $b_2 = 0$ and a twofold rotation symmetry for arbitrary $b_1$, $b_2$. This
model has zero-energy corner states. Figure~\ref{fig:Aexample3d}c shows the
support of these zero-energy corner states for the parameter choice $\vb = (0.4,0)$.

\subsubsection{Class AII with $s=4$, $t=d_{\parallel}+1 \mod 4$}

\begin{figure}
\includegraphics[width=0.75\columnwidth]{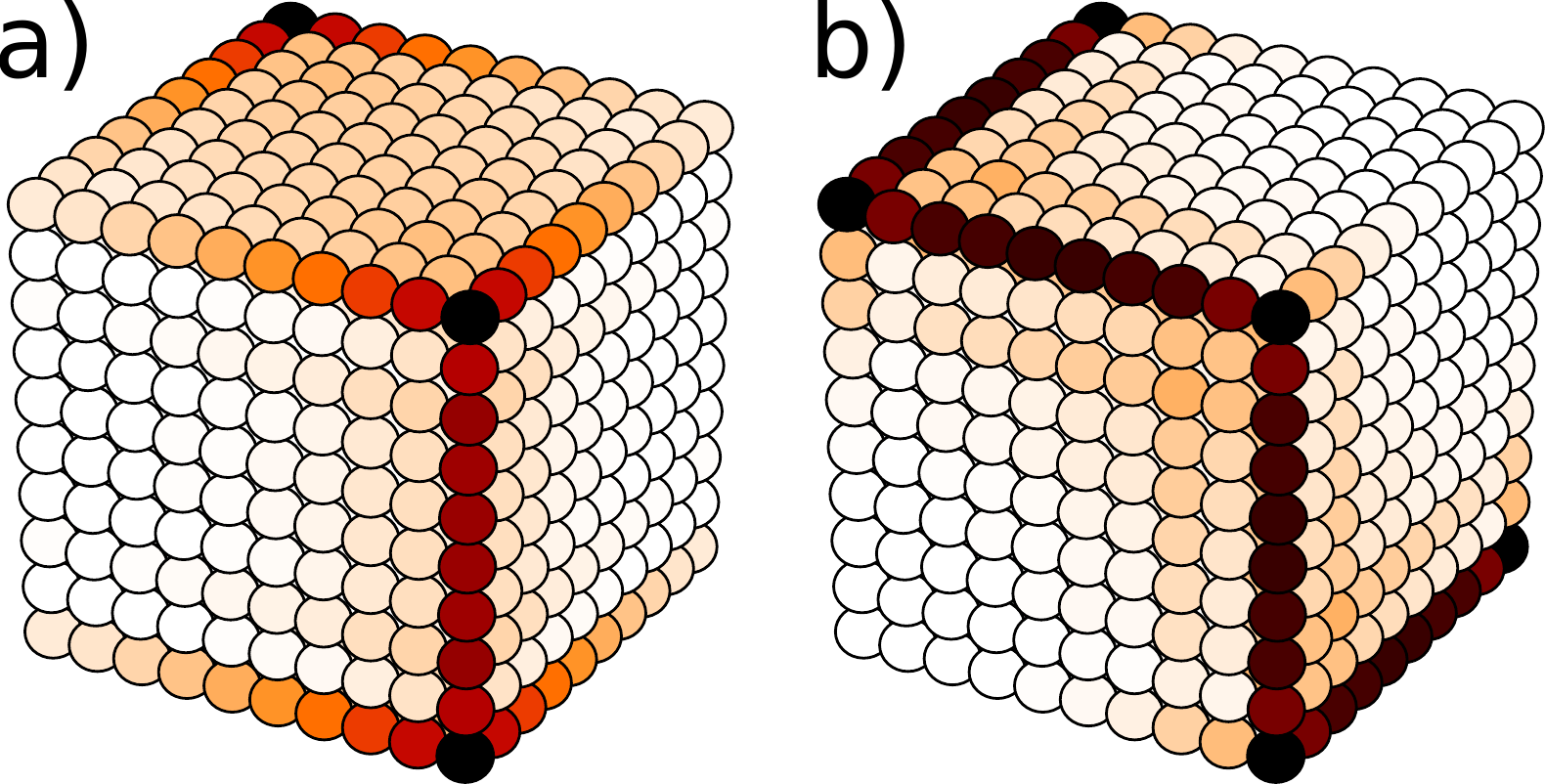}\\
\caption{\label{fig:AIIt13dR} Support of helical hinge modes of the tight-binding Hamiltonian~(\ref{eq:HdG}) with $\bm\Gamma$ and $\vB$ given by Eq.~(\ref{eq:AIIt13dG}) and $\vb = (0.4,0.4,0)$ (a) and $\vb = (0.4,0.4,0.4)$ (b). For the example shown in panel (a) the top and bottom surfaces are invariant with respect to the twofold rotation symmetry, which explains the presence of gapless surface modes at the top and bottom surface. The twofold rotation symmetry is broken in panel (b), which has inversion symmetry only.}
\end{figure}

This example applies to the classes AII$^{\mathcal{M_-}}$, AII$^{\mathcal{CR_+}}$, and AII$^{\mathcal{I_+}}$, which all have a $\ZZ$ bulk crystalline classification, with purely crystalline component $K' = 2 \ZZ$.\cite{lu2014,turner2012,chiu2013,morimoto2013,shiozaki2014,trifunovic2017} We use $U_\mathcal{T} = \sigma_2$, represent the (spinful) mirror operation by $U_\mathcal{M}=\sigma_3\tau_3$, rotation antisymmetry by $U_\mathcal{CR} = \sigma_1\tau_2$, and inversion as $U_I=\tau_3$. The lattice Hamiltonian is specified by
\begin{align}
 \label{eq:AIIt13dG}
  \Gamma_0 =&\, \tau_3,\ \ \bm\Gamma =(\sigma_3\tau_1,\sigma_2\tau_1,\sigma_1\tau_1),\nonumber \\
	\vB =&\, (\sigma_3\tau_0\rho_2,\sigma_2\tau_0\rho_2,\sigma_1\tau_0\rho_2),
\end{align}
where mirror symmetry forces $b_2=b_3=0$ and rotation antisymmetry forces $b_3=0$. In addition to the spatial symmetries mentioned above, the model has a mirror symmetry with $k_2 \to - k_2$ if $b_1 = b_3 = 0$, represented by $\sigma_2 \tau_3$, a mirror symmetry with $k_3 \to -k_3$ if $b_1 = b_2 = 0$, represented by $\sigma_1 \tau_3$, and rotation antisymmetries around the $x_1$ axis (if $b_1 = 0$) and $x_2$ axis (if $b_2=0$), represented by $\sigma_3 \tau_2$ and $\sigma_2 \tau_2$, respectively. The model with mirror symmetry has a single helical mode located at the mirror-symmetric sample hinges.\cite{langbehn2017} The helical modes persist upon turning on all perturbations $b_j B_j$, $j=1,2,3$, leaving inversion as the only symmetry of the model. Figure \ref{fig:AIIt13dR} shows the helical hinge modes for two different choices of $\textbf{b}$. The existence of hinge modes in the presence of inversion symmetry is consistent with Refs.\ \onlinecite{fang2018,khalaf2018}, where the same symmetry class was considered. The case of a spinful mirror symmetry was analyzed previously in Refs.\ \onlinecite{schindler2018,langbehn2017}.

\subsubsection{Class AII with $s=4$, $t=d_{\parallel}$}

This example applies to the classes AII$^{\mathcal{CM_-}}$, AII$^{\mathcal{R_-}}$, and AII$^{\mathcal{CI_+}}$, which all have a $\ZZ_2^2$ bulk crystalline classification,\cite{lu2014,turner2012,chiu2013,morimoto2013,shiozaki2014,trifunovic2017} with purely crystalline component $K' = \ZZ_2$. Here we again represent time-reversal as $U_\mathcal{T} = \sigma_2$, and use $U_\mathcal{CM} = \sigma_1\tau_3$, $U_\mathcal{R} = \sigma_3$, and $U_\mathcal{CI} = \tau_3$ to represent the mirror antisymmetry, spinful rotation symmetry, and inversion antisymmetry. We choose the matrices of the tight-binding Hamiltonian as 
\begin{align}
	\label{eq:AIIt03dG}
  \Gamma_0 = \tau_1\rho_3,\ \ 
  \bm\Gamma =(\sigma_1\tau_3\rho_3,\sigma_2\tau_3,\sigma_3\tau_3), \nonumber \\
	\vB =(\sigma_0\tau_2\rho_2,-\sigma_3\tau_2\rho_1,\sigma_2\tau_2\rho_1),
\end{align}
where mirror antisymmetry forces $b_2=b_3=0$ and rotation symmetry forces $b_3=0$. The model has additional mirror antisymmetries with $k_2 \to -k_2$ (if $b_1=b_3 = 0$) and $k_3 \to -k_3$ (if $b_1 = b_2 = 0$), represented by $\sigma_2\tau_3$ and $\sigma_3\tau_3$, respectively, and rotation symmetries around the $x_1$ axis (if $b_1 = 0$) and $x_2$ axis (if $b_2=0$), represented by $\sigma_1$ and $\sigma_2$, respectively. A numerical diagonalization gives results that are indistinguishable from those of Fig.\ \ref{fig:AIIt13dR}. The existence of hinge modes in the presence of spinful twofold rotation symmetry is consistent with Ref.\ \onlinecite{khalaf2018}, where the same symmetry class was considered.


\subsubsection{Antiunitary symmetry: Class A with $s=4-2 d_\parallel \mod 8$}

\begin{figure}
\includegraphics[width=0.9\columnwidth]{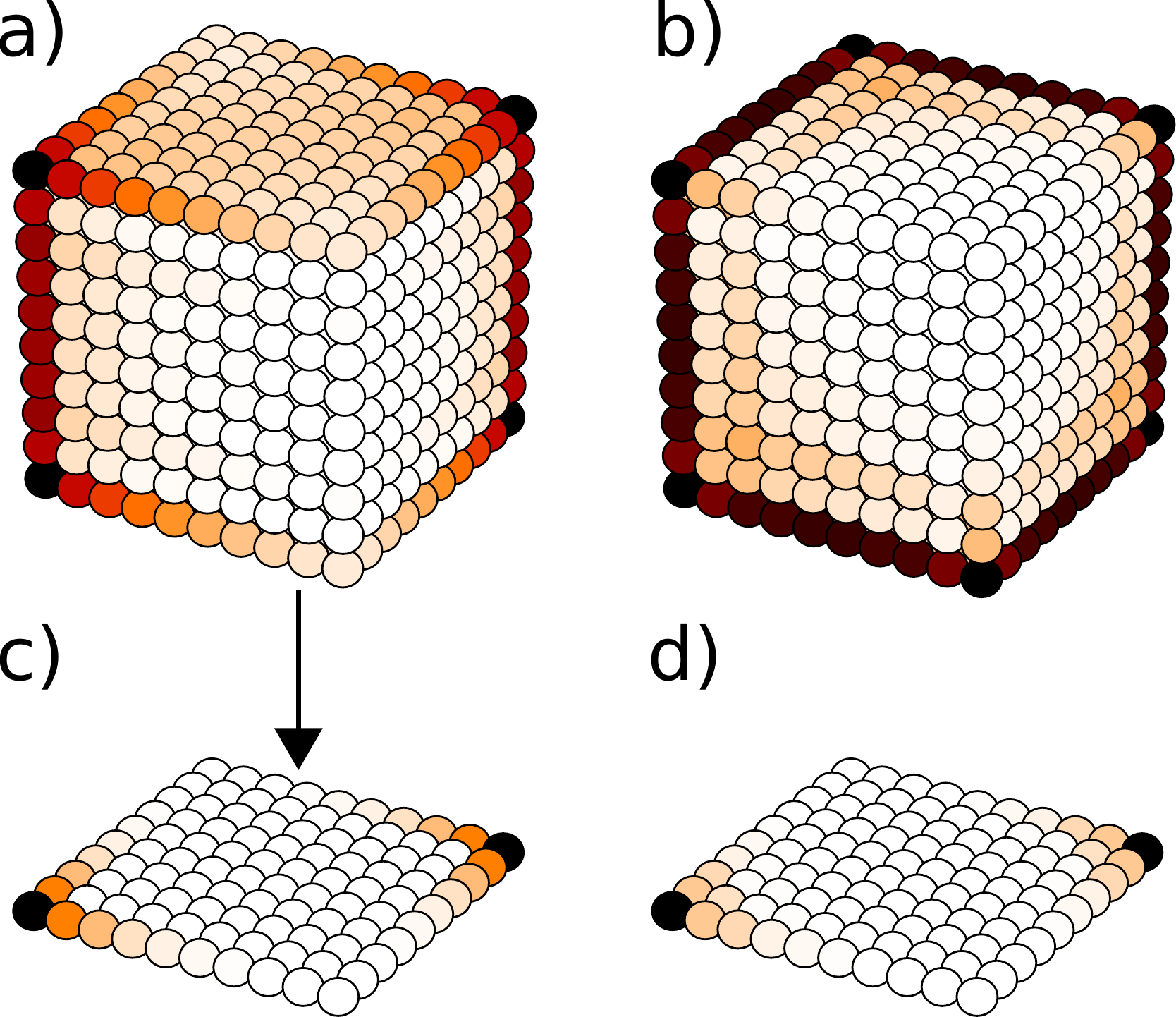}\\
\caption{\label{fig:AR} Support of the zero-energy states of the tight-binding Hamiltonian~(\ref{eq:HdG}) with $\bm\Gamma$ and $\vB$ given by Eq.~(\ref{eq:GammaC2T}) and $\vb = (0.4,-0.4,0)$ (a) and $\vb = (0.4,-0.4,0.4)$ (b). For the example shown in panel (a) the top and bottom surfaces are invariant with respect to the twofold rotation symmetry, which explains the presence of gapless surface modes at the top and bottom surface. The twofold rotation symmetry is broken in panel (b), which only has inversion symmetry. Panel (c) shows the support of the zero-energy corner modes of the two-dimensional Hamiltonian in class AIII$^{\mathcal{T^+R}_-}$ obtained by dimensional reduction of the three-dimensional model, with parameter $\vb = (0.4,-0.4,0)$. For comparison, panel (d) shows the support of the zero-energy corner modes obtained for the two-dimensional tight-binding model specified by Eqs.\ (\ref{eq:HdG}) and (\ref{eq:AIIIt1Gd2}) with $\vb = (0.4,-0.4)$.}
\end{figure}

This example applies to the classes A$^{\mathcal{P^+M}}$, A$^{\mathcal{T^+R}}$, and A$^{\mathcal{P^-I}}$. We represent the symmetry operations using $U_{\mathcal{PM}} = \tau_3$, $U_{\mathcal{TR}} = \sigma_1$, and $U_{\mathcal{PI}} = \sigma_2$ and consider a tight-binding Hamiltonian of the form (\ref{eq:HdG}) with
\begin{align}
  \Gamma_0 &= \sigma_2 \tau_0,\ \
  \bm \Gamma = (\sigma_1 \tau_1,\sigma_1 \tau_3,\sigma_3\tau_0),\nonumber \\
  \vB &= (\sigma_2\tau_3,-\sigma_2\tau_1,\sigma_0\tau_2),
  \label{eq:GammaC2T}
\end{align}
where the antiunitary mirror antisymmetry requires that $b_2=b_3=0$ and the twofold antiunitary rotation symmetry requires that $b_3=0$. Figures \ref{fig:AR}a and b show the hinge states for two example lattice structures with $m=-1$ and $\vb = (0.4, -0.4, 0)$ and $b = (0,4,-0.4,0.4)$, respectively.

Upon performing the reflection-matrix dimensional reduction scheme of Sec.\ \ref{sec:3} the model defined by the choice (\ref{eq:GammaC2T}) can be used to generate a two-dimensional Hamiltonian in classes AIII$^{\mathcal{T^+M}_+}$ and AIII$^{\mathcal{T^+R}_-}$ with $U_{\cal C} = \rho_3$, $U_{\mathcal{TM}} = \sigma_0 \tau_3$, and $U_{\mathcal{TR}} = \sigma_1 \rho_1$. Figure \ref{fig:AR}d shows the support of the zero-energy corner states of the two-dimensional Hamiltonian that is obtained this way. 

The model that is obtained using the dimensional reduction scheme is an eight-band model. This is not the minimal number of bands for which a nontrivial second-order topological insulator in the classes AIII$^{\mathcal{T^+M}_+}$ and AIII$^{\mathcal{T^+R}_-}$ exists. An example of a minimal model is given by a two-dimensional tight-binding Hamiltonian of the form (\ref{eq:HdG}) with
\begin{equation}
  \Gamma_0 = \sigma_2 \tau_0,\ \ \bm \Gamma = (\sigma_1 \tau_1, \sigma_1 \tau_3),\ \ \vB = (\sigma_2 \tau_3,-\sigma_2 \tau_1),
  \label{eq:AIIIt1Gd2}
\end{equation}
which has a chiral symmetry $U_{\cal C} = \sigma_3$, mirror symmetry $U_{\mathcal{TM}} = \sigma_3 \tau_3$ for $b_2 = 0$ and a twofold rotation symmetry $U_{\mathcal{TR}} = \sigma_1$ for arbitrary $b_1$, $b_2$. This model has zero-energy corner states, as shown in Fig.\ \ref{fig:AR}d for the parameter choice $m = -1$, $\vb = (0.4,-0.4)$.


\section{Conclusion}

In this work we extend the construction scheme introduced by Langbehn~\textit{et
al.}~\cite{langbehn2017} for second-order topological insulators and superconductors with mirror reflection symmetry to the larger class of topological insulators and superconductors stabilized by any order-two crystalline symmetry or antisymmetry, unitary or antiunitary. The order-two crystalline symmetries include mirror reflection, twofold rotation, and inversion. 

For the mirror-symmetric topological crystalline insulator and superconductors we showed that a topologically nontrivial bulk implies that either all boundaries have gapless modes, in which case the topological crystalline insulator or superconductor is a strong topological insulator or superconductor which does not rely on the crystalline symmetry for its protection, or it is a second-order topological insulator, with zero-energy states at mirror-symmetric corners or gapless modes at mirror-symmetric hinges. Moreover, we showed that there is a ``corner-to-bulk correspondence'' or ``hinge-to-bulk correspondence'', according to which the classification of possible protected corner or hinge states modulo lattice termination effects is identical to the that of the bulk topology, after removal of the strong topological phases. On the other hand, no complete corner-to-bulk correspondence or hinge-to-bulk correspondence exists for topological crystalline phases protected by a twofold rotation symmetry or by inversion symmetry, since these symmetries no not allow for symmetry-invariant corners or hinges in two and three dimensions. Instead, there is a partial correspondence, which relates the parity of the number of corner states or hinge modes to the bulk topology.

For topological crystalline phases in which the number $d_{\parallel}$ of inverted spatial dimensions is smaller than the spatial dimension $d$, such as phases protected by mirror reflection for $d \ge 2$ or twofold rotation for $d \ge 3$, there is a bulk-to-boundary correspondence, which uniquely links the bulk topology with the boundary states on a symmetry-invariant boundary. The corner-to-bulk correspondence or hinge-to-bulk correspondence for those phases shows that they may have protected states at corners or hinges, too, but it does not provide information beyond what is already known from considering symmetry-invariant boundaries. 
This is different for topological crystalline phases with $d_{\parallel} = d$, such as two-fold rotation symmetry for $d=2$ or inversion symmetry for $d=3$, for which there are no symmetry-invariant boundaries and, hence, no (first-order) bulk-to-boundary correspondence. In this case the $\ZZ_2$ sum rule for the number of corner states or hinge modes that we derive here provides a unique boundary signature of a nontrivial topological crystalline phase for a case in which no other boundary signatures are known to exist.\cite{fang2018,khalaf2018} Correspondingly, the demonstration that a nontrivial topological crystalline phase implies the existence of protected corner states or hinge modes cannot start from a theory of gapless boundary modes, as it does for $d_{\parallel} < d$,\cite{langbehn2017} but, instead, must start from the gapped bulk, as is done in Ref.\ \onlinecite{khalaf2018} and Sec.\ \ref{sec:6} for specific examples, or, as a general construction, by dimensional reduction from a hypothetical higher-dimensional topological crystalline phase for which symmetry-invariant boundaries exist. This is the route we take in Sec.\ \ref{sec:5}, using a real-space dimensional reduction scheme based on the scattering matrix.\cite{fulga2012,trifunovic2017} 

It is important to stress that, although crystalline symmetries are key to our construction of second-order topological phases, second-order topological phases are immune to weak perturbations that break the crystalline symmetry, as long as the boundary gaps are preserved.\cite{langbehn2017} In our description, this stability is reflected in the use of two classification schemes: An {\em extrinsic} classification scheme, which classifies corner states or hinge modes with respect to continuous transformations of the Hamiltonian that preserve both bulk and boundary gaps, and an {\em intrinsic} classification, which allows transformations of the Hamiltonian in which the boundary gap is closed, as long as the bulk gap is preserved. 
The intrinsic classification depends on the bulk topology only, and is
independent of the lattice termination. On the other hand, it is the extrinsic
classification, with the possible inclusion of local symmetry-breaking
perturbations, that captures the robustness of the phenomena associated with a
second-order topological phase to weak symmetry-breaking perturbations.

Not all two-dimensional materials with corner states or all three-dimensional
materials with gapless hinge modes are in a second-order topological phase ---
just like not all materials with a gapped bulk and gapless boundary states are
topological. For a second-order topological phase it is necessary that the
corner states or hinge modes have a topological protection. A classification of
the type that we present here is a key prerequisite to determine whether a true
topological protection can exist, or whether the existence of corner states or
hinge modes in a given model is merely a matter of coincidence. For example,
the existence of zero-energy corner modes always requires that the Hamiltonian
satisfy an {\em antisymmetry}, ruling out a second-order phase in a
two-dimensional lattice model with {\em symmetries} only --- in contrast to
recent claims in the literature.\cite{ezawa2018,ezawa2018b,xu2018}

The phenomenology of a second-order topological phase --- the existence of protected zero-energy corner states or gapless hinge modes on an otherwise gapped boundary --- is not the only possible manifestation of a nontrivial bulk topology if the standard bulk-to-boundary correspondence does not apply. As pointed out in Refs.\ \onlinecite{lau2016,vanmiert2017,benalcazar2017,benalcazar2017b,ezawa2018}, a nontrivial bulk crystalline topology may also manifest itself through a nontrivial quantized electric multipole moment or through the existence of fractional end or corner charges. (Note that a corner charge is different from a zero-energy corner state: A zero-energy corner state implies a degeneracy of the many-body ground state, whereas a corner charge implies the local accumulation of charge in an otherwise non-degenerate many-particle ground state.) If the Hamiltonian possesses an antisymmetry, as is the case for certain models considered in the literature,\cite{lau2016,benalcazar2017} a nontrivial electric multipole moment and protected zero-energy corner states can exist simultaneously, but this need not always be case. A counterexample is the ``breathing pyrochlore lattice'' of Ref.\ \onlinecite{ezawa2018}, for which the nontrivial bulk topology manifests itself through a quantized bulk polarization, whereas the zero-energy corner states of Ref.\ \onlinecite{ezawa2018} lack topological protection.

Only few materials have been proposed as realizations of second-order phases.
Examples are strained SnTe\cite{schindler2018} or odd-parity superconducting
order in doped nodal-loop materials\cite{shapourian2018topological} both with
mirror symmetry, and bismuth~\cite{schindler2018b} with inversion symmetry.
Simultaneously, the phenomenology of second-order phases has been reproduced
experimentally in artificial ``materials'', such as electrical\cite{imhof2018}
or microwave\cite{peterson2017} circuits, or coupled mechanical
oscillators.\cite{serra-garcia2018} We hope that the complete classification
presented here will help to identify new material candidates for the
solid-state realizations of second-order topological insulators and
superconductors.

\acknowledgements

We thank Andrei Bernevig, Eslam Khalaf, and Felix von Oppen for stimulating discussions. We acknowledge support by project A03 of the CRC-TR 183 and by the priority programme SPP 1666 of the German Science Foundation (DFG). A recent article by Eslam Khalaf on the classification of inversion-symmetric higher-order topological insulators and superconductors contains closely related results. We thank Eslam Khalaf for making his manuscript available to us prior to publication.

\label{sec:7}
\appendix

\section{Reflection-matrix-based dimensional reduction scheme}
\label{app:1}

In this appendix we describe details of the reflection-matrix based dimensional
reduction scheme. We first review how this method works in the absence of
crystalline symmetry, following the original article by Fulga 
{\em et al.},\cite{fulga2012} and then show how to include order-two crystalline symmetries
with $d_{\parallel} < d$, generalizing the analysis of Ref.\
\onlinecite{trifunovic2017}. The reflection-matrix based dimensional reduction
scheme leaves $d_{\parallel}$ unchanged, so that the minimal dimension it can
achieve is $d=d_{\parallel}$. The main text discusses how the reflection-matrix
based dimensional reduction scheme can also be applied to second-order
topological insulators and superconductors.

\subsection{Altland-Zirnbauer classes without crystalline symmetries}

The key step in the method of Ref.~\onlinecite{fulga2012} is the
construction of a $(d-1)$-dimensional gapped Hamiltonian $H_{d-1}$ for each
$d$ dimensional gapped Hamiltonian $H_d$. The Hamiltonians $H_d$ and
$H_{d-1}$ have different symmetries, but the same (strong) topological
invariants. Fulga {\em et al.} show how the Hamiltonian $H_{d-1}$ can be
constructed from the reflection matrix $r_d$ if a gapped system with Hamiltonian
$H_{d}$ is attached to an ideal lead with a $(d-1)$-dimensional cross section.

\begin{figure}
\includegraphics[width=0.8\columnwidth]{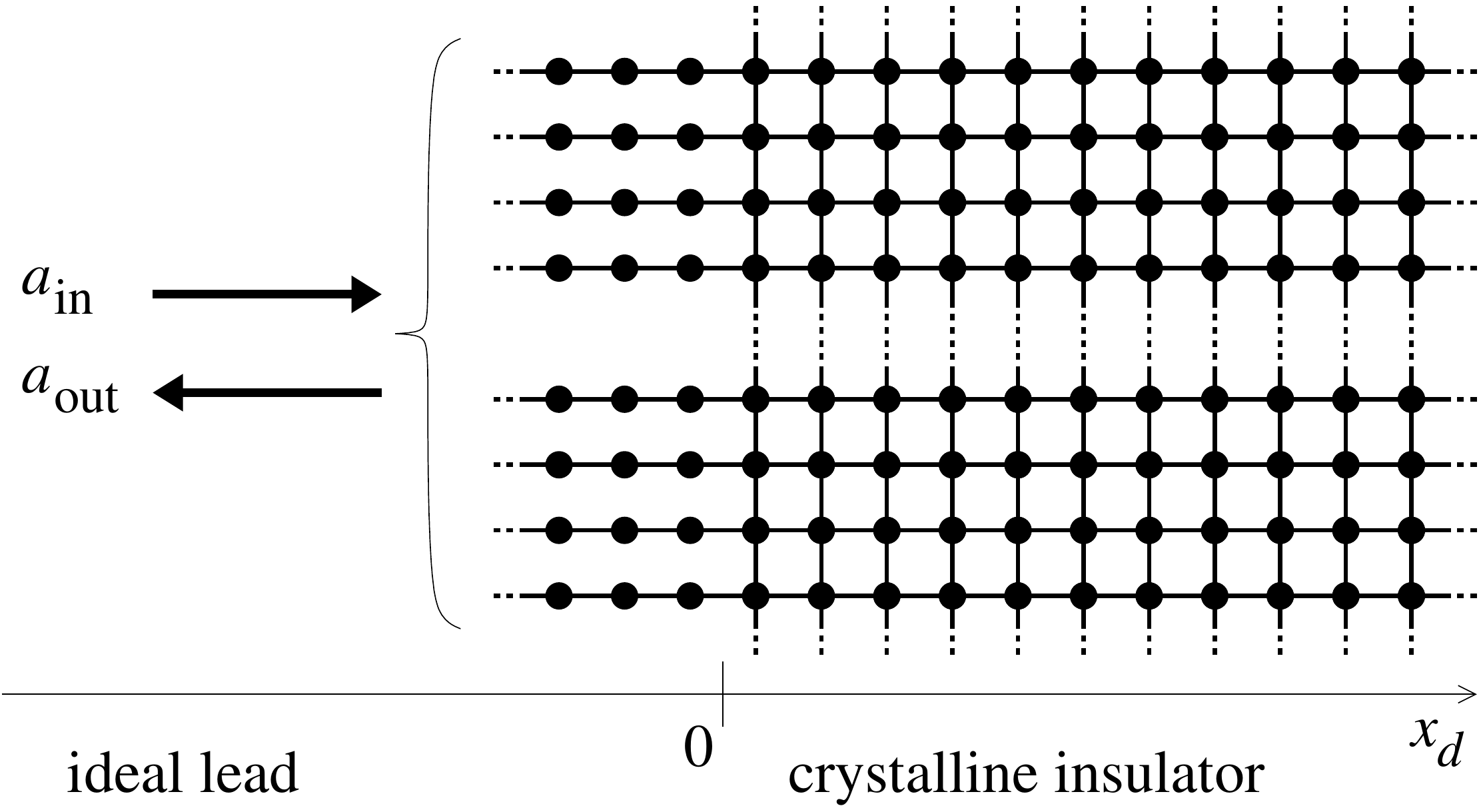}
\caption{\label{fig:setup} Schematic picture of a $d$-dimensional gapped crystalline insulator occupying the half space $x_d > 0$, with periodic boundary conditions applied along the remaining $(d-1)$-dimension, coupled to an ideal lead with a $(d-1)$-dimensional cross section. The reflection matrix $r_d(\vk_{\perp})$ relates the amplitudes $a_{\rm out}(\vk_{\perp})$ and $a_{\rm in}(\vk_{\perp})$ of outgoing and incoming modes in the lead.}
\end{figure}

To be specific, following Ref.\ \onlinecite{fulga2012} we consider a $d$-dimensional gapped insulator with Hamiltonian $H_d(\vk) = H_d(\vk_{\perp},k_d)$, occupying the half space $x_d > 0$ and periodic boundary conditions in the transverse directions, see Fig.~\ref{fig:setup}. The half space $x_d < 0$ consists of an ideal lead with transverse modes labeled by the $d-1$ dimensional wavevector $\vk_{\perp}$. The amplitudes $a_{\rm out}(\vk_{\perp})$ and $a_{\rm in}(\vk_{\perp})$ of outgoing and incoming modes are related by the reflection matrix $r_d(\vk_{\perp})$,
\begin{equation}
  a_{\rm out}(\vk_{\perp}) = r_d(\vk_{\perp}) a_{\rm in}(\vk_{\perp}).
\end{equation}
Since $H_d$ is gapped, $r_d(\vk_{\perp})$ is unitary. Time-reversal symmetry, particle-hole antisymmetry, or chiral antisymmetry pose additional constraints on $r_d(\vk_{\perp})$. These follow from the action of these symmetries on the amplitudes $a_{\rm in}$ and $a_{\rm out}$, 
\begin{align}
  {\cal T} a_{\rm in}(\vk_{\perp}) & = Q_{{\cal T}}\, a_{\rm out}^*(-\vk_{\perp}), \nonumber \\   {\cal T} a_{\rm out}(\vk_{\perp}) &=\, V_{{\cal T}} a_{\rm in}^*(-\vk_{\perp}), \\
  {\cal P} a_{\rm in}(\vk_{\perp}) & = V_{{\cal P}}\, a_{\rm in}^*(-\vk_{\perp}), \nonumber \\  {\cal P} a_{\rm out}(\vk_{\perp}) &= Q_{{\cal P}}\, a_{\rm out}^*(-\vk_{\perp}), \\
  {\cal C} a_{\rm in}(\vk_{\perp}) & = Q_{{\cal C}}\, a_{\rm out}(\vk_{\perp}), \nonumber \\   {\cal C} a_{\rm out}(\vk_{\perp}) & = V_{{\cal C}}\, a_{\rm in}(\vk_{\perp}),
\end{align}
where $V_{{\cal T}}$, $Q_{{\cal T}}$, $V_{{\cal P}}$, $Q_{{\cal P}}$, $V_{{\cal
C}}$, and $Q_{{\cal C}}$ are $\vk_{\perp}$-independent unitary matrices that satisfy $V_{\cal T} Q_{{\cal T}}^* = Q_{\cal T} V_{{\cal T}}^* = {\cal
T}^2$,  $V_{{\cal P}} V_{{\cal P}}^* = Q_{{\cal P}} Q_{{\cal P}}^* = {\cal
P}^2$, and $Q_{{\cal C}} V_{{\cal C}} = {\cal C}^2 = 1$. Systems with both
time-reversal symmetry and particle-hole antisymmetry also have a chiral antisymmetry, with
$Q_{{\cal C}} = V_{{\cal P}} Q_{{\cal T}}^* = {\cal T}^2 {\cal P}^2 Q_{{\cal
T}} Q_{{\cal P}}^*$ and $V_{{\cal C}} = Q_{{\cal P}} V_{{\cal T}}^*= {\cal T}^2
{\cal P}^2 V_{{\cal T}} V_{{\cal P}}^*$. For the reflection matrix
$r_d(\vk_{\perp})$ the presence of time-reversal symmetry,
particle-hole antisymmetry, and/or chiral antisymmetry leads to the constraints
\begin{align}
  r_d(\vk_{\perp}) = Q_{{\cal T}}^{\rm T} r_d(-\vk_{\perp})^{\rm T} V_{{\cal T}}^*,
  \label{eq:rtrs} \\
  r_d(\vk_{\perp}) = Q_{{\cal P}}^{\rm T} r_d(-\vk_{\perp})^{*} V_{{\cal P}}^*,
  \label{eq:rphs} \\
  r_d(\vk_{\perp}) = Q_{{\cal C}}^{\dagger} r_d(\vk_{\perp})^{\dagger} V_{{\cal C}}.
  \label{eq:rchiral}
\end{align}

The effective Hamiltonian $H_{d-1}$ is constructed out of $r_d(\vk_{\perp})$ in different ways, depending on the presence or absence of chiral symmetry. With chiral symmetry one sets
\begin{equation}
  H_{d-1}(\vk) \equiv Q_{{\cal C}} r_d(\vk),
  \label{eq:Hdchiral}
\end{equation}
using Eq.\ (\ref{eq:rchiral}) to verify that $H_{d-1}$ is indeed hermitian. (Recall that $V_{{\cal C}} = Q_{{\cal C}}^{\dagger}$ since $Q_{{\cal C}} V_{{\cal C}} = {\cal C}^2 = 1$.) Equation (\ref{eq:Hdchiral}) simplifies to Eq.\ (\ref{eq:HdchiralF}) of the main text if the basis of scattering states is chosen such that $Q_{\cal C} = V_{\cal C} = 1$. Without chiral symmetry one defines $H_{d-1}$ as
\begin{equation}
  H_{d-1}(\vk) = 
  \begin{pmatrix}
    0 & r_d(\vk)\\
    r^\dagger_d(\vk) & 0
  \end{pmatrix},
  \label{eq:Hdnonchiral}
\end{equation}
which is manifestly hermitian and satisfies a chiral symmetry with $U_{\cal C} = \sigma_3 = \mbox{diag}\,(1,-1)$. Hence, for the complex classes the dimensional reduction procedure $H_{d} \to H_{d-1}$ maps a Hamiltonian with chiral symmetry to one without, and vice versa, corresponding to the period-two sequence 
$$
  \mbox{A} \xrightarrow{d-1} \mbox{AIII} \xrightarrow{d-1} \mbox{A}.
$$%
Bulk-boundary correspondence implies that the bulk, which is described by the
Hamiltonian $H_d(\vk)$, and the boundary, which determines the reflection
matrix $r_d(\vk_{\perp})$, have the same topological classification. Since
$r_d(\vk_{\perp})$ is in one-to-one correspondence with the Hamiltonian
$H_{d-1}(\vk_{\perp})$, this implies that $H_d$ and $H_{d-1}$ have the same
topological classification.

Central point in the construction of Ref.\ \onlinecite{fulga2012} is that if the Hamiltonian $H_{d}$ possesses an additional antiunitary symmetry and/or antisymmetry, placing it in one of the real Altland-Zirnbauer classes labeled $s=0,1,\ldots,7$, then $H_{d-1}$ possesses an antiunitary symmetry and/or antisymmetry, too, such that it is in Altland Zirnbauer class $s-1$.\cite{fulga2012,trifunovic2017} Hence, for the real Altland-Zirnbauer classes, the reflection-matrix based dimensional reduction scheme generates the period-eight sequence
\begin{align}
  \mbox{CI} &\xrightarrow{d-1}
  \mbox{C} \xrightarrow{d-1}
  \mbox{CII} \xrightarrow{d-1}
  \mbox{AII} \xrightarrow{d-1}
  \mbox{DIII}  \nonumber\\ & \xrightarrow{d-1}
  \mbox{D} \xrightarrow{d-1}
  \mbox{BDI} \xrightarrow{d-1}
  \mbox{AI} \xrightarrow{d-1}
  \mbox{CI},
  \label{eq:8seq_standard}
\end{align}
which is the well-known period-eight Bott periodicity known from the classification of topological insulators and superconductors.\cite{schnyder2008,kitaev2009,stone2011,wen2012,abramovici2012,kennedy2016}

\subsection{With order-two crystalline symmetries}

Bulk-boundary correspondence continues to exist in the presence of an order-two crystalline symmetry with $d_{\parallel} < d$, if the sample surface is left invariant under the symmetry operation. (For $d_{\parallel} = d$ there are no such invariant surfaces.) Correspondingly, the reflection-matrix based dimensional reduction scheme may be used in the presence of such crystalline symmetries, too, as was shown for the case of reflection symmetry by two of us in Ref.\ \onlinecite{trifunovic2017}. 

Labeling the coordinates as in Sec.~\ref{sec:2}, the coordinate $x_d$ is left invariant by the crystalline symmetry operation ${\cal S}$ if $d_{\parallel} < d$. Hence, taking the same geometry as above, the lead and the lead-crystal interface are mapped to themselves under ${\cal S}$. We now discuss the four cases of unitary symmetry, unitary antisymmetry, antiunitary symmetry, and antiunitary antisymmetry separately.

{\em Unitary symmetry.---} As with the non-spatial symmetries, the action of a unitary symmetry operation ${\cal S}$ on the amplitudes $a_{\rm in}$ and $a_{\rm out}$ of incoming and outgoing states in the leads involves multiplication with $\vk_{\perp}$-independent unitary matrices,
\begin{align}
  {\cal S} a_{\rm in}(\vk_{\perp}) & = V_{{\cal S}}\, a_{\rm in}(S \vk_{\perp}), \nonumber \\   {\cal S} a_{\rm out}(\vk_{\perp}) &=\, Q_{{\cal S}} a_{\rm out}(S \vk_{\perp}),
\end{align}
where $S \vk_{\perp} = (-k_1,\ldots,-k_{d_{\parallel}},k_{d_{\parallel}+1},\ldots,k_{d-1})$ denotes the action of the symmetry operation on the mode vector $\vk$ and the matrices $V_{{\cal S}}$ and $Q_{{\cal S}}$ satisfy $V_{{\cal S}}^2 = Q_{{\cal S}}^2 = {\cal S}^2 = 1$. The presence of the order-two crystalline symmetry leads to a constraint on the reflection matrix,
\begin{equation}
  r_d(\vk_{\perp}) = Q_{{\cal S}}^{\dagger} r_d(S \vk_{\perp}) V_{{\cal S}}.
\end{equation}
The algebraic relations involving the matrices $Q_{{\cal S}}$, $V_{{\cal S}}$
depend on whether the symmetry operation ${\cal S}$ commutes or anticommutes with the non-spatial symmetry operations ${\cal T}$, ${\cal P}$, and ${\cal C}$,
$Q_{{\cal T}} Q_{{\cal S}}^* = \eta_{{\cal T}} V_{{\cal S}} Q_{{\cal T}}$,
$V_{{\cal T}} V_{{\cal S}}^* = \eta_{{\cal T}} Q_{{\cal S}} V_{{\cal T}}$,
$V_{{\cal P}} V_{{\cal S}}^* = \eta_{{\cal P}} V_{{\cal S}} V_{{\cal P}}$,
$Q_{{\cal P}} Q_{{\cal S}}^* = \eta_{{\cal P}} Q_{{\cal S}} Q_{{\cal P}}$,
$Q_{{\cal C}} Q_{{\cal S}} = \eta_{{\cal C}} V_{{\cal S}} Q_{{\cal C}}$, and
$V_{{\cal C}} V_{{\cal S}} = \eta_{{\cal C}} Q_{{\cal S}} V_{{\cal C}}$.

{\em Unitary antisymmetry.---} An order-two unitary antisymmetry ${\cal CS}$ also exchanges incoming and outgoing modes, such that one has
\begin{align}
  {\cal CS} a_{\rm in}(\vk_{\perp}) & = Q_{{\cal CS}}\, a_{\rm out}(S \vk_{\perp}), \nonumber \\   {\cal CS} a_{\rm out}(\vk_{\perp}) &=\, V_{{\cal CS}} a_{\rm in}(S \vk_{\perp}),
\end{align}
with $S \vk_{\perp}$ defined a above. For an antisymmetry operation ${\cal CS}$ the matrices $V_{{\cal CS}}$ and $Q_{{\cal CS}}$ satisfy $V_{{\cal CS}} Q_{{\cal CS}} = Q_{{\cal SC}} V_{{\cal CS}} = ({\cal CS})^2 = 1$. The presence of the crystalline unitary antisymmetry ${\cal CS}$ implies that the reflection matrix satisfies
\begin{equation}
  r_d(\vk_{\perp}) = Q_{{\cal CS}}^{\dagger} r_d(S \vk_{\perp})^{\dagger} V_{{\cal CS}},
\end{equation}
and the matrices $Q_{{\cal CS}}$ and $V_{{\cal CS}}$ satisfy the algebraic relations
$Q_{{\cal T}} V_{{\cal CS}}^* = \eta_{{\cal T}} Q_{{\cal CS}} V_{{\cal T}}$, 
$V_{{\cal T}} Q_{{\cal CS}}^* = \eta_{{\cal T}} V_{{\cal CS}} Q_{{\cal T}}$, 
$V_{{\cal P}} Q_{{\cal CS}}^* = \eta_{{\cal P}} Q_{{\cal CS}} Q_{{\cal P}}$, 
$Q_{{\cal P}} V_{{\cal CS}}^* = \eta_{{\cal P}} V_{{\cal CS}} V_{{\cal P}}$, 
$Q_{{\cal CS}} V_{{\cal C}} = \eta_{{\cal C}} Q_{{\cal C}} V_{{\cal CS}}$, 
$V_{{\cal CS}} Q_{{\cal C}} = \eta_{{\cal C}} V_{{\cal C}} Q_{{\cal CS}}$.

{\em Antiunitary symmetry.---} The action of an order-two antiunitary symmetry $\mathcal{T^{\pm} S}$ on the scattering amplitudes is represented by unitary matrices $V_{\cal TS}$ and $Q_{\cal TS}$,
\begin{align}
  {\cal TS} a_{\rm in}(\vk_{\perp}) & = Q_{{\cal TS}}\, a_{\rm out}^*(-S \vk_{\perp}), \nonumber \\   {\cal TS} a_{\rm out}(\vk_{\perp}) &=\, V_{{\cal TS}} a_{\rm in}^*(-S \vk_{\perp}),
\end{align}
with $V_{{\cal TS}} Q_{{\cal TS}}^* = Q_{{\cal TS}} V_{{\cal TS}}^* = ({\cal TS})^2 = \pm 1$. The presence of the order-two crystalline antiunitary symmetry leads to a constraint on the reflection matrix,
\begin{equation}
  r_d(\vk_{\perp}) = Q_{{\cal TS}}^{\rm T} r_d(-S \vk_{\perp})^{\rm T} V_{{\cal TS}}^*,
\end{equation}
and the matrices $Q_{{\cal TS}}$ and $V_{{\cal TS}}$ satisfy the algebraic relations 
$Q_{{\cal T}} V_{{\cal TS}}^* = \eta_{{\cal T}} Q_{{\cal TS}} V_{{\cal T}}^*$, 
$V_{{\cal T}} Q_{{\cal TS}}^* = \eta_{{\cal T}} V_{{\cal TS}} Q_{{\cal T}}^*$, 
$V_{{\cal P}} Q_{{\cal TS}}^* = \eta_{{\cal P}} Q_{{\cal TS}} Q_{{\cal P}}^*$, 
$Q_{{\cal P}} V_{{\cal TS}}^* = \eta_{{\cal P}} V_{{\cal TS}} V_{{\cal P}}^*$, 
$Q_{{\cal TS}} V_{{\cal C}}^* = \eta_{{\cal C}} Q_{{\cal C}} V_{{\cal TS}}$, 
$V_{{\cal TS}} Q_{{\cal C}}^* = \eta_{{\cal C}} V_{{\cal C}} Q_{{\cal TS}}$.

{\em Antiunitary antisymmetry.---} Finally, for an antiunitary antisymmetry $\mathcal{P^{\pm}S}$ one has
\begin{align}
  {\cal PS} a_{\rm in}(\vk_{\perp}) & = V_{{\cal PS}}\, a_{\rm in}^*(-S \vk_{\perp}), \nonumber \\   {\cal PS} a_{\rm out}(\vk_{\perp}) &=\, Q_{{\cal PS}} a_{\rm out}^*(-S \vk_{\perp}),
\end{align}
with $V_{{\cal PS}} V_{{\cal PS}}^* = Q_{{\cal PS}} Q_{{\cal PS}}^* = ({\cal PS})^2 = \pm 1$. The reflection matrix satisfies
\begin{equation}
  r_d(\vk_{\perp}) = Q_{{\cal PS}}^{\rm T} r_d(-S \vk_{\perp})^{*} V_{{\cal PS}}^*
\end{equation}
and the algebraic relations involving the matrices $Q_{{\cal PS}}$ and $V_{{\cal PS}}$ are 
$Q_{{\cal T}} Q_{{\cal PS}}^* = \eta_{{\cal T}} V_{{\cal PS}} Q_{{\cal T}}^*$,
$V_{{\cal T}} V_{{\cal PS}}^* = \eta_{{\cal T}} Q_{{\cal PS}} V_{{\cal T}}^*$,
$V_{{\cal P}} V_{{\cal PS}}^* = \eta_{{\cal P}} V_{{\cal PS}} V_{{\cal P}}^*$,
$Q_{{\cal P}} Q_{{\cal PS}}^* = \eta_{{\cal P}} Q_{{\cal PS}} Q_{{\cal P}}^*$,
$Q_{{\cal C}} Q_{{\cal PS}} = \eta_{{\cal C}} V_{{\cal PS}} Q_{{\cal C}}^*$, and
$V_{{\cal C}} V_{{\cal PS}} = \eta_{{\cal C}} Q_{{\cal PS}} V_{{\cal C}}^*$.

To see how the presence of an order-two crystalline symmetry or antisymmetry
affects the dimensional
reduction scheme we first consider the complex classes A and AIII. We start from a Hamiltonian in Shiozaki-Sato symmetry class A$^{\cal S}$, $(s,t) = (0,0)$, which is represented by a Hamiltonian $H_d$ in symmetry class A with a unitary symmetry ${\cal S}$. Constructing a Hamiltonian $H_{d-1}$ in class AIII as described above, we find that the unitary symmetry ${\cal S}$ imposes the additional constraint
\begin{equation}
  H_{d-1}(\vk_{\perp}) = U_{\cal S}^{\dagger} H_{d-1}(S \vk_{\perp}) U_{\cal S},
\end{equation}
on $H_{d-1}$, with
\begin{equation}
  U_{\cal S} = 
  \begin{pmatrix}
    Q_{{\cal S}} & 0 \\
    0 & V_{{\cal S}} 
  \end{pmatrix}.
\end{equation}
Since $U_{\cal S}$ commutes with $\sigma_3$ and $U_{\cal S}^2 = 1$, we conclude that dimensional reduction maps the class A$^{\cal S}$ to class AIII$^{{\cal S}_+}$. In the classification of Shiozaki and Sato this class is labeled $(s,t) = (1,0)$. Similarly, if $H_d$ is a Hamiltonian in Shiozaki class $(s,t) = (0,1)$, which is represented by a unitary antisymmetry ${\cal CS}$, the mapped Hamiltonian $H_{d-1}$ satisfies the additional symmetry
\begin{equation}
  U_{\cal CS}^{\dagger} H_{d-1}(S \vk_{\perp}) U_{\cal CS}
\end{equation}
with
\begin{equation}
  U_{\cal CS} = 
  \begin{pmatrix}
    0 & V_{{\cal CS}}  \\
    Q_{{\cal CS}} & 0
  \end{pmatrix}.
\end{equation}
This is a unitary symmetry operation that anticommutes with the chiral operation $\sigma_3$, so that the mapped Hamiltonian is in Shiozaki-Sato class AIII$^{{\cal S}_-}$, $(s,t) = (1,1)$. Finally, starting with a Hamiltonian with symmetry of type $(s,t) = (1,t)$, represented by a class AIII Hamiltonian with an order-two crystalline symmetry ${\cal S}$ commuting ($\eta_{\cal C} = 1)$ or anticommuting ($\eta_{\cal C} = -1$) with ${\cal C}$ for $t=0$, $1$, respectively, the mapped Hamiltonian satisfies the constraint
\begin{equation}
  H_{d-1}(\vk_{\perp}) = \eta_{\cal C} V_{\cal S}^{\dagger} H_{d-1}(S \vk_{\perp}) V_{\cal S},
\end{equation}
which is a symmetry of Shiozaki-Sato type $(s,t) = (0,t)$, $t=0,1$. (It is a unitary symmetry for $\eta_{\cal C} = 1$ and an antisymmetry for $\eta_{\cal C} = -1$.)

A similar procedure can be applied to the remaining complex Shiozaki-Sato classes, which are labeled by a single integer $s=0,1,\ldots,7$. Starting with a Hamiltonian of Shiozaki classes $s=0$ and $s=4$ (classes A$^{\mathcal{T^+S}}$ and A$^{\mathcal{T^-S}}$, antiunitary symmetry ${\cal TS}$ squaring to $1$ and $-1$, respectively), we find that the mapped Hamiltonian $H_{d-1}$ satisfies the constraint
\begin{equation}
  H_{d-1}(\vk_{\perp}) = U_{\cal TS}^{\dagger} H_{d-1}(-S\vk_{\perp})^* U_{\cal TS}
\end{equation}
with
\begin{equation}
  U_{\cal TS} = 
  \begin{pmatrix}
    0 & V_{{\cal TS}}^*  \\
    Q_{{\cal TS}}^* & 0
  \end{pmatrix}.
\end{equation}
Hence, $H_{d-1}$ satisfies an antiunitary symmetry that anticommutes with the chiral operation ${\cal C}$ and squares to $1$ or $-1$, so that the mapped Hamiltonian is in Shiozaki classes $s=7$ and $s=3$, respectively. Similarly, for symmetry classes $s=2$ and $s=6$, corresponding to an antiunitary antisymmetry squaring to $1$ or $-1$, respectively, we find that $H_{d-1}$ satisfies the constraint
\begin{equation}
  H_{d-1}(\vk_{\perp}) = U_{\cal PS}^{\dagger} H_{d-1}(-S\vk_{\perp})^* U_{\cal PS}
\end{equation}
with
\begin{equation}
  U_{\cal PS} = 
  \begin{pmatrix}
    Q_{{\cal PS}}^{*} & 0  \\
    0 & V_{{\cal PS}}^{*}
  \end{pmatrix}.
\end{equation}
This is an antiunitary symmetry that commutes with the chiral operation and squares to $1$ or $-1$, corresponding to symmetry classes $s=1$ and $s=5$, respectively. Finally, for the remaining symmetry classes we may start from an antiunitary symmetry squaring to $\pm 1$ and find that the mapped Hamiltonian satisfies
\begin{equation}
  H_{d-1}(\vk_{\perp}) = \eta_{\cal C} (Q_{\cal C} V_{\cal TS})^{\dagger} H_{d-1}(-S \vk_{\perp})^* (Q_{\cal C} V_{\cal TS}),
\end{equation}
which is an antiunitary symmetry (for $\eta_{\cal C} = 1$) or antisymmetry (for $\eta_{\cal C}=-1$) that squares to $\pm \eta_{\cal C}$, so that under dimensional reduction the class $s=1$, $3$, $5$, $7$ is mapped to $s=0$, $2$, $4$, and $6$, respectively.

For the real classes we may proceed in the same way. One finds that under dimensional reduction the Shiozaki-Sato symmetry class $(s,t)$ is mapped to $(s-1,t)$, with $s$ modulo $8$. The derivation is identical to that given in Ref.\ \onlinecite{trifunovic2017} for the case of mirror reflection symmetry. 

\section{Classification of mirror-symmetric corners of two-dimensional crystals}

\label{app:2}

In this Appendix we explain the origin of the entries in Tables~\ref{tab:4}--\ref{tab:6}. Throughout we will use the convention that the chiral operation ${\cal C}$ squares to one.

\subsection{Complex classes with antiunitary symmetries and antisymmetries}

{\em Class A$^{\mathcal{T^+M}}$, $s=0$.---} The topological crystalline phases coincide with the strong topological phases of Altland-Zirnbauer class A. No protected zero-energy corner states can persist in the trivial strong phase.

{\em Class AIII$^{\mathcal{T^+M_+}}$, $s=1$.---} Since the antiunitary mirror reflection operation ${\cal TM}$ commutes with the chiral operation ${\cal C}$, corner state have a well-defined parity $\sigma_{\cal C}$ under ${\cal C}$ and can be chosen to be mapped to themselves under the antiunitary mirror reflection operation ${\cal TM}$. Two corner states with opposite $\sigma_{\cal C}$ can be gapped out by a reflection-symmetric mass term, so that we may use the (extrinsic) integer topological index $N = N_+ - N_-$ to characterize the zero-energy states at a corner. 

A decoration of the edges by a topologically nontrivial one-dimensional chain
leads to the addition of two zero-energy states $\ket{\rm L}$ and $\ket{\rm R} =
{\cal TM} \ket{\rm L}$ placed symmetrically around the corner as in
Fig.~\ref{fig:decoration}. Since ${\cal TM}$ commutes with ${\cal C}$, these
corner states have the same value of $\sigma_{\cal C}$. Moreover, the linear
combinations $\ket{\rm L} + \ket{\rm R}$ and $i(\ket{\rm L} - \ket{\rm R})$ map
to themselves under ${\cal TM}$, so that they meet the classification criteria
for corner states formulated above. Hence, by changing the lattice termination
we may change $N_+$ or $N_-$ and, hence, $N$, by two. The parity of $N$ remains
unchanged under such a change of termination, which corresponds to an intrinsic $\ZZ_2$
topological index. 

If the antiunitary mirror reflection symmetry is broken locally near the corner these
conclusions do not change. We may still define $N = N_+ - N_-$ as a topological
invariant, which can not change without closing a boundary gap or the bulk gap,
and by changing the lattice termination one may still change add pairs of zero
modes to the corner, so that $N \mod 2$ is the appropriate invariant if
topological equivalence is defined with respect to transformations that possibly
close boundary gaps.

{\em Class A$^{\mathcal{P^+M}}$, $s=2$.---}
In this symmetry class the antiunitary reflection antisymmetry ${\cal PM}$ may protect a single zero-energy state at a mirror-symmetric corner. A pair of zero-energy states can, however, be gapped out by a mirror-antisymmetric perturbation. To see this, consider two zero modes $\ket{1}$ and $\ket{2}$, for which we may assume that they are both invariant under ${\cal PM}$. Then $i (\ket{1} \bra{2} - \ket{2} \bra{1})$ is a local perturbation that obeys the mirror reflection antisymmetry and gaps out the states $\ket{1}$ and $\ket{2}$. We conclude that a mirror-symmetric corner is described by a $\ZZ_2$ index.

{\em Class AIII$^{\mathcal{T^-M_-}}$, $s=3$.---}
The bulk phase is always topologically trivial.\cite{shiozaki2014} However, a single pair of corner states can be obtained by symmetrically decorating mirror-related edges with topologically nontrivial one-dimensional chains, as in Fig.~\ref{fig:decoration}. To see this, denote states $\ket{\rm L}$ and $\ket{\rm R} = {\cal TM} \ket{\rm L}$, as in Fig.~\ref{fig:decoration}. Since $({\cal TM})^2=-1$ the states $\ket{\rm L}$ and $\ket{\rm R}$ form a Kramers pair under the antiunitary mirror reflection operation, $\ket{\rm L} = - {\cal TM} \ket{\rm R}$. A single such pair of zero-energy states can not be gapped out by a perturbation that respects the antiunitary mirror reflection symmetry. 

{\em Class A$^{\mathcal{T^-M}}$, $s=4$.---}
The nontrivial topological crystalline insulator phases in this symmetry class are also strong topological phases, {\em i.e.,} they have protected edge modes on all edges, not only on mirror-symmetric edges. A second-order topological insulator phase with gapped edges and protected corner states does not exist for this symmetry class. 

{\em Class AIII$^{\mathcal{T^-M_+}}$, $s=5$.---}
The bulk phase is topologically trivial.\cite{shiozaki2014} However, (multiple) pairs of corner states can be obtained by symmetrically decorating mirror-related edges with topologically nontrivial one-dimensional chains, as in Fig.~\ref{fig:decoration}. To see this, denote states $\ket{\rm L}$ and $\ket{\rm R} = {\cal TM} \ket{\rm L}$, as in Fig.~\ref{fig:decoration}. The states $\ket{\rm L}$ and $\ket{\rm R}$ have the same parity under the chiral operation ${\cal C}$, since the antiunitary mirror reflection operation ${\cal TM}$ commutes with ${\cal C}$. Antisymmetry of the Hamiltonian under ${\cal C}$ protects corner states of equal parity, corresponding to a $2 \ZZ$ topological index.

{\em Class A$^{\mathcal{P^-M}}$, $s=6$.---}
In this symmetry class the bulk phase is topologically trivial. Alternatively, one can see that no protected zero-energy corner states can be consistent with the existence of an antiunitary mirror reflection antisymmetry ${\cal PM}$ with $({\cal PM})^2 = -1$: Such corner states would have to appear in pairs $\ket{0}$, ${\cal PM}\ket{0}$, which can be gapped out by the mass term $\ket{0}\bra{0} {\cal PM} + {\cal PM} \ket{0}\bra{0} $, which obeys the required antisymmetry under ${\cal PM}$.

{\em Class AIII$^{\mathcal{T^+M_-}}$, $s=7$.---}
This symmetry class is topologically trivial as a bulk phase and no corner states can be obtained by symmetrically decorating mirror-related edges with topologically nontrivial one-dimensional chains. To see, we again denote these end states $\ket{\rm L}$ and $\ket{\rm R} = {\cal TM} \ket{\rm L}$, as in Fig.~\ref{fig:decoration}. The states $\ket{\rm L}$ and $\ket{\rm R}$ have opposite parity under the chiral operation ${\cal C}$, since the antiunitary mirror reflection operation ${\cal TM}$ anticommutes with ${\cal C}$. The Hamiltonian $\ket{\rm R}\bra{\rm L} + \ket{\rm L} \bra{\rm R}$ anticommutes with ${\cal C}$, commutes with ${\cal TM}$, and gaps out the zero modes $\ket{\rm L}$ and $\ket{\rm R}$.

\subsection{Real classes}

We represent the Shiozaki-Sato classes using unitary mirror reflection symmetries ${\cal M}$ or antisymmetries ${\cal CM}$ squaring to one.

{\em Class AI$^{\mathcal{M_+}}$, $(s,t) = (0,0)$.---} 
This class has a topologically trivial bulk phase and does not allow for protected corner modes.

{\em Class BDI$^{\mathcal{M_{++}}}$, $(s,t) = (1,0)$.---}
In a mirror-symmetric corner, corner states can be chosen to have well-defined parities $\sigma_{\cal C}$ and $\sigma_{\cal M}$ with respect to the chiral operation ${\cal C}$ and mirror reflection ${\cal M}$. We use $N_{\sigma_{\cal C} \sigma_{\cal M}}$ to denote the number of corner states with the corresponding parities. No mass terms can be added that gap out states with the same parity $\sigma_{\cal C}$. Local mass terms may gap out pairs of corner states with different $\sigma_{\cal C}$, but only if they have the same value of $\sigma_{\cal M}$; corner states with different $\sigma_{\cal C}$ {\em and} different $\sigma_{\cal M}$ are protected. As a result, $N_{++} - N_{-+}$ and $N_{+-} - N_{--}$ are two independent topological invariants describing a mirror-symmetric corner. This gives rise to an extrinsic $\ZZ^2$ topological index.

Upon changing the lattice termination while preserving the global mirror reflection symmetry, {\em e.g.}, by ``glueing'' a topologically nontrivial one-dimensional chain to the crystal edges as in Fig.~\ref{fig:decoration}, a pair of corner states with the same parity $\sigma_{\cal C}$ and opposite parities $\sigma_{\cal M}$ can be added to a corner. Such changes of the lattice termination change the invariants $N_{++} - N_{-+}$ and $N_{+-} - N_{--}$, but leaves their difference $N_{++} - N_{-+} - N_{+-} + N_{--}$ unaffected. Hence, if crystals that differ only by lattice termination are considered equivalent, the relevant topological invariant is $N_{++} - N_{-+} - N_{+-} + N_{--}$, corresponding to an intrinsic $\ZZ$ topological index.

If mirror reflection symmetry is broken locally at the corner, corner states can be characterized by their parity $\sigma_{\cal C}$ only. Using $N_{\sigma_{\cal C}}$ to denote the number of corner states with parity $\sigma_{\cal C}$, $N_+ - N_-$ is a topological invariant --- corresponding to an extrinsic $\ZZ$ classification ---, which is defined modulo $2$ only if crystals that differ only lattice termination are considered equivalent.

{\em Class D$^{\mathcal{M_+}}$, $(s,t) = (2,0)$.---}
In this class zero-energy corner states can be chosen to be particle-hole symmetric ({\em i.e.}, they are Majorana states) and with well-defined parity $\sigma_{\cal M}$ under mirror reflection ${\cal M}$. We use $N_{\sigma_{\cal M}}$ to denote the number of corner states at parity $\sigma_{\cal M}$. The numbers $N_+$ and $N_-$ are defined modulo two only, since two zero modes of the same parity can be gapped out by a mirror-symmetric mass term. This gives an extrinsic $\ZZ_2^2$ topological classification of mirror-symmetric corners. 

A change of lattice termination, {\em e.g.}, by the addition of topologically nontrivial one-dimensional superconductors on mirror-related edges, adds two zero modes of opposite mirror parity to the corner. This reduces the extrinsic $\ZZ_2^2$ classification to an intrinsic $\ZZ_2$ classification in case that crystals differ only by their lattice termination are considered equivalent. Without local mirror reflection symmetry at the corner, any pair of Majorana zero modes can gap out, corresponding to a $\ZZ_2$ classification.

{\em Class DIII$^{\mathcal{M_{++}}}$, $(s,t) = (3,0)$.---}
Since particle-hole conjugation ${\cal P}$ and time-reversal ${\cal T}$ anticommute with the chiral operation ${\cal C}$ --- recall that we require that ${\cal C}^2 = 1$ --- zero-energy corner states always appear in Kramers pairs $\ket{0}$ and ${\cal T}\ket{0}$, which have opposite parities under ${\cal C}$. Since both states of such a Kramers pair have the same mirror parity $\sigma_{\cal M}$, we may characterize the corner states with the help of the number $N_{\sigma_{\cal M}}$ of Kramers pairs of corner states of mirror parity $\sigma_{\cal M}$. Mirror reflection symmetry forbids the gapping out of Kramers pairs at opposite mirror parity $\sigma_{\cal M}$, but allows two Kramers pairs at same $\sigma_{\cal M}$ to annihilate. As a result, both $N_+$ and $N_-$ are defined modulo two only, giving rise to a $\ZZ_2^2$ topological classification.

A change of lattice termination, {\em e.g.}, by the addition of topologically nontrivial one-dimensional superconductors on mirror-related edges, adds two Kramers pairs of zero-energy states of opposite mirror parity to the corner, thus reducing the extrinsic $\ZZ_2^2$ classification to an intrinsic $\ZZ_2$ classification. Without local mirror reflection symmetry at the corner, any two Kramers pairs of Majorana zero modes can gap out, corresponding to a $\ZZ_2$ classification.

{\em Class AII$^{\mathcal{M_+}}$, $(s,t) = (4,0)$.---}
This class has a topologically trivial bulk phase and does not allow for protected corner modes.

{\em Class CII$^{\mathcal{M_{++}}}$, $(s,t) = (5,0)$.---}
For Altland-Zirnbauer class CII the chiral operation ${\cal C}$ commutes with particle-conjugation ${\cal P}$ and time reversal ${\cal T}$, so that a corner hosts Kramers pairs of zero modes at the same parity $\sigma_{\cal C}$ under the chiral operation ${\cal C}$. Both states in such a Kramers pair have the same mirror parity $\sigma_{\cal M}$, which allows us to count the number $N_{\sigma_{\cal C},\sigma_{\cal M}}$ of Kramers pairs with the corresponding parities $\sigma_{\cal C}$ and $\sigma_{\cal M}$. Two Kramers pairs with opposite $\sigma_{\cal C}$ but the same $\sigma_{\cal M}$ may be gapped by a local mirror reflection-symmetric perturbation to the Hamiltonian, giving rise to integer topological invariants $N_{++} - N_{-+}$ and $N_{+-} - N_{--}$. This corresponds to a $2\ZZ^2$ extrinsic topological classification.

A change of lattice termination, {\em e.g.}, by the addition of topologically nontrivial one-dimensional chains on mirror-related edges, adds two Kramers pairs of zero modes of opposite parity $\sigma_{\cal M}$ to the corner. This leaves $N_{++} - N_{-+} - N_{+-} + N_{--}$ as the only integer invariant, corresponding to a $2\ZZ$ classification. 

Without local mirror reflection symmetry at the corner, Kramers pairs corner states are characterized by their parity $\sigma_{\cal C}$ only. The difference $N_+ - N_-$ of the number of zero-energy Kramers doublets with the corresponding parities $\sigma_{\cal C}$ is an integer topological invariant. If crystals that differ only lattice termination are considered equivalent, this difference is defined modulo two only, leading to a $\ZZ_2$ topological invariant.

{\em Classes C$^{\mathcal{M_+}}$, $(s,t) = (6,0)$, and CI$^{\mathcal{M_{++}}}$, $(s,t) = (7,0)$.---}
These classes have a topologically trivial bulk phase and do not allow for protected corner modes.

{\em Class AI$^{\mathcal{CM_-}}$, $(s,t) = (0,1)$.---}
This class has a topologically trivial bulk phase and does not allow for protected corner modes. This conclusion holds despite the presence of a mirror reflection antisymmetry ${\cal CM}$. Since ${\cal CM}$ anticommutes with the time-reversal operation, corner modes can not be simultaneously eigenstates of ${\cal CM}$ and of the time-reversal operation ${\cal T}$. Hence, corner modes appear as pairs, which can be chosen such that the two states $\ket{\pm}$ in the pair are invariant under ${\cal T}$ and ${\cal CM} \ket{\pm} = \pm i \ket{\mp}$. Then the local perturbation $\ket{+}\bra{-} + \ket{-}\bra{+}$ satisfies time-reversal symmetry and mirror reflection antisymmetry and gaps out these two corner states.

{\em Class BDI$^{\mathcal{M_{+-}}}$, $(s,t) = (1,1)$.---}
This class has a topologically trivial bulk phase. To see whether stable corner states may be induced by a suitably chosen lattice termination, we consider adding two topologically nontrivial one-dimensional chains in a symmetric fashion to two symmetry-related crystal edges, as in Fig.~\ref{fig:decoration}. The chains have zero energy end states $\ket{\rm L}$ and $\ket{\rm R} = {\cal M} \ket{\rm L}$, which may be chosen to be invariant under time reversal. Since the mirror reflection operation ${\cal M}$ anticommutes with ${\cal C}$, the states $\ket{\rm L}$ and $\ket{\rm R}$ have opposite parity under ${\cal C}$. The Hamiltonian $\ket{\rm L} \bra{\rm R} + \ket{\rm R} \bra{\rm L}$ is mirror reflection symmetric, satisfies the symmetry constraints of class BDI and gaps out the states $\ket{\rm L}$ and $\ket{\rm R}$. We conclude that no stable corner states may be induced on top of an otherwise trivial bulk by suitably adapting the lattice termination.

{\em Class D$^{\mathcal{CM_{+}}}$, $(s,t) = (2,1)$.---}
Particle-hole symmetric ({\em i.e.}, Majorana) corner states can be counted according to their parity $\sigma_{\cal CM}$ under the mirror reflection antisymmetry. Since a pair of corner states $\ket{\pm}$ with opposite parity $\sigma_{\cal CM}$ can be gapped by the local perturbation $i(\ket{+} \bra{-} - \ket{-} \bra{+})$, whereas corner states with equal parity $\sigma_{\cal CM}$ are protected by the mirror reflection antisymmetry, the difference $N_+ - N_-$ of the numbers $N_{\sigma_{\cal C}}$ of corner states with parity $\sigma_{\cal C}$ is a well-defined topological invariant. This number remains unchanged if one-dimensional topological superconductors are ``glued'' to mirror-related edges, since this procedure adds a pair of zero-energy states with opposite $\sigma_{\cal CM}$. We conclude that there is a $\ZZ$ topological classification.

If the mirror reflection symmetry is broken locally at the corner, any pair of Majorana states can be gapped out by a local perturbation, and one arrives at a $\ZZ_2$ topological classification.

{\em Class DIII$^{\mathcal{M_{-+}}}$, $(s,t) = (3,1)$.---}
Since time-reversal ${\cal T}$ anticommutes with ${\cal C}$, zero-energy corner states appear as Kramers pairs with opposite parity with respect to the chiral operation ${\cal C}$. We denote such a Kramers pair as $\ket{+}$ and $\ket{-} = {\cal T} \ket{+}$, where the sign $\pm$ refers to the ${\cal C}$-eigenvalue. Since mirror reflection ${\cal M}$ anticommutes with ${\cal C}$, these states can not be chosen to simultaneously be ${\cal M}$-eigenstates. However, multiple Kramers pairs of zero modes can always be organized in such a way that ${\cal M}$ acts within a single Kramers pair. Since ${\cal M}$ anticommutes with ${\cal C}$ and ${\cal M}^2 = 1$, one has ${\cal M}\ket{\pm} = e^{\pm i \phi} \ket{\mp}$, where we may fix the phases of the basis kets $\ket{\pm}$ such that $\phi = 0$. Whereas a single such Kramers pair is topologically protected, two Kramers pairs $\ket{\pm,1}$ and $\ket{\pm,2}$ can be gapped out by the local perturbation $i(\ket{+,1} \bra{-,2} - \ket{-,2} \bra{+,1} - \ket{+,2} \bra{-,1} + \ket{-,1} \bra{+,2})$, which obeys all relevant symmetries. We conclude that the only invariant is the parity of the number of zero-energy Kramers pairs, which gives a $\ZZ_2$ topological classification.

{\em Class AII$^{\mathcal{CM_{-}}}$, $(s,t) = (4,1)$.---}
A corner may host Kramers pairs of zero modes, which may also be chosen to have a well-defined parity under the mirror reflection antisymmetry ${\cal CM}$. Since ${\cal CM}$ anticommutes with time-reversal ${\cal T}$, the two states in the Kramers pair have opposite ${\cal CM}$-parity. Denoting the two members of a Kramers pair by $\ket{\pm}$, time-reversal symmetry forbids perturbations that have a nonzero matrix element between the states $\ket{+}$ and $\ket{-}$, whereas mirror reflection antisymmetry forbids perturbations that have nonzero matrix elements between $\ket{+}$ and $\ket{+}$ and between $\ket{-}$ and $\ket{-}$. We conclude that a single such Kramers pair is protected by the combination of time-reversal symmetry and mirror reflection antisymmetry. A pair of such Kramers pairs can, however, be gapped out: Denoting the states in the two Kramers pairs by $\ket{\pm,1}$ and $\ket{\pm,2}$, such a pair of Kramers pairs is gapped out by the local perturbation $i(\ket{+,1} \bra{-,2} - \ket{-,2} \bra{+,1} - \ket{+,2} \bra{-,1} + \ket{-,1} \bra{+,2})$. As a result, we find that this class has a $\ZZ_2$ topological index. If mirror reflection (anti)symmetry is locally broken at the crystal corner, a Kramers pair can obtain a finite energy and no protected zero-energy corner states exist.

{\em Class CII$^{\mathcal{M_{+-}}}$, $(s,t) = (5,1)$.---}
This class has a topologically trivial bulk phase and does not admit corner states. To see this, note that a mirror reflection operator with ${\cal M}^2 = 1$ represents a hermitian operator which satisfies all symmetry requirements for this class: It commutes with time-reversal ${\cal T}$ and itself, and anticommutes with particle-hole conjugation ${\cal P}$. Hence ${\cal M}$ is a valid term in the Hamiltonian, which gaps out any mirror-symmetric configuration of corner states.


{\em Class C$^{\mathcal{CM_{+}}}$, $(s,t) = (6,1)$.---}
Corner states appear as pairs related by particle-hole conjugation ${\cal P}$. Since the mirror reflection antisymmetry ${\cal CM}$ commutes with ${\cal P}$, the two states in the Kramers pair have the same mirror eigenvalue $\sigma_{\cal CM}$. Multiple Kramers pairs with the same $\sigma_{\cal CM}$ can not be gapped out by a mirror reflection-antisymmetric Hamiltonian, but Kramers pairs of opposite parity $\sigma_{\cal CM}$ may be mutually gapped out by a local mirror reflection-antisymmetric Hamiltonian. (For example, in a representation in which ${\cal P} = \sigma_2 K$ and ${\cal CM} = \tau_3$, $\tau_2$ gaps out two pairs of Kramers pairs at opposite ${\cal CM}$-parity.) We conclude that the difference $N = N_+ - N_-$ between the numbers of Kramers pairs with ${\cal CM}$-parity $\sigma_{\cal CM}$ is a well-defined topological invariant, giving a $\ZZ$ topological classification. Since Altland-Zirnbauer class C does not allow one-dimensional chains with protected zero-energy end states, this conclusion does not depend on whether freedom of lattice termination plays a role in the topological classification. No corner states can be stabilized in the absence of mirror reflection antisymmetry.

{\em Classes CI$^{\mathcal{M_{-+}}}$, $(s,t) = (7,1)$, and  AI$^{\mathcal{M_{-}}}$, $(s,t) = (0,2)$.---}
These classes have a topologically trivial bulk phase and do not allow for protected corner modes.

{\em Class BDI$^{\mathcal{M_{--}}}$, $(s,t) = (1,2)$.---}
This class has a topologically trivial bulk phase. To see whether stable corner states may be induced by a suitably chosen lattice termination, we consider adding two topologically nontrivial one-dimensional chains in a symmetric fashion to two symmetry-related crystal edges, as in Fig.~\ref{fig:decoration}. The chains have zero energy end states $\ket{\rm L}$ and $\ket{\rm R} = {\cal M} \ket{\rm L}$, which may be chosen to be invariant under time reversal and particle-hole conjugation. Since the mirror reflection operation ${\cal M}$ commutes with ${\cal C}$, the states $\ket{\rm L}$ and $\ket{\rm R}$ have equal parity $\sigma_{\cal C}$ under ${\cal C}$. Taking symmetric and antisymmetric linear combinations of the states $\ket{\rm L}$ and $\ket{\rm R}$, one obtains a corner state doublet with opposite parity under ${\cal M}$, but equal $\sigma_{\cal C}$. Multiple doublets of this type with the same $\sigma_{\cal C}$ cannot be gapped out by a local perturbation, whereas two corner state doublets with opposite $\sigma_{\cal C}$ can. Hence, $N_+ - N_-$ is a valid integer topological invariant, where $N_{\sigma_{\cal C}}$ is the number of zero-energy doublets of ${\cal C}$-parity $\sigma_{\cal C}$.

{\em Class D$^{\mathcal{M_{-}}}$, $(s,t) = (2,2)$.---}
This class has a topologically trivial bulk phase. No zero-energy corner states can be induced by a suitably chosen lattice termination. To see this, we consider a mirror reflection symmetry ${\cal M}$ that squares to one, so that ${\cal M}$ is represented by a hermitian operator. Since ${\cal M}$ anticommutes with particle-hole conjugation ${\cal P}$, ${\cal M}$ is itself a valid perturbation to the Hamiltonian which gaps out any set of corner states.

{\em Class DIII$^{\mathcal{M_{--}}}$, $(s,t) = (3,2)$.---} Since ${\cal T}^2=-1$ and ${\cal T}$ anticommutes with ${\cal C}$, corner modes consist of Kramers Majorana pairs of opposite parity under the chiral
operation ${\cal C}$. Since the product ${\cal M C}$ commutes with the
time-reversal operation ${\cal T}$ and with the chiral operation ${\cal C}$,
both states in a Kramers pair have the same ``mixed parity'' $\sigma_{\cal MC}$
under ${\cal MC}$. Two Kramers pairs of equal mixed parity $\sigma_{\cal MC}$
can not be gapped out by a mirror-symmetric perturbation, since ${\cal MC}$
anticommutes with the Hamiltonian. Two Kramers pairs of opposite mixed parity
$\sigma_{\cal MC}$ can be gapped out by a local perturbation satisfying ${\cal
T}$ and ${\cal M}$ symmetries and ${\cal C}$ antisymmetry. (For example, the two
Kramers pairs $\ket{\sigma_{\cal MC},\sigma_{\cal C}}$ with $\sigma_{\cal MC}$
and $\sigma_{\cal C} = \pm$ and $\ket{\sigma_{\cal MC},-} = {\cal
T}\ket{\sigma_{\cal MC},+}$, are gapped out by the perturbation
$\ket{+,+}\bra{-,-} + \ket{-,-}\bra{+,+} - \ket{+,-} \bra{-,+} - \ket{-,+}
\ket{+,-}$.) Denoting the number of zero-energy Kramers pairs with mixed parity
$\sigma_{\cal MC}$ by $N_{\sigma_{\cal MC}}$, we thus find that $N_+ - N_-$ is a
valid integer topological invariant. This invariant can not be changed by
changing the lattice termination, since addition of two one-dimensional
topological superconductors on mirror-related crystal edges as in
Fig.~\ref{fig:decoration} leads to the addition of two Kramers doublets with
opposite mixed parities $\sigma_{\cal MC}$. If the mirror symmetry is broken
locally at the corner, any pair of Majorana Kramers doublets can gap out, and
the $\ZZ$ topological classification is reduced to a $\ZZ_2$ classification.

{\em Class AII$^{\mathcal{M_{-}}}$, $(s,t) = (4,2)$.---}
The bulk crystalline phase is a strong topological phase, and no stable zero-energy states can be induced by a suitably chosen lattice termination in the trivial bulk phase.

{\em Class CII$^{\mathcal{M_{--}}}$, $(s,t) = (5,2)$.---}
Corners allow Kramers doublets with equal ${\cal C}$ parity $\sigma_{\cal C}$
but opposite ${\cal M}$ parity $\sigma_{\cal M}$. Two doublets at the same
${\cal C}$-parity $\sigma_{\cal C}$ can not be gapped out, but two doublets with
opposite ${\cal C}$ can. (For example, the two Kramers doublets
$\ket{\sigma_{\cal C},\sigma_{\cal M}}$ with $\sigma_{\cal C}$ and $\sigma_{\cal
M} = \pm$ and $\ket{\sigma_{\cal C},-} = {\cal T}\ket{\sigma_{\cal C,+}}$, are
gapped out by the perturbation $\ket{+,+}\bra{-,+} + \ket{-,+}\bra{+,+} +
\ket{-,-} \bra{+,-} + \ket{+,-} \ket{-,-}$.) Denoting the number of Kramers
pairs with ${\cal C}$-parity $\sigma_{\cal C}$ by $N_{\sigma_{\cal C}}$, we thus
find that $N_+ - N_-$ is a well-defined integer topological invariant.

A change of lattice termination, {\em e.g.}, by the addition of topologically nontrivial one-dimensional chains on mirror-related edges, adds two Kramers pairs of zero modes of the same parity $\sigma_{\cal C}$ to the corner. Taking symmetric and antisymmetric linear combinations these can be reorganized into two Kramers pairs $\ket{\sigma_{\cal C},\pm}$ of the type discussed above. Since changing the lattice termination allows one to change $N_+ - N_-$ by an even number, it is only the modulo two part if this invariant which is determined by the bulk band structure.
The above analysis does not change if the mirror reflection symmetry is broken locally at the corner.

{\em Class C$^{\mathcal{M_{-}}}$, $(s,t) = (6,2)$.---}
This class has a topologically trivial bulk phase and does not allow for protected corner modes.

{\em Class CI$^{\mathcal{M_{--}}}$, $(s,t) = (7,2)$.---}
Corner states appear in doublets related by particle-hole conjugation ${\cal
P}$. Such doublets have opposite parity under the chiral operation, since ${\cal
P}$ and the chiral operation ${\cal C}$ anticommute for this class. The product
${\cal M C}$ commutes with ${\cal P}$ and ${\cal C}$, so that both states in a
doublet have the same mixed parity $\sigma_{\cal MC}$ under ${\cal MC}$. Two doublets
of equal mixed parity $\sigma_{\cal MC}$ can not be gapped out by a
mirror-symmetric perturbation, since ${\cal MC}$ anticommutes with the
Hamiltonian. Two doublets of opposite mixed parity $\sigma_{\cal MC}$ can be
gapped out by a local perturbation satisfying ${\cal M}$ symmetry and ${\cal P}$
and ${\cal C}$ antisymmetries. (For example, the two doublets $\ket{\sigma_{\cal
MC},\sigma_{\cal C}}$ with $\sigma_{\cal MC}$ and $\sigma_{\cal C} = \pm$ and
$\ket{\sigma_{\cal MC},-} = {\cal P}\ket{\sigma_{\cal MC},+}$, are gapped out by
the perturbation $\ket{+,+}\bra{-,-} + \ket{-,-}\bra{+,+} + \ket{+,-} \bra{-,+}
+ \ket{-,+} \ket{+,-}$.) Denoting the number of zero-energy Kramers pairs with
mixed parity $\sigma_{\cal MC}$ by $N_{\sigma_{\cal MC}}$, we thus find that
$N_+ - N_-$ is a valid integer topological invariant. This invariant can not by
changed by changing the lattice termination, since the Altland-Zirnbauer class
CI does not allow a nontrivial one-dimensional phase with protected end states.

{\em Class AI$^{\mathcal{CM_{+}}}$, $(s,t) = (0,3)$.---}
The mirror reflection antisymmetry ${\cal CM}$ allows for the protection of corner states at a mirror-symmetric corner. Corner states can be chosen to be real and with well-defined parity $\sigma_{\cal CM}$ under mirror reflection. Corner states of equal parity can not be gapped out because of the mirror reflection antisymmetry; corner states with opposite parity can be gapped out. Hence $N = N_+ - N_-$ is an appropriate topological invariant, with $N_{\sigma_{\cal CM}}$ the number of corner states with ${\cal CM}$-parity $\sigma_{\cal CM}$.

{\em Class BDI$^{\mathcal{M_{-+}}}$, $(s,t) = (1,3)$.---}
This class has a trivial bulk phase. To see whether stable corner states may be induced by a suitably chosen lattice termination, we consider adding two topologically nontrivial one-dimensional chains in a symmetric fashion to two symmetry-related crystal edges, as in Fig.~\ref{fig:decoration}. The chains have zero energy end states $\ket{\rm L}$ and $\ket{\rm R} = {\cal M} \ket{\rm L}$, which may be chosen to be invariant under particle-hole conjugation since the mirror reflection operation ${\cal M}$ commutes with particle-hole conjugation ${\cal P}$. A pair of zero-energy states $\ket{\rm L,R}$ is protected by the combination of ${\cal P}$ antisymmetry and ${\cal M}$ symmetry. Two such doublets $\ket{{\rm L,R},1}$ and $\ket{{\rm L,R},2}$, however, can be gapped out by the local perturbation $i(\ket{{\rm L},1} \bra{{\rm R},2} - \ket{{\rm R},2} \bra{{\rm L},1} - \ket{{\rm L},2} \bra{{\rm R},1} + \ket{{\rm R},1} \bra{{\rm L},2})$, which obeys ${\cal P}$ and ${\cal C}$ antisymmetries and ${\cal M}$ symmetry. We conclude that the only invariant is the parity of the number of such zero-energy doublets, which gives a $\ZZ_2$ topological classification. If mirror reflection symmetry is broken locally at the corner, the ${\cal M}$-induced protection of a single doublet disappears, and even a single doublet of zero-energy corner states can be gapped out.

{\em Class D$^{\mathcal{CM_{-}}}$, $(s,t) = (2,3)$.---}
This class is a strong topological phase, which has doublets of chiral Majorana
modes at edges. A single chiral Majorana mode is not compatible with the
symmetries, since such mode would have to be invariant under ${\cal P}$ and ${\cal CM}$, which is not possible since ${\cal P}$ and ${\cal CM}$ anticommute.
Nevertheless, by a suitable
choice of lattice termination, a protected pair of Majorana zero modes can be
localized at a mirror-symmetric corner in the topologically trivial bulk
phase. To see this we consider adding two one-dimensional superconductors with
Majorana end states $\ket{\rm L}$ and $\ket{\rm R}$ to mirror-related
crystal edges of an otherwise topologically trivial bulk crystal, as in
Fig.~\ref{fig:decoration}. The end states $\ket{\rm L,R}$ are chosen invariant
under particle-hole conjugation ${\cal P}$. Since ${\cal CM}$ anticommutes with
${\cal P}$ we have $\ket{\rm R} = i {\cal CM}\ket{\rm L}$. A zero-energy doublet
$\ket{\rm L,R}$ is then protected by the combination of ${\cal CM}$ and ${\cal
P}$ antisymmetries. Two such doublets, however, can be gapped out by a local
perturbation, which results in a $\ZZ_2$ topological classification.

{\em Class DIII$^{\mathcal{M_{+-}}}$, $(s,t) = (3,3)$.---}
This class has a trivial bulk phase and cannot host protected corner states. (In a representation in which ${\cal M}^2 = 1$ the mirror reflection operation ${\cal M}$ is represented by a hermitian operator, which satisfies ${\cal P}$ antisymmetry and ${\cal T}$ symmetry and gaps out any set of zero-energy states localized at the corner.)

{\em Class AII$^{\mathcal{CM_{+}}}$, $(s,t) = (4,3)$.---}
This class allows mirror-protected zero-energy Kramers pairs at corners.
Since the mirror reflection antisymmetry ${\cal CM}$ commutes with the time-reversal
operator ${\cal T}$, such Kramers pairs have the same parity $\sigma_{\cal CM}$
under mirror reflection. Reflection antisymmetry protects zero-energy Kramers pairs
with equal parity $\sigma_{\cal CM}$, but allows the mutual gapping out of Kramers pairs with opposite $\sigma_{\cal CM}$. Hence, $N = N_+ - N_-$ is a valid integer topological index for this class, with $N_{\sigma_{\cal CM}}$ the number of zero-energy corner states with ${\cal CM}$ parity $\sigma_{\cal CM}$.

{\em Class CII$^{\mathcal{M_{-+}}}$, $(s,t) = (5,3)$.---}
This class has a topologically trivial bulk phase, and does not allow protected zero-energy states at corners. To see this, we consider the addition of two topologically nontrivial one-dimensional chains in a symmetric fashion to two symmetry-related crystal edges, as in Fig.~\ref{fig:decoration}. We denote the doublets at the two chains by $\ket{\rm L}$, $\ket{\rm L'} = {\cal P}\ket{\rm L}$, $\ket{\rm R} = {\cal M} \ket{\rm L}$ and $\ket{\rm R'} = {\cal P} \ket{\rm R} = {\cal P} {\cal M} \ket{\rm L}$. Since ${\cal M}$ anticommutes with ${\cal C}$, doublets at the ends of the left and right chains have opposite parity under the chiral operation ${\cal C}$. These four states can be gapped out by the perturbation $i(\ket{\rm L}\bra{\rm R'} - \ket{\rm R'} \bra{\rm L} + \ket{\rm R}\bra{\rm L'} - \ket{\rm L'}\bra{\rm R})$. 

{\em Class C$^{\mathcal{CM_{-}}}$, $(s,t) = (6,3)$.---}
This class is a strong topological phase, which has chiral modes at edges. No corner modes can be constructed in the trivial bulk phase, because the Altland-Zirnbauer class C is topologically trivial in one dimension.

{\em Class CI$^{\mathcal{M_{+-}}}$, $(s,t) = (7,3)$.---}
This class is topologically trivial and does not allow for protected zero-energy corner states.

\section{Edge-to-corner correspondence for two-dimensional mirror-symmetric crystals} \label{app:3}

A nontrivial mirror-symmetric topological crystalline bulk phase implies the existence of protected gapless states on mirror-symmetric edges. If the topological crystalline insulator or superconductor is not in a strong topological phase, these edge states can be gapped out for edges that are not invariant under the mirror operation. In that case protected zero-energy states remain at mirror-symmetric corners. The main text discusses this scenario in detail for the complex Altland-Zirnbauer classes with unitary mirror symmetries and antisymmetries. In this appendix we give details for the complex classes with antiunitary mirror symmetries and antisymmetries and for the real Altland-Zirnbauer classes. For completeness, we repeat the discussion of those mirror-symmetric topological phases that were already contained in Ref.\ \onlinecite{langbehn2017}. 

Throughout this appendix we will use $x$ as a coordinate along a mirror-symmetric edge, see Fig.\ \ref{fig:edge_deformation}a, or along an edge that is symmetrically deformed from a mirror-symmetric edge, with a mirror-symmetric corner located at $x=0$, see Fig.\ \ref{fig:edge_deformation}b. Further, $v$ is a constant with the dimension of velocity, and we use $\sigma_j$, $\tau_j$, $j=0,1,2,3$ to refer to Pauli matrices acting on different spinor degrees of freedom, and $\openone_N$ to denote the $N \times N$ unit matrix. Edge Hamiltonians are always given in the simplest possible form, after a suitable basis transformation and after rescaling of energies and coordinates.

\subsection{Complex Altland-Zirnbauer classes with antiunitary symmetries and antisymmetries}

{\em Class A$^{\mathcal{T^+M}}$, $s=0$.---} Representing ${\cal TM}$ by complex conjugation $K$, this phase allows chiral edge modes with Hamiltonian $H_{\rm edge} = -i v \openone_N \partial_x$. This is a strong topological phase, which does not allow localized zero-energy states at corners.

{\em Class AIII$^{\mathcal{T^+M_+}}$, $s=1$.---} We represent the chiral operation ${\cal C}$ using $U_{\cal C} = \sigma_3$ and the antiunitary mirror reflection operation using $U_{\cal TM} = 1$, so that the bulk Hamiltonian $H(k_x,k_y)$ satisfies the constraints $H(k_x,k_y) = -\sigma_3 H(k_x,k_y) \sigma_3 = H^*(k_x,k_y)$. A nontrivial mirror-symmetric edge is described by the edge Hamiltonian 
\begin{equation}
  H_{\rm edge} = -i v \sigma_1 \partial_x.
\end{equation}
This edge allows a unique mass term $m\sigma_2$, which is odd under ${\cal TM}$. The intersection of two mirror-related edges represents a domain wall with respect to such a mass term and hosts a protected zero-energy mode. The chiral parity $\sigma_{\cal C}$ of such a corner state depends on the sign of $m$ far away from the corner, such that $\sigma_{\cal C}$ is negative if $m(x)$ is positive for $x \to \infty$. The $\ZZ$ (extrinsic) classification of corner states follows from the observation that corner states at equal parity $\sigma_{\cal C}$ can not mutually gap out.

{\em Class A$^{\mathcal{P^+M}}$, $s=2$.---} We represent the antiunitary mirror antisymmetry ${\cal PM}$ by complex conjugation $K$, so that $H(k_x,k_y) = - H^*(k_x,-k_y)$. The edge Hamiltonian at a mirror-symmetric edge is
\begin{equation}
  H_{\rm edge} = -i v \sigma_2 \partial_x.
\end{equation}
Upon deforming the edge away symmetrically around a corner at $x=0$, two mass terms $m_1(x) \sigma_1 + m_2(x) \sigma_3$ are allowed, with $m_{1,2}(x) = -m_{1,2}(-x)$. Such a Hamiltonian hosts a zero mode symmetrically located around the corner at $x=0$. A mirror-symmetry-breaking perturbation near $x=0$ can however push this state away from zero energy.

{\em Class A$^{\mathcal{T^-M}}$, $s=2$.---} We represent ${\cal TM}$ by $\sigma_2 K$, where $K$ is complex conjugation. A mirror-symmetric edge can host multiple Kramers pairs of chiral modes, described by the edge Hamiltonian $H_{\rm edge} = -i v \sigma_0 \openone_N \partial_x$. This is a strong topological phase which does not allow for localized states at corners.

\subsection{Real classes}

{\em Class BDI$^{\mathcal{M_{++}}}$, $(s,t)=(1,0)$.---} 
We use $U_{\mathcal{T}} = 1$, $U_{\mathcal{P}} = \sigma_3$, $U_{\cal C} = \sigma_3$, and $U_{\mathcal{M}} = \sigma_3 \tau_3$ to represent time-reversal, particle-hole conjugation, chiral operation, and mirror reflection. The integer topological invariant $N$ for class BDI$^{\mathcal{M_{++}}}$ counts the difference of the number of helical edge states with positive and negative mixed parity $\sigma_{\cal MC}$ at zero energy. For a minimal mirror-invariant edge with $N \ge 0$, all edge states have the same (positive) mixed parity, so that effectively we may set $U_{\cal M} = \sigma_3$. With a suitable choice of basis and after rescaling the edge Hamiltonian takes the form
\begin{equation}
  H_{\rm edge} = -i v \sigma_2 \openone_N \partial_x.
\end{equation}
The unique mass term $m \sigma_1$, with $m$ a $N \times N$ hermitian matrix, is odd under reflection. The intersection of two mirror-related edges represents a domain wall with respect to such a mass term and hosts $N$ protected zero-energy modes. The parity $\sigma_{\cal C}$ of these modes depends on the sign of the eigenvalues of the matrix $m(x)$ away from the corner at $x=0$, such that a positive eigenvalue for $x \to \infty$ corresponds to a corner state with positive $\sigma_{\cal C}$. This reproduces the $\ZZ^2$ (extrinsic) classification of corner states derived in App.\ \ref{app:2}. 

{\em Class D$^{\mathcal{M_{+}}}$, $(s,t)=(2,0)$.---} We choose the unitary matrices $U_{\cal P} = 1$ and $U_{\cal M} = \sigma_1$ to represent particle-hole conjugation and mirror reflection. The bulk topological crystalline phase has a $\ZZ_2$ classification, for which the nontrivial phase has counterpropagating edge modes at a mirror-symmetric edge described by the edge Hamiltonian $H_{\rm edge} = - i v \sigma_3 \partial_x$. This edge Hamiltonian has a unique mass term $m \sigma_2$. The intersection between two mirror-related edges represents a domain wall and hosts a localized zero-energy state. The mirror parity $\sigma_{\cal M}$ depends on the sign of $m$ far away from the corner at $x=0$, such that a positive value of $m$ for $x \to \infty$ corresponds to a positive mirror parity $\sigma_{\cal M}$. Since no matrix elements may exist between two corner state with opposite mirror parity $\sigma_{\cal M}$, this reproduces the $\ZZ_2^2$ (extrinsic) classification of corner states derived in App.\ \ref{app:2}. 

{\em Class DIII$^{\mathcal{M_{++}}}$, $(s,t)=(3,0)$.---} We set $U_{\cal T} = \sigma_2$, $U_{\cal P} = 1$, and $U_{\cal M} = \sigma_2 \tau_2$. With a suitable choice of basis, a mirror-invariant edge in the nontrivial topological crystalline phase has a pair of counter-propagating Majorana modes with Hamiltonian
\begin{equation}
  H_{\rm edge} = -i v \sigma_3 \tau_0 \partial_x.
\end{equation}
The unique mass term  $m \sigma_1 \tau_2$ is odd under ${\cal M}$. As a result, intersection between two mirror-related edges represents a domain wall and hosts a Kramers pair localized zero-energy states. The mirror parity $\sigma_{\cal M}$ depends on the sign of $m$ far away from the corner at $x=0$. The $\ZZ_2^2$ (extrinsic) classification of corner states derived in App.\ \ref{app:2} follows upon noting that no matrix elements may exist between two corner state with opposite mirror parity $\sigma_{\cal M}$. 

{\em Class CII$^{\mathcal{M_{++}}}$, $(s,t)=(5,0)$.---} We set $U_{\cal T} = \sigma_2$, ${\cal U}_{\rm P} = \sigma_2 \tau_3$, so that $U_{\cal C} = \tau_3$. The $2 \ZZ$ bulk classification of this symmetry class counts the difference $N$ of ``edge quartets'' with positive and negative ``mixed parity'' $\sigma_{\cal CM}$. For a minimal edge all edge modes have the same mixed parity, so that effectively ${\cal M}$ may be represented by $U_{\cal M} = \tau_3$. A minimal edge has Hamiltonian
\begin{equation}
  H_{\rm edge} = -i v \partial_x \sigma_0 \tau_2 \openone_N.
\end{equation}
The unique mass term gapping out such edge modes is $m \tau_1$, with $m$ a real symmetric $N \times N$ matrix. This mass term is odd under mirror reflection, ensuring the existence of $N$ Kramers pairs of corner states at the intersection between two mirror-related edges. Both states in such a Kramers pair have the same parity $\sigma_{\cal C}$, which is determined by the sign of the eigenvalues of $m$ far away from the corner at $x=0$. This corresponds to the $2\ZZ^2$ (extrinsic) classification of corner states derived in Sec.\ \ref{app:2}. 

{\em Class D$^{\mathcal{CM_{+}}}$, $(s,t)=(2,1)$.---} We represent particle-hole conjugation ${\cal P}$ by complex conjugation and the mirror antisymmetry ${\cal CM}$ by $U_{\cal CM} = \sigma_3$. We use $N_{{\rm L}\sigma_{\cal CM}}$ and $N_{{\rm R}\sigma_{\cal CM}}$ to denote the numbers of left-moving and right-moving edge modes with mirror parity $\sigma_{\cal M}$ at zero energy, respectively. Since edge modes moving in opposite directions and with opposite mirror parity can mutually gap out, the differences $N_{{\rm R}+} - N_{{\rm L}-}$ and $N_{{\rm R}-} - N_{{\rm L}+}$ are topological invariants, giving a $\ZZ^2$ classification of edge states. The sum $N_{{\rm R}+} - N_{{\rm L}-} + N_{{\rm R}-} - N_{{\rm L}+}$ is a strong topological invariant. For a second-order topological superconductor phase, we are interested in the case $N_{{\rm R}+} - N_{{\rm L}-} + N_{{\rm R}-} - N_{{\rm R}+} = 0$, a minimal realization of which has $N_{{\rm L}-} = N_{{\rm R}-} =0$ and $N = N_{{\rm R}+} = N_{{\rm L}+}$. With a suitable choice of basis and after rescaling, the corresponding edge Hamiltonian reads
\begin{equation}
  H_{\rm edge} = -i v \tau_3 \openone_N \partial_x,
\end{equation}
where $\tau_3$ is a Pauli matrix in the left mover--right mover basis. The unique mass term $m \tau_2$ is odd under the mirror antisymmetry, so that the intersection between two mirror-related edges hosts $N$ Majorana corner states. All $N$ corner states have the same mirror parity, so that no further classification is possible. This is consistent with the $\ZZ$ (extrinsic) classification derived in App.\ \ref{app:2}. 

{\em Class DIII$^{\mathcal{M_{-+}}}$, $(s,t)=(3,1)$.---} Here we choose the representations $U_{\cal T} = \sigma_2$ and $U_{\cal P} = \sigma_1$, so that $U_{\cal C} = \sigma_3$. Although in the most general case the representation of ${\cal M}$ requires the introduction of additional spinor degrees of freedom, the generators for the nontrivial topological phases can be constructed using the simpler representation $U_{\cal M} = \sigma_1$. The two generators of the $\ZZ_2^2$ topological crystalline classification have edge Hamiltonians $H_{{\rm edge},1} = -i v \sigma_2 \partial_x$ and $H_{{\rm edge},2} = -i v \sigma_2 \tau_3 \partial_x$. The former edge Hamiltonian represents a strong topological phase and is not compatible with a second-order topological phase. The latter edge Hamiltonian has a unique mass term $m \sigma_2 \tau_2$, which is odd under mirror reflection. As a result, the intersection of two mirror-related edges hosts a Kramers pair of Majorana zero modes. 

{\em Class AII$^{\mathcal{CM_{-}}}$, $(s,t)=(4,1)$.---} 
We represent ${\cal T}$ by $\sigma_2 K$ and ${\cal CM}$ by $\sigma_3$.
The two generators of the $\ZZ_2^2$ topological crystalline classification have edge Hamiltonians $H_{{\rm edge},1} = -i v \sigma_3 \partial_x$ and $H_{{\rm edge},2} = -i v \tau_2 \sigma_0 \partial_x$, where $x$ is the coordinate along the mirror-symmetric edge and the Pauli matrix $\tau_2$ acts on a separate spinor degree of freedom. The former edge Hamiltonian $H_{{\rm edge},1}$ describes a strong phase in which the edge state is protected by time-reversal symmetry alone and can not be gapped out. The latter Hamiltonian $H_{{\rm edge},2}$ has two mass terms $m_1 \tau_1 \sigma_0 + m_2 \tau_3 \sigma_0$, which are both odd under mirror reflection. Such a Hamiltonian hosts a zero mode symmetrically located around a mirror-symmetric corner at $x=0$. A local perturbation near the corner at $x=0$ that breaks the mirror symmetry can move this state away from zero energy.

{\em Class C$^{\mathcal{CM_{+}}}$, $(s,t)=(6,1)$.---}  We set $U_{\cal P} = \sigma_2$. This phase allows a strong topological phase with doublets of particle-hole conjugated co-propagating chiral edge modes. Pairs of counterpropagating doublets are prevented from mutually gapping out if they have the same parity under ${\cal CM}$. Hence, within the relevant subspace we may represent ${\cal CM}$ by the identity, $U_{\cal CM} = 1$. The edge Hamiltonian for $N$ such pairs of counterpropagating doublets reads
\begin{equation}
  H_{\rm edge} = -i v \sigma_1 \tau_2 \partial_x \openone_N,
\end{equation}
where $\tau_2$ is a Pauli matrix acting on a different spinor degree of freedom than the $\sigma$ matrices. Upon deforming the edge away symmetrically around a corner at $x=0$, four mass terms $m_1(x) \rho_0 \sigma_2 + m_2(x) \rho_0 \sigma_3 + m_3(x) \sigma_1 \rho_1 + m_4(x) \sigma_1 \rho_3$ are allowed under a global reflection symmetry, with $m_j(x) = -m_j(-x)$, $j=1,2,3,4$. Such a Hamiltonian hosts $N$ doublets of zero modes symmetrically located around the corner at $x=0$.

{\em Class DIII$^{\mathcal{M_{--}}}$, $(s,t)=(3,2)$.---} We choose the representations $U_{\cal T} = \sigma_2$ and $U_{\cal P} = \sigma_1$, so that $U_{\cal C} = \sigma_3$. The $\ZZ$ bulk topological invariant $N$ is the difference of the numbers of helical edge doublets with positive and negative ``mixed parity'' $\sigma_{\cal MC}$. For a ``minimal'' edge with $N \ge 0$ all modes have the same (positive) mixed parity and we may represent $U_{\cal M} = \sigma_3$. 
Only topological crystalline phases with an even number $N$ of pairs of helical modes 
can be used for the construction of
a second-order topological insulator, since a single helical Majorana mode
corresponds to a strong topological phase. With a suitable rescaling and basis choice, an edge with $N$ pairs of helical modes is described by the edge Hamiltonian
\begin{equation}
  H_{\rm edge} = -i v \sigma_1 \tau_3 \openone_{N/2} \partial_x.
\end{equation}
This edge Hamiltonian has the unique mass term $m \tau_2 \sigma_1$, where $m$ is a real symmetric $N/2 \times N/2$ matrix. The mass term is odd under mirror reflection, ensuring the existence of $N/2$ Majorana Kramers pairs at a mirror-symmetric corner. 

{\em Class AII$^{\mathcal{M_{-}}}$, $(s,t)=(4,2)$.---} We represent time-reversal and mirror symmetry using $U_{\cal T} = \sigma_2$ and $U_{\cal M} = \sigma_3$, respectively. The bulk has a $\ZZ_2$ topological crystalline classification, the generator of which has edge Hamiltonian $H_{\rm edge} = -i v \sigma_3 \partial_x$, with $x$ a coordinate along a mirror symmetric edge. This is a strong topological phase.

{\em Class CII$^{\mathcal{M_{--}}}$, $(s,t)=(5,2)$.---} We choose $U_{\cal T} = \sigma_2$, ${\cal U}_{\rm P} = \sigma_2 \tau_3$, and $U_{\cal M} = \sigma_3$. With a suitable choice of basis, the nontrivial topological crystalline phase has edge Hamiltonian
\begin{equation}
  H_{\rm edge} = -i v \sigma_1 \tau_1 \partial_x.
\end{equation}
The unique mass term $m \sigma_1 \tau_2$ for this Hamiltonian is odd under mirror reflection, ensuring the existence of a Kramers pair of zero-energy states at a mirror-symmetric corner. A pair of corner states has a well defined parity $\sigma_{\cal C}$ with respect to the chiral operation ${\cal C}$, which depends on the asymptotic sign of the mass $m$ far away from the corner. Multiple corner doublets with the same $\sigma_{\cal C}$ cannot gap out, consistent with the $2\ZZ$ (extrinsic) classification of corner states derived in App.\ \ref{app:2}.

{\em Class CI$^{\mathcal{M_{--}}}$, $(s,t)=(7,2)$.---} 
We represent ${\cal T}$ by $\sigma_1 K$,
${\cal P}$ by $\sigma_2 K$, so that $U_{\cal C} = \sigma_3$.
An edge allows multiple pairs of counterpropagating states with support on orbitals with the same parity under the product ${\cal MC}$, so one may represent ${\cal M}$ by $U_{\cal M} = \sigma_3$ on a minimal edge. With a suitable basis transformation and after rescaling, an edge with $N$ such pairs of counterpropagating modes is described by the edge Hamiltonian 
\begin{equation}
  H_{\rm edge} = -i v \sigma_1 \tau_2 \partial_x \openone_N,
\end{equation}
where $\tau_2$ is a Pauli matrix acting on an additional spinor degree of
freedom. Upon deforming the edge away symmetrically
around a corner at $x=0$, three mass terms $m_1(x) \sigma_1 \tau_1 + m_2(x)
\sigma_2 +m_3(x) \sigma_1\tau_3$ are allowed under a global reflection symmetry,
with $m_i$ real symmetric matrices satisfying $m_i(x) = -m_i(-x)$, $i=1,2,3$.
Such a Hamiltonian hosts $2N$ zero-energy Majorana states symmetrically
located around the corner at $x=0$.

{\em Class AI$^{\mathcal{CM_{+}}}$, $(s,t)=(0,3)$.---} 
We represent ${\cal T}$ by complex conjugation $K$. An edge allows multiple pairs of counterpropagating states with support on orbitals with the same mirror parity, so that we may represent the mirror antisymmetry ${\cal CM}$ using $U_{\cal CM} = 1$ for a minimal edge. The corresponding edge Hamiltonian reads 
\begin{equation}
  H_{\rm edge} = -i v \tau_2 \partial_x \openone_N,
\end{equation}
where $\tau_2$ is a Pauli matrix acting on an additional spinor degree of freedom. Upon deforming the edge away symmetrically around a corner at $x=0$, two mass terms $m_1(x) \tau_1 + m_2(x) \tau_3$ are allowed under a global reflection symmetry, with $m_1$ and $m_2$ real symmetric matrices satisfying $m_1(x) = -m_1(-x)$ and $m_2(x) = -m_2(-x)$. Such a Hamiltonian hosts $N$ zero modes symmetrically located around the corner at $x=0$.

{\em Class D$^{\mathcal{CM_{-}}}$, $(s,t)=(2,3)$.---} We choose $U_{\cal P} = 1$ and $U_{\cal CM} = \sigma_2$. These symmetries allow a chiral edge Hamiltonian $H_{\rm edge} = -i v \sigma_0 \openone_N \partial_x$, with $x$ a coordinate along the edge and $\openone_N$ the $N \times N$ identity matrix. Such an edge represents a strong topological phase.

{\em Class AII$^{\mathcal{CM_{+}}}$, $(s,t)=(4,3)$.---} 
We represent ${\cal T}$ by $\sigma_2 K$.
An edge allows multiple pairs of helical modes with support on orbitals with the
same ${\cal CM}$ parity, so that we may represent ${\cal CM}$ using $U_{\cal CM} = 1$ for a minimal model. An insulator with an odd number of such helical edge modes is a strong topological insulator. With a suitable choice of basis, a ``minimal'' edge with an even number $N$ of helical modes is described by
the edge Hamiltonian 
\begin{equation}
  H_{\rm edge} = -i v \sigma_1 \tau_0 \partial_x \openone_N/2,
\end{equation}
where $\tau_0$ the $2 \times 2$ identity matrix acting an additional spinor degrees of freedom. Upon deforming the edge away symmetrically around a corner at $x=0$, two mass terms $m_1(x) \tau_2 \sigma_2 + m_2(x) \tau_2 \sigma_3$ are allowed under a global reflection symmetry, with $m_1$ and $m_2$ real symmetric $N/2
\times N/2$ matrices satisfying $m_1(x) = -m_1(-x)$ and $m_2(x) = -m_2(-x)$. Such
a Hamiltonian hosts $N/2$ Kramers pairs of zero modes symmetrically located around
the corner at $x=0$.

{\em Class C$^{\mathcal{CM_{-}}}$, $(s,t)=(6,3)$.---} We set $U_{\cal P} = \sigma_2$ and $U_{\cal CM} = \sigma_3$. An edge allows multiple pairs of chiral modes, described by the edge Hamiltonian $H_{\rm edge} = -i v \sigma_1 \openone_N \partial_x$, where $x$ is a coordinate along the edge and $\openone_N$ the $N \times N$ identity matrix. This is a strong topological phase.

\section{Surface-to-hinge correspondence with twofold rotation symmetry}
\label{app:4}

In this appendix we give details for the surface-to-hinge correspondence for topological crystalline insulators and superconductors with twofold rotation symmetry or antisymmetry and with mirror symmetry or antisymmetry, starting from a symmetry characterization of the gapless surface states on symmetry-invariant surfaces. The general idea underlying the surface-to-hinge correspondence is the same as for edge-to-corner correspondence with mirror-symmetric edges and corners, see Sec.\ \ref{sec:4b} and App.\ \ref{app:3}. The low-energy theory of the surface states is given in terms of one or multiple Dirac cones that are compatible with the non-spatial and spatial symmetries of the corresponding Shiozaki-Sato class.\cite{shiozaki2014} Tilting the surface away from the invariant direction, as in Fig.\ \ref{fig:tilted_surface}, allows for mass terms which must be odd under twofold rotation or mirror reflection --- because otherwise they would be allowed for the symmetry-invariant surface orientation. If the mass term is unique, the intersection of surfaces with opposite sign of the mass constitutes a domain wall, hosting a gapless hinge state. If the mass term is not unique, a mirror-symmetric hinge will still host a gapless mode, but there is no protection for gapless hinge modes in the rotation-symmetric case.

Throughout this appendix, $x$ and $y$ are coordinates on a (eventually tilted) rotation-invariant or mirror-symmetric surface, where the mirror reflections sends $x \to -x$, $\openone_N$ is the $N \times N$ unit matrix, and $\sigma_i$, $\tau_i$, $\rho_i$, and $\mu_i$, $i=0,1,2,3$, are Pauli matrices acting on different spinor degrees of freedom. We will restrict our discussion to symmetry classes with a nontrivial bulk topological crystalline phase, see Ref.\ \onlinecite{shiozaki2014} and Tables \ref{tab:SS2} and \ref{tab:SS3}. 

\subsection{Complex classes with antiunitary symmetries and antisymmetries}

{\em Classes A$^\mathcal{T^+R}$, $s=0$, and A$^\mathcal{P^+M}$, $s=2$.---} We choose $U_{\mathcal{TR}}=\sigma_{1}$ and $U_{\mathcal{PM}}=\sigma_3$ to represent the magnetic point group symmetry $\mathcal{TR}$ and mirror antisymmetry $\mathcal{PM}$, respectively. These symmetries can protect a single gapless surface state with a Dirac-like dispersion $H_{\rm surface} = -iv (\sigma_{1}\partial_{x} + \sigma_{2}\partial_{y})$ (with a suitable choice of basis). The unique mass term $m \sigma_3$ is odd under ${\cal TR}$ and ${\cal PM}$. A hinge at the intersection of crystal surfaces with opposite signs of $m$ host a gapless hinge mode.

{\em Classes AIII$^\mathcal{T^{+}R_{+}}$, $s=1$, and AIII$^{\mathcal{T^-M_-}}$, $s=3$.---} We choose $U_{\cal C} = \sigma_3$, $U_{\cal TR} = 1$, $U_{\cal TM} = \sigma_2 \tau_3$. The nontrivial phase hosts a pair of Dirac cones with dispersion $H_{\rm surface} = -iv \sigma_{1} (\tau_1\partial_{x} +\tau_3\partial_{y})$  (with a suitable choice of basis). There are two mass terms that may gap the Dirac cones if the surface is tilted away from the invariant direction, $m_1 \sigma_2 \tau_0 + m_2 \sigma_1 \tau_2$, where both $m_1$ and $m_2$ must change sign upon shifting to the rotated/mirror-reflected tilt direction. With two mass terms, there is a protected hinge mode at a mirror-symmetric hinge, but not generically in the presence of the twofold rotation symmetry ${\cal TR}$. 

{\em Classes AIII$^\mathcal{T^{-}R_{-}}$, $s=3$ and AIII$^\mathcal{T^{-}M_{+}}$, $s=5$.---} We use $U_{\cal C} = \sigma_3$, $U_{\cal TR} = \sigma_2$, $U_{\cal TM} = \sigma_3 \tau_2$ to represent the operations ${\cal C}$, ${\cal TR}$, and ${\cal TM}$, respectively. The twofold rotation symmetry is compatible with pairs of Dirac cones with dispersion $\propto -i v \tau_2 (\sigma_1 \partial_x \pm \sigma_2 \partial_{y})$, which defines the chirality $\pm$. The $2 \ZZ$ bulk topological crystalline index for this symmetry class counts the difference $N = N_+ - N_-$ of such Dirac cones with positive and negative chirality. For a ``minimal'' surface all surface Dirac cones have the same chirality, so that after rescaling and with suitable choice of basis the surface Hamiltonian reads $H_{\rm surface} = -i v \tau_2 (\sigma_1 \partial_x + \sigma_2 \partial_{y}) \openone_N$. Since such surface states are protected by chiral antisymmetry alone, this represents a strong topological phase.

{\em Classes AIII$^\mathcal{{T^{+}R}_{-}}$, $s=7$, and AIII$^{\mathcal{T^+M_+}}$, $s=1$.---}
\label{app:4_cc_a_4} 
Like the previous case, this is a strong phase, with gapless surface states on all surfaces. We choose $U_{\cal C} = \sigma_3$, $U_{\cal TR} = \sigma_1$ and $U_{\cal M} = 1$. The integer bulk topological crystalline index counts the difference $N = N_+-N_-$ of surface Dirac cones with dispersion $\propto -i v (\partial_x \sigma_1 \pm \partial_y \sigma_2)$. For a ``minimal'' surface all surface Dirac cones have the same chirality, so that after rescaling and with suitable choice of basis the surface Hamiltonian reads $H_{\rm surface} = -i v (\partial_x \sigma_{1} + \partial_{y}\sigma_{2}) \openone_N$. Such surface states are protected by chiral antisymmetry alone.

\subsection{Real classes}
\label{app:4_rc}

{\em Classes BDI$^{\mathcal{R}_{++}}$, $(s,t)=(1,0)$, and BDI$^{\mathcal{M_{-+}}}$, $(s,t)=(1,3)$.---} We represent time-reversal and particle-hole conjugation using $U_{\mathcal{T}}=\sigma_{0}$, $U_{\mathcal{P}}=\sigma_{3}$, $U_{\cal C} = \sigma_3$, $U_{\cal R} = \sigma_3 \rho_3$, and $U_{\cal M} = \sigma_2$. A symmetry-invariant surface may host multiple pairs of Dirac cones with dispersion $\propto -i v (\sigma_1 \tau_2 \partial_x \pm \sigma_2 \tau_0 \partial_y)$, which defines the ``mirror chirality'' $\pm$ for class BDI$^{\mathcal{M_{-+}}}$. The integer invariant $N$ counts the number of such pairs of Dirac cones, weighted by the parity under ${\cal RC}$ (for class BDI$^{\mathcal{R}_{++}}$) or by mirror chirality (for class BDI$^{\mathcal{M_{-+}}}$). A minimal surface with $N \ge 0$ has pairs of Dirac cones of the same mirror chirality or the same ${\cal RC}$-parity, so that effectively we may use $U_{\cal R} = \sigma_3$ to represent ${\cal R}$. The corresponding surface Hamiltonian is
\begin{equation}
  H_{\rm surface} = -i v (\sigma_1\tau_2\partial_{x} +\sigma_2 \tau_0 \partial_{y}) \openone_N.
\end{equation}
Two mass terms $m_1 \sigma_1 \tau_1 + m_2 \sigma_1 \tau_3$, with $m_1$ and $m_2$ $N \times N$ real symmetric matrices, are allowed upon tilting the surface away from the symmetry-invariant orientation. These mass terms are odd under ${\cal R}$ and ${\cal M}$. Since there are two such mass terms, there are no protected hinge modes for the rotation-symmetric case. However, there are protected hinge modes at mirror-symmetric hinges in the mirror-symmetric case.

{\em Classes DIII$^{\mathcal{R}_{++}}$, $(s,t)=(3,0)$, and DIII$^{\mathcal{M}_{+-}}$, $(s,t)=(3,3)$.---} We set $U_{\mathcal{T}}=\sigma_{2}$, $U_{\mathcal{P}}=\sigma_{1}$, $U_{\cal M} = \sigma_1 \tau_2$, and $U_{\cal R} = \tau_3$. Without rotation or mirror symmetry, there are protected surface states with dispersion $-i v (\sigma_1 \partial_x \pm \sigma_2 \partial_y)$, which defines the chirality $\pm$. The integer topological invariant $N$ counts the number of such surface Dirac cones, weighted by the chirality. One such Dirac cone is not compatible with ${\cal R}$ or ${\cal M}$ symmetry on a symmetry-invariant surface, but two Dirac cones with the same chirality are, the dispersion for a pair of Dirac cones being $-i v \tau_1 (\sigma_1 \partial_x \pm \sigma_2 \partial_y)$. Since they have the same chirality, such a pair of Dirac cones is protected by ${\cal T}$ and ${\cal P}$ alone. A phase with multiple such pairs of Dirac cones is a strong topological phase with gapless surface states for surfaces of arbitrary orientation.

{\em Classes CII$^{\mathcal{R}_{++}}$, $(s,t) = (5,0)$, and CII$^{\mathcal{M}_{-+}}$, $(s,t)=(5,3)$.---} We let time-reversal be represented by $U_{\mathcal{T}}=i\sigma_{2}$ and particle-hole by $U_{\mathcal{P}}=i\sigma_2\tau_{3}$, so that $U_{\cal C} = \tau_3$. We further set $U_{\cal M} = \tau_3$ and $U_{\cal R} = \tau_3 \rho_3$. A symmetry-invariant surface admits pairs of gapless surface states with Dirac-like dispersion $\propto -i(\sigma_0 \tau_2 \partial_x \pm \sigma_1 \tau_1 \partial_y)$, which defines the mirror chirality $\pm$ for class CII$^{\mathcal{M}_{-+}}$. The integer invariant $N$ counts the number of such pairs of Dirac cones, weighted by the parity under ${\cal RC}$ (for class CII$^{\mathcal{R}_{++}}$) or by mirror chirality (for class CII$^{\mathcal{M_{-+}}}$). A minimal surface with $N \ge 0$ has pairs of Dirac cones of the same mirror chirality or the same ${\cal RC}$-parity, so that effectively we may use $U_{\cal R} = \tau_3$ to represent ${\cal R}$. A single pair of Dirac cones is protected by ${\cal T}$ and ${\cal P}$ symmetry alone, corresponding to a strong topological phase with gapless surface states on all surfaces. A purely crystalline phase requires an even number $N$ of pairs of surface Dirac cones, so that the corresponding surface Hamiltonian is
\begin{equation}
  H_{\rm surface} = -i v(\tau_1 \sigma_1 \partial_x + \tau_2 \sigma_0 \partial_y) \mu_0 \openone_{N/2}.
\end{equation}
Such a Hamiltonian admits two mass terms $m_1 \tau_1 \sigma_2 \mu_2 + m_2 \tau_1 \sigma_3 \mu_2$, where $m_1$ and $m_2$ are $N/2 \times N/2$ real symmetric matrices, which change sign under mirror reflection and twofold rotation. Since there are two mass terms, the rotation-symmetric class CII$^{\mathcal{R}_{++}}$ does not have protected hinge modes, whereas the mirror-symmetric class CII$^{\mathcal{M}_{-+}}$ has protected gapless modes at mirror-symmetric hinges.

{\em Classes CI$^{\mathcal{R}_{++}}$, $(s,t)=(7,0)$, and CI$^{\mathcal{M}_{+-}}$, $(s,t) = (7,3)$.---} We set $U_{\mathcal{T}}=\sigma_{1}$ and $U_{\mathcal{P}}=\sigma_{2}$, so that $U_{\cal C} = \sigma_3$. We further set $U_{\cal R} = \tau_3$ and $U_{\cal M} = \sigma_1 \tau_3$. A symmetry-invariant surface admits surface states with dispersion $-i \tau_2 (\sigma_1 \partial_x \pm \sigma_2 \partial_y)$, where $\pm$ defines the chirality. The integer invariant counts the number of such pairs of surface Dirac cones, weighted by chirality. Since such pairs of surface Dirac cones do not rely on crystalline symmetries for their protection this is a strong phase, which has gapless surface states on surfaces of arbitrary orientation.

{\em Classes D$^{\mathcal{CR}_{+}}$, $(s,t) = (2,2)$, and D$^{\mathcal{M}_{+}}$, $(s,t) = (2,1)$.---} We set $U_{\cal P} = 1$, $U_{\cal CR} = \tau_3$, and $U_{\cal M} = \sigma_3$. A symmetry-invariant surface admits surface states with dispersion $\propto -i (\sigma_1 \partial_x \pm \sigma_3 \partial_y)$, which defines the mirror chirality $\pm$ (for class D$^{\mathcal{M}_{+}}$). This class has an integer topological invariant $N$, which counts the differences of the number of Dirac cones with positive and negative ${\cal CR}$-parity at zero energy or mirror chirality, as appropriate. On a minimal surface with $N \ge 0$ all surface Dirac cones have the same mirror chirality or ${\cal CR}$-parity, so that we may effectively represent ${\cal CR}$ by $U_{\cal CR} = 1$. The corresponding surface Hamiltonian reads
\begin{equation}
  H_{\rm surface} = -i v (\sigma_1 \partial_x + \sigma_3 \partial_y) \openone_N.
\end{equation}
Such a surface admits a unique mass term $\sigma_2 m$, with $m$ an $N \times N$ real symmetric matrix, which changes sign under mirror reflection or under the rotation antisymmetry. Correspondingly, a mirror-symmetric hinge admits gapless modes. With twofold rotation antisymmetry, gapless hinge modes are guaranteed to exist if $N$ is odd.

{\em Classes DIII$^{\mathcal{R}_{-+}}$, $(s,t) = (3,1)$, and $DIII^{\mathcal{M}_{++}}$, $(s,t) = (3,0)$.--- } We set $U_{\cal T} = \sigma_2$ and $U_{\cal P} = \sigma_1$, so that $U_{\cal C} = \sigma_3$, $U_{\cal R} = \sigma_1 \tau_3$, $U_{\cal M} = \tau_3$. The crystalline bulk phase has a $\ZZ_2$ topological classification, for which the nontrivial phase has a surface state with Hamiltonian
\begin{equation}
  H_{\rm surface} = -i v (\sigma_1 \tau_1 \partial_x + \sigma_2 \tau_0 \partial_y)
\end{equation}
at a symmetry-invariant surface. There is a unique mass term $m \sigma_1 \tau_2$, which is odd under twofold rotation and under mirror reflection. We conclude that the conditions for the existence of gapless hinge modes are met.

{\em Classes AII$^{\mathcal{CR}_{-}}$, $(s,t) = (4,1)$ and AII$^{\mathcal{M}_{+}}$, $(s,t) = (4,0)$.---} For a minimal model we choose $U_{\cal T} = \sigma_2$, $U_{\cal CR} = \sigma_1$, and $U_{\cal M} = \tau_3$. The crystalline bulk phase has a $\ZZ_2$ topological classification, for which the nontrivial phase has a surface state with Hamiltonian
\begin{equation}
  H_{\rm surface} = -i v \sigma_1 (\tau_1 \partial_x + \tau_3 \partial_y)
\end{equation}
The model admits a unique mass term $m \sigma_1 \tau_2$, which changes sign under the twofold rotation antisymmetry operation and under mirror reflection. Correspondingly, this model admits a helical gapless hinge mode.

{\em Classes C$^{\mathcal{CR}_{+}}$, $(s,t) = (6,1)$ and C$^{\mathcal{M}_{+}}$, $(s,t) = (6,0)$.---} We set $U_{\cal P} = \sigma_2$, $U_{\cal CR} = \rho_3$, and $U_{\cal M} = \tau_2 \sigma_3$. The surface admits pairs of surface states with a dispersion $-i v \tau_2 (\sigma_1 \partial_x \pm \sigma_3 \partial_y)$, which defines the mirror chirality (for class C$^{\mathcal{M}_{+}}$). The integer topological invariant $N$ for class C$^{\mathcal{CR}_{+}}$ counts the difference of the number of surface Dirac cones with $\mathcal{CR}$-eigenvalue $1$ and $-1$ on a symmetry-invariant surface, and we may use $U_{\cal CR} = 1$ to represent ${\cal CR}$ on a minimal surface with $N \ge 0$. With mirror symmetry, $N$ counts the number of pairs of surface Dirac cones weighted by mirror chirality. In both cases a minimal surface with $N \ge 0$ has Hamiltonian
\begin{equation}
  H_{\rm surface} = -i v \tau_2 (\sigma_1 \partial_x + \sigma_3 \partial_y) \openone_N.
\end{equation}
Such a surface has a unique mass term $m \tau_2 \sigma_2$, with $m$ a real symmetric $N \times N$ matrix which is odd under ${\cal CR}$ and ${\cal M}$. Correspondingly, this mirror-symmetric model admits helical gapless hinge modes at a mirror-symmetric hinge for all $N$, whereas the rotation-antisymmetric model has gapless hinge modes if $N$ is odd.

{\em Classes DIII$^{\mathcal{R}_{--}}$, $(s,t) = (3,2)$ and DIII$^{\mathcal{M}_{-+}}$, $(s,t) = (3,1)$.---} We set $U_{\cal T} = \sigma_2$, $U_{\cal P} = \sigma_1$, $U_{\cal C} = \sigma_3$, $U_{\cal R} = \sigma_3 \tau_3$, and $U_{\cal M} = \sigma_1$. These classes admit surface states with Dirac dispersion $-i v (\sigma_2 \partial_x \pm \sigma_1 \partial_y)$, which defines the chirality $\pm$. Such a surface state is compatible with ${\cal R}$ and ${\cal M}$ symmetries, but protected by chiral antisymmetry ${\cal C}$ alone. The corresponding strong integer index counts their number, weighted by chirality. A pair of surface states of opposite chirality, with dispersion $-i v (\sigma_2 \rho_3 \partial_x \pm \sigma_1 \rho_0 \partial_y)$, where the sign $\pm$ defines the mirror chirality for class DIII$^{\mathcal{M}_{-+}}$, is protected by rotation or mirror symmetry. The associated integer topological index $N$ counts the number of such pairs of surface Dirac cones, weighted by ${\cal RC}$-parity (for class DIII$^{\mathcal{R}_{--}}$) or by mirror chirality (for class DIII$^{\mathcal{M}_{-+}}$). This allows one to effectively set $U_{\cal R} = \sigma_3$ for a minimal surface with $N \ge 0$. The corresponding surface Hamiltonian reads
\begin{equation}
  H_{\rm surface} = -i v (\sigma_2 \rho_3 \partial_x + \sigma_1 \rho_0 \partial_y) \openone_N.
\end{equation}  
The surface Hamiltonian admits a unique mass term $m \sigma_2 \rho_2$, with $m$ a real symmetric $N \times N$ matrix which changes sign under the twofold rotation antisymmetry operation and under mirror reflection. Correspondingly, this mirror-symmetric model admits helical gapless hinge modes at a mirror-symmetric hinge for all $N$, whereas the rotation-antisymmetric model has gapless hinge modes if $N$ is odd.

{\em Classes AII$^{\mathcal{R}_{-}}$, $(s,t)=(4,2)$, and AII$^{\mathcal{CM}_{-}}$, $(s,t)=(4,1)$} We set $U_{\cal T} = \sigma_2$, $U_{\cal R} = \sigma_3$, and $U_{\cal CM} = \sigma_1$. These classes have a $\ZZ_2^2$ classification, with purely crystalline part $\ZZ_2$. A generator for the strong phase has a surface state with Dirac dispersion $-i v (\sigma_1 \partial_x + \sigma_2 \partial_y)$, which is protected by time-reversal symmetry alone. The generator for the purely crystalline topological phase has a pair of surface Dirac cones with surface Hamiltonian
\begin{equation}
  H_{\rm surface} = -i v (\sigma_1 \tau_0 \partial_x + \sigma_2 \tau_3 \partial_y).
\end{equation}
This surface Hamiltonian has a unique mass term $m \sigma_2 \tau_2$, which is odd under ${\cal R}$ or ${\cal M}$. We conclude that these classes admits a protected hinge mode.

{\em Classes CII$^{\mathcal{R}_{--}}$, $(s,t) = (5,2)$ and CII$^{\mathcal{M}_{+-}}$, $(s,t) = (5,1)$.---} We set $U_{\cal T} = \sigma_2$, $U_{\cal P} = \sigma_2 \tau_3$, $U_{\cal C} = \tau_3$, $U_{\cal R} = \sigma_3$ and $U_{\cal M} = \sigma_2 \tau_2$. These classes have a $\ZZ_2^2$ classification, with purely crystalline part $\ZZ_2$. On a symmetry-invariant surface, the generator for the strong phase has a pair of surface Dirac cones with dispersion $-i v \tau_1(\sigma_2 \partial_x + \sigma_1 \partial_y)$, which is compatible with ${\cal R}$ and ${\cal M}$ symmetries, but does not rely on those symmetry for its protection. The nontrivial purely crystalline phase has two pairs of surface Dirac cones with Hamiltonian
\begin{equation}
  H_{\rm surface} = -i v \tau_1(\sigma_1 \rho_0 \partial_x + \sigma_2 \rho_3 \partial_y).
\end{equation}
This surface Hamiltonian admits two mass terms $m_1 \sigma_2 \tau_1 \rho_2 + m_2 \sigma_1 \tau_2 \rho_1$, which is odd under ${\cal R}$ or ${\cal M}$. We conclude that class CII$^{\mathcal{M}_{+-}}$ admits a protected hinge mode along mirror-symmetric hinges, whereas class CII$^{\mathcal{R}_{--}}$ does not allow protected hinge modes.

{\em Classes CI$^{\mathcal{R}_{--}}$, $(s,t)=(7,2)$, and CI$^{\mathcal{M}_{-+}}$, $(s,t)=(7,1)$.---} We choose $U_{\cal T} = \sigma_1$, $U_{\cal P} = \sigma_2$, $U_{\cal C} = \sigma_3$, $U_{\cal R} = \sigma_3 \rho_3$, $U_{\cal M} = \sigma_2 \tau_2$. These classes admit pairs of surface states with dispersion $-i v \tau_2(\sigma_1 \partial_x \pm \sigma_2 \partial_y)$, which defines the chirality $\pm$. Such a surface state is compatible with ${\cal R}$ and ${\cal M}$ symmetries, but protected by chiral antisymmetry ${\cal C}$ alone. The corresponding strong integer index counts their number, weighted by chirality. Two pairs of surface states of opposite chirality, with dispersion $-i v \tau_2 (\sigma_1 \mu_3 \partial_x \pm \sigma_2 \mu_0 \partial_y)$, where the sign $\pm$ defines the mirror chirality for class CI$^{\mathcal{M}_{-+}}$, are protected by rotation or mirror symmetry. The associated integer topological index $N$ counts the number of such pairs of surface Dirac cones, weighted by ${\cal RC}$-parity (for class CI$^{\mathcal{R}_{--}}$) or by mirror chirality (for class CI$^{\mathcal{M}_{-+}}$). This allows one to effectively set $U_{\cal R} = \sigma_3$ for a minimal surface with $N \ge 0$. The corresponding surface Hamiltonian reads
\begin{equation}
  H_{\rm surface} = -i v \tau_2 (\sigma_1 \mu_3 \partial_x + \sigma_2 \mu_0 \partial_y) \openone_N.
\end{equation}  
The surface Hamiltonian admits four mass terms $m_1 \sigma_1 \tau_2 \mu_2 + m_2 \sigma_2 \tau_1 \mu_1 + m_3 \sigma_2 \tau_3 \mu_1 + m_4 \sigma_1 \tau_0 \mu_1$, with $m_1$, $m_2$, $m_3$, and $m_4$ real symmetric $N \times N$ matrices which change sign under the twofold rotation antisymmetry operation and under mirror reflection. Correspondingly, this mirror-symmetric model admits helical gapless hinge modes at a mirror-symmetric hinge for all $N$, but the rotation-symmetric model has no protected hinge states.

{\em Classes AI$^{\mathcal{CR}_{+}}$, $(s,t) = (0,3)$, and AI$^{\mathcal{M }_{-}}$, $(s,t) = (0,2)$.---} We choose $U_{\cal T} = 1$, $U_{\cal CR} = \rho_3$, and $U_{\cal M} = \sigma_2 \tau_3$ to represent time reversal, twofold rotation antisymmetry, and mirror reflection symmetry, respectively. A symmetry-invariant surface admits pairs of surface states with a dispersion $-i v \sigma_2 (\tau_1 \partial_x \pm \tau_3 \partial_y)$, which defines the mirror chirality (for class AI$^{\mathcal{M}_{-}}$). The integer topological invariant $N$ counts the number of such pairs of surface Dirac cones, weighted by $\mathcal{CR}$-parity or by mirror chirality, as appropriate. On a minimal surface with $N \ge 0$ we may use $U_{\cal CR} = 1$ to represent ${\cal CR}$. The corresponding surface Hamiltonian reads
\begin{equation}
  H_{\rm surface} = -i v \sigma_2 (\tau_1 \partial_x + \tau_3 \partial_y) \openone_N.
\end{equation}
Such a surface has three mass terms $m_1 \sigma_1 \tau_1 + m_2 \sigma_2 \tau_2 + m_3 \sigma_3 \tau_0$, with $m_1$, $m_2$, and $m_3$ real symmetric $N \times N$ matrices. Correspondingly, this mirror-symmetric model admits helical gapless hinge modes at a mirror-symmetric hinge for all $N$, but the rotation-antisymmetric model has no protected hinge states.

{\em Classes AII$^{\mathcal{CR}_{+}}$, $(s,t)=(4,3)$ and AII$^{\mathcal{M}_{-}}$, $(s,t)=(4,2)$.---} We set $U_{\cal T} = \sigma_2$, $U_{\cal CR} = \tau_3$, and $U_{\cal M} = \sigma_2$. This phase allows surface Dirac cones on symmetry-invariant surfaces with dispersion $-i v (\sigma_1 \partial_x \pm \sigma_2 \partial_y)$, which defines the mirror chirality for class AII$^{\mathcal{M}_{-}}$. The integer invariant $N$ counts the number of such surface Dirac cones, weighted by ${\cal CR}$-parity or mirror chirality, as appropriate. Odd values of $N$ correspond to strong phases, which have gapless surface states irrespective of the surface orientation. For even $N$ one has a purely crystalline phase. For a minimal model with $N \ge 0$ one may effectively use $U_{\cal R} = 1$ to represent twofold rotation. The corresponding surface Hamiltonian is
\begin{equation}
  H_{\rm surface} = -i v \rho_0 (\sigma_1 \partial_x + \sigma_2 \partial_y) \openone_{N/2}.
\end{equation}
There is a unique mass term $m \sigma_3 \rho_2$, with $m$ an $N/2 \times N/2$ matrix, which is odd under ${\cal CR}$ or ${\cal M}$. Correspondingly, a mirror-symmetric hinge has $N/2$ protected helical modes, whereas there are protected hinge modes in the presence of twofold rotation antisymmetry if $N/2$ is odd.

{\em Classes CII$^{\mathcal{R}_{-+}}$, $(s,t) = (5,3)$, and CII$^{\mathcal{M}_{--}}$, $(s,t) = (5,2)$.---} We set $U_{\mathcal{T}}=\sigma_{2}$, $U_{\mathcal{P}}=\sigma_{2} \tau_{3}$, $U_{\cal C} = \tau_3$, $U_{\cal R} = \sigma_3 \tau_1$, and $U_{\cal M} = \sigma_1$. These classes have a $\ZZ_2$ classification, for which the nontrivial phase has a pair of Dirac cones with dispersion $-i \tau_1 (\sigma_1 \partial_x + \sigma_2 \partial_y)$ on a symmetry-invariant surface. Such a pair of Dirac cones is protected by time-reversal symmetry and particle-hole antisymmetry alone, so that this is a strong topological phase, which has gapless modes on all surfaces.

{\em Classes C$^{\mathcal{CR}_{-}}$, $(s,t) = (6,3)$, and C$^{\mathcal{M}_{-}}$, $(s,t)=(6,2)$.---} We choose $U_{\cal P} = \tau_2$, $U_{\cal CR} = \tau_3$, and $U_{\cal M} = \tau_3 \sigma_3$. These classes have a $\ZZ_2$ classification, for which the nontrivial phase has a pair of Dirac cones with dispersion $-i \tau_0 (\sigma_1 \partial_x + \sigma_3 \partial_y)$ on a symmetry-invariant surface. Such a surface admits a unique mass term $m \sigma_2 \tau_0$, which is odd under ${\cal CR}$ and ${\cal M}$. We conclude that the conditions for gapless hinge modes on a mirror-symmetric hinge or with rotation-symmetric crystal termination at surfaces are met.

\bibliography{refs}

\end{document}